\documentclass[pdftex,man,a4paper]{mgragh} 
\usepackage[utf8]{inputenc}  
\usepackage[T1]{fontenc}
\usepackage{amsmath}
\usepackage{amssymb} 
\usepackage{amsthm} 
\usepackage{array} 
\usepackage[intoc]{nomencl}
\usepackage{bigfoot}
\usepackage[bookmarks=false]{hyperref}
\usepackage{cite}
\newcommand{\citep}[1]{\cite{#1}}

\begin{document}

\title{Modelling beam transport and biological effectiveness to develop treatment planning for ion beam radiotherapy}
\author{Leszek Grzanka}
\uczelniaNazwaA{The Henryk Niewodniczański}
\uczelniaNazwaB{Institute of Nuclear Physics}
\uczelniaImienia{Polish Academy of Sciences}
\rok{2013}
\promotor{prof. dr hab. Paweł Olko}
\copromotor{}

\maketitle

\newpage

\slowakluczowe{radiobiologia, radioterapia hadronowa}

\frontmatter


\vspace*{12 cm}

\hspace*{4cm}
\begin{minipage}{0.7\textwidth}
{\scriptsize \emph{Clara est, et quæ numquam marcescit sapientia, \\et facile videtur ab his qui diligunt eam, et invenitur ab his qui quærunt illam.}}

{\scriptsize \emph{Præoccupat qui se concupiscunt, ut illis se prior ostendat.}}

{\scriptsize \emph{Qui de luce vigilaverit ad illam, non laborabit: \\assidentem enim illam foribus suis inveniet.}}

{\scriptsize \emph{Cogitare ergo de illa sensus est consummatus: \\et qui vigilaverit propter illam, cito securus erit.}}

{\scriptsize \emph{Quoniam dignos se ipsa circuit quærens, \\et in viis ostendit se illis hilariter, et in omni providentia occurrit illis.}}

{\scriptsize \emph{Initium enim illius verissima est disciplinæ concupiscentia.}}

{\scriptsize \emph{Cura ergo disciplinæ, dilectio est: et dilectio, custodia legum illius est: \\custoditio autem legum, consummatio incorruptionis est:}}

{\scriptsize \emph{incorruptio autem facit esse proximum Deo.}}

{\hspace*{5cm} \em \scriptsize The Book of Wisdom, 6: 12-20}
\end{minipage}

\pagenumbering{gobble}
\clearpage

\vspace{3 cm}
\hspace{3 cm}
{\LARGE \bf Acknowledgements}
\vspace{3 cm}

During the long and winding road to complete my PhD thesis I had the chance of
meeting many inspiring persons, contributing to the presented work in many ways.

I would like to thank my Supervisors: Professor Paweł Olko and Professor Michael
Waligórski for their knowledge, experience and guidance, and the time spent on
valuable discussions and reading the many drafts of this thesis.

I am grateful to Dr. Steffen Greilich and Dr. Marta Korcyl with whom I have been
working on the libamtrack library project.
Considerable progress in this project was triggered by my cooperation with the
German Cancer Research Center (DKFZ) in Heidelberg. 
I am especially grateful to Dr. Steffen Greilich for enormous amount of support
and knowledge sharing  and also to the administrative director of DKFZ, Prof. Josef Puchta 
for financial support during my stays in Heidelberg.

This project has been supported by a three-year doctoral scholarship program
\emph{Doctus}, co-funded by the European Social Fund.
The necessary financial support was also provided by the Institute of Nuclear
Physics of the Polish Academy of Sciences (IFJ PAN). I am grateful to both these
institutions for financing my Ph.D. project.
Part of the computing within this project was performed using the Cracow Cloud
infrastructure at the IFJ PAN and at the Academic Computer Centre ACK Cyfronet
AGH.

In parallel to work on the thesis project, I had the opportunity of
participating in the TOTEM experiment at CERN, Geneva. I thank Hubert
Niewiadomski and the whole TOTEM collaboration for sharing many ideas, resources
needed to conduct scientific research and for covering expenses related to my
stays at the CERN laboratory.

Let me also express my gratitude to all my friends, especially those encountered
during the period of my PhD studies, for their support, patience and time spent
together. I also thank my family, especially my Parents. When I was a child they
helped me to develop my curiosity, which is still guiding me.

\tableofcontents

\mainmatter

\begin{streszczenie}
Radioterapia z użyciem wiązki jonów węgla jest nową techniką, mającą
szczególne zastosowanie w leczeniu radioopornych guzów o trudnej lokalizacji.
Planowanie leczenia, gdzie na podstawie zaleceń klinicznych fizyk medyczny
dokonuje optymalizacji prze\-strzen\-ne\-go rozkładu inaktywacji komórek nowotworowych
przez odpowiednie napromienienie objętości leczonej, jest jedną z podstawowych
procedur radioterapii. Zasadniczą trudnością w planowaniu leczenia wiązkami
jonowymi jest konieczność prawidłowego uwzględnienia zmian względnej
skuteczności biologicznej jonów (WSB) w obszarze poszerzonego piku Bragga. W
przypadku radioterapii jonowej, inaczej niż w radioterapii konwencjonalnej
prowadzonej wiązkami fotonów i elektronów, uzyskanie jednorodnego rozkładu dawki
w obszarze leczonym nie oznacza, że uzyskano jednorodny rozkład inaktywacji
komórek nowotworowych, z powodu zmian WSB dla dawek promieniowania
jonowego.

W ramach tej pracy opracowany został algorytm mający zastosowanie w systemach planowania
leczenia radioterapii jonami węgla. Algorytm ten składa się z części radiobiologicznej, odpowiedzialnej za 
obliczenie rozkładu dawki i inaktywacji komórek, oraz z modelu transportu wiązki jonów węgla,
opartego o symulacje metodą Monte Carlo.
Do obliczeń rozkładu inaktywacji komórek zastosowany został model struktury śladu (model Katza), 
pozwalający opracować wydajną obliczeniowo metodę przewidywania przeżywalności komórek w zadanym
mieszanym polu jonów węgla i cząstek wtórnych. Model Katza wraz z modelem wiązki zostały użyte w procedurze
optymalizacji wejściowego spektrum energii-fluencji
wiązki węglowej w taki sposób aby na wybranym obszarze otrzymać zadany rozkład
dawki lub przeżywalności.

Poprawność przewidywań przy pomocy przygotowanego algorytmu została została zweryfikowana przez porównanie z
opublikowanych danych przeżywalności komórek jajnika chomika chińskiego (CHO) naświetlanych
in-vitro wiązką jonów węgla.

Systemy planowania leczenia używane obecnie w radioterapii jonami węgla oparte są o
model efektu lokalnego (LEM). Zastosowanie modelu Katza daje możliwości porównania 
planów leczenia opracowanych przy pomocy różnych modeli.
Otwartość implementacji przygotowanych rozwiązań stwarza możliwości do rozwoju i
szerokiej współpracy z innymi grupami badawczymi pracującymi nad tematyką radioterapii jonowej.

\end{streszczenie}

\begin{abstract}
Radiation therapy with carbon ions is a novel technique of cancer radiotherapy,
applicable in particular to treating radioresistant tumours at difficult
localisations.  Therapy planning, where the medical physicist, following the
medical prescription, finds the optimum distribution of cancer cells to be
inactivated by their irradiation over the tumour volume, is a basic procedure of
cancer radiotherapy. The main difficulty encountered in therapy planning for ion
radiotherapy is to correctly account for the enhanced radiobiological
effectiveness of ions in the Spread Out Bragg Peak (SOBP) region over the
tumour volume. In this case, unlike in conventional radiotherapy with photon
beams, achieving a uniform dose distribution over the tumour volume does not
imply achieving uniform cancer cell inactivation.

In this thesis, an algorithm of the basic element (kernel) of a treatment
planning system (TPS) for carbon ion therapy is developed. The algorithm
consists of a radiobiological part which suitably corrects for the enhanced
biological effect of ion irradiation of cancer cells, and of a physical beam
transport model. In the radiobiological component, Katz’s track structure model
of cellular survival is applied, after validating its physical assumptions and
improving some aspects of this model. The Katz model offers fast and accurate
predictions of cell survival in mixed fields of the primary carbon ions and of
their secondary fragments. The physical beam model was based on available
tabularized data, prepared earlier by Monte Carlo simulations. Both components
of the developed TPS kernel are combined within an optimization tool, allowing
the entrance energy-fluence spectra of the carbon ion beam to be selected in
order to achieve a pre-assumed uniform (flat) depth-survival profile over the
SOBP region, assuring uniform cancer cell inactivation over the tumour depth.

Implementations of all the relevant codes developed in this thesis are contained
in the freely available libamtrack code library.

The developed TPS kernel is successfully benchmarked against a published
data set of CHO (Chinese Hamster Ovary) cell survival curves, after irradiation
of these cells in-vitro by carbon ion beams.

The developed 1-dimensional kernel of a carbon ion therapy planning system could
be expanded to a realistic full-dimensional system, also for proton
radiotherapy. Application of Katz’s radiobiological model in this kernel offers
an interesting alternative to the presently used ion planning systems based on
the Local Effect Model, due to the robustness and simplicity of the Katz model
and to the efficient computational techniques applied. Open-source coding and
the general availability of the libamtrack library may stimulate other research
groups to cooperate in further development of results obtained in this thesis.
\end{abstract}

\begin{wstep}
Among the three basic techniques of treating primary cancers, \textit{radiotherapy} is
applied most frequently, alone or in combination with surgery or chemotherapy.
Applying a high dose of ionizing radiation sterilises the rapidly multiplying
cancer cells and stops their further multiplication. Since healthy cells
surrounding the tumour volume will also be sterilised (or inactivated) by this
high absorbed dose, optimising radiotherapy relies on delivering the prescribed
dose precisely and uniformly to the tumour volume while sparing to the extent
possible the neighbouring healthy tissues. In modern \textit{teleradiotherapy} external
conventional beams of megavolt X-rays or electrons, generated by medical linear
accelerators, are applied to treat the tumour volume located at some depth
within the patient’s body. Careful adjustment of the beam direction, beam
collimation and accurate calculations of beam transport and dose deposition in
the patient’s body are necessary to optimally deliver the therapeutic dose.
Dedicated computer software systems, so-called \textit{Treatment Planning Systems} (TPS)
are used for this purpose. The role of the TPS is to enable the medical
physicist to properly adjust the energy, directions and collimation of the X-ray
or electron beams to achieve conformal distribution of the medically prescribed
dose to the tumour (target) volume. As input, the conventional TPS incorporates
\textit{accelerator-specific data} (reference dosimetry related to the physical
specifications of the medical accelerator, such as beam energy, beam orientation
and geometry, or beam modification by collimator settings, etc.), and
\textit{patient-specific data} (three-dimensional volume representations of patient’s
tissues and of the treated volumes, together with their local density
specifications). Patient-specific data are usually obtained from a series of
computed tomography (CT) images of the relevant part of the patient’s body,
including respective volume distributions of Hounsfield numbers which enable
locally deposited dose in these volume elements to be calculated. The TPS
incorporates a \textit{physical model of beam transport} through the patient’s body,
where, basing on the patient data, absorbed dose deposited locally in the
treated area and in the target volume can be calculated. In these calculations,
local variation of absorber density (e.g., in bone, lung or in soft tissues) and
superposition of the effect of applying the beams from several angles, can be
accounted for. Advanced therapy planning systems are also able to use “inverse
planning” techniques which seek an optimum dose distribution in the target
volume, satisfying some pre-set conditions for this optimisation.

In conventional radiotherapy where megavolt photon or electron beams are
applied, uniform distribution of dose absorbed in the tumour volume implies a
uniform level of inactivation of tumour cells in this volume. Because tumour
cells rapidly proliferate, they are often deprived of oxygen supply, making them
more radioresistant than the neighbouring healthy tissues (through the so-called
\textit{oxygen effect}). Application of fractionated schemes of conventional radiotherapy
(where, typically, some 60 Gy is delivered to the tumour volume in 30 daily
fractions of 2 Gy each) allows better sparing of the healthy tissues against
tumour cells, due to radiobiological considerations. By performing treatment
planning using several beam shaping techniques, such as multileaf collimators,
irradiation from many angles or applying intensity modulation of the photon beam
fluence (IMRT) and advanced “inverse planning” techniques, accurate conformation
of the dose delivered to the tumour (target) volume can be achieved, however
with some enhancement of dose to the neighbouring regions.

In 1946 Robert Wilson \citep{wilson1946radiological_ref71}
 drew attention to the possibility of applying beams
of protons or of heavier ions in cancer radiotherapy. There are two main
advantages to this proposal: the well-defined proton range and the dramatic
increase of dose deposited at its distal range, known as the \textit{Bragg peak}. In the
case of ions heavier than proton, apart from these two advantages, the other
important advantages are in the possibility of achieving an enhancement of the
biological effect per deposited dose, known as the enhanced \textit{Radiobiological
Effectiveness} (RBE) of such ions. Additionally, heavier ions may have the
capability of eliminating the oxygen effect, i.e. are able to effectively
sterilise, on a per dose basis, also the oxygen-deprived or radio-resistant
cancer cells. A general feature distinguishing between X-ray or electron beams
and beams of protons or of heavier ions is their different stopping power, or
Linear Energy Transfer (LET). Thus, beams of photons or electrons are “low-LET”
radiations, while beams of protons or heavier ions are called “high-LET”
radiations, to underscore the importance of ionisation density in evaluating
differences between the radiobiological properties of X-ray or electron beams
and of ion beams.

In the 1950’s the technique of culturing cell lines (\textit{in vitro})\citep{puck1956action_ref76} was developed
, which enabled detailed studies to be undertaken of the radiobiological
properties of “high-LET” ion beams, mainly in terms of the cellular survival
biological endpoint. In parallel, biophysical models of radiation action on
cells were developed. Based on microdosimetry considerations \citep{icru1983report}, 
cellular survival versus dose of low-LET (X-ray)
radiation became described by an exponential expression with terms linear and
quadratic with dose (the so-called “alpha-beta” or linear-quadratic
description), where a purely exponential survival (with an “alpha” term only)
dose dependence could be observed for some high-LET radiations. Within this
formalism, the RBE could then be introduced and defined (see Chapter 1). RBE was
found to be a complicated function of ion LET, of the survival level and of the
intrinsic radiosensitivity of the cellular system. LET alone was found not to be
a good general predictor of RBE over a range of ion species. In particular, ions
of different charges and of the same LET values showed different RBE values,
indicating the importance of track structure in these considerations. However,
representations of individual cell lines by their linear-quadratic parameters
and by their alpha/beta ratios were also found to be useful in predicting the
outcome of fractionated conventional low-LET radiotherapy and in describing
early and late effects in tissues irradiated by photon and electron beams
\citep{fowler1989linear_ref72}, \citep{fowler201021years_ref73} 
and thus became commonly accepted in clinical radiotherapy, though for
reasons other than high-LET RBE modelling considerations.

To exploit the properties of high-LET radiations in radiotherapy, beams of \textit{fast
neutrons} were developed and applied clinically in the 1960’s \citep{wambersie1994development_ref77}. 
Also, at that
time, radiotherapy using beams of heavy ions, up to xenon, began at the Bevalac
accelerator in Berkeley (USA) \citep{bevelac1980_ref75}. While these early trials were not entirely
successful, they paved the way for the development of modern ion radiotherapy,
indicating that energetic carbon ions (of some 300 MeV/amu) may be the most
convenient “heavy” ion beam for radiotherapy applications, from clinical and
radiobiological considerations. The alternative “light” ion beams for
radiotherapy are beams of protons of energy of about 260 MeV. As illustrated in
Fig. \ref{fig:intro_depthprofiles}, such energies of these ion beams, resulting in their range in water of
about 25 cm in water, allow tumour volumes to be reached in all parts of the
patient’s body, demonstrating advantageous depth-dose characteristics against
conventional megavolt X-ray radiotherapy. 

\begin{figure}[h!]
 \centering
 \includegraphics[width=8cm,bb=0 0 435 322]{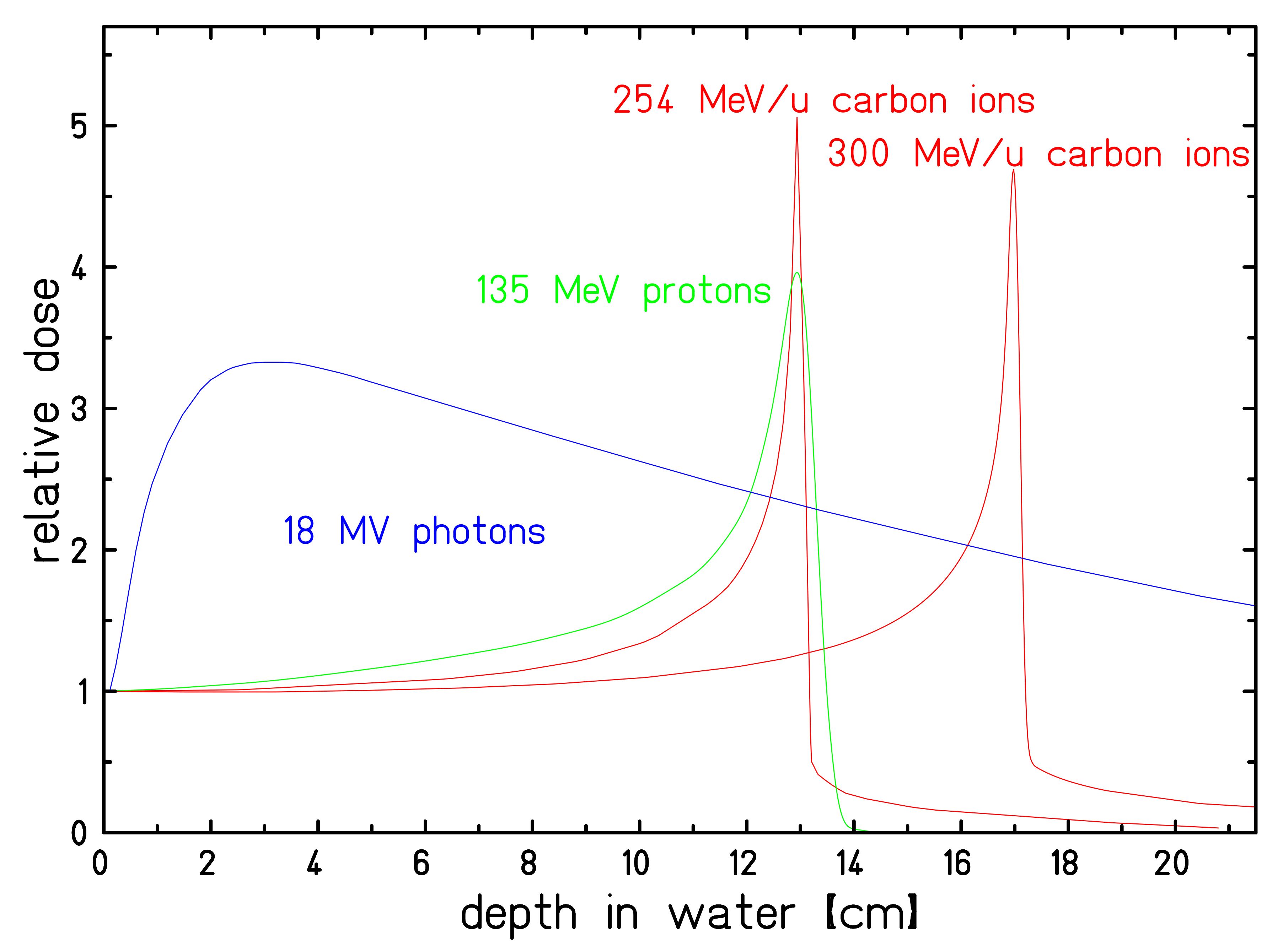}
 \caption{Relative depth-dose profiles of photons, protons and carbon ions. The
Bragg peak of proton and carbon beams is clearly visible. Note that all dose
profiles have been normalized to the same entrance dose. Reprinted from Haettner
\citep{Haettner1997}.}
 \label{fig:intro_depthprofiles}
\end{figure}

Since the width of the Bragg peak is usually much less than the size of the
tumour volume, techniques of spreading-out of the Bragg peak have to be employed
\citep{RussoPhD_ref52}. A non-trivial superposition of several Bragg peaks of beams of a range of
energies is required for this purpose, as may be seen in Fig. 2, different for
proton and carbon beams. Notable in this figure is the difference in the depth
distributions of the “physical depth dose” and of the “biological depth dose”,
which represents here a measure of cellular survival with depth and which arises
from the enhanced RBE of the ion beams, being higher for the carbon beam. This
implies that the ion beam treatment planning system has also not only to contain
the physical dose component, but also to include a radiobiology component in
order to represent the overall biological effect (e.g., via RBE) of the
spread-out ion beam composition. Notably, as seen in Fig. 2, achieving a uniform
distribution of physical dose over the tumour volume will not result in a
uniform distribution of the biological effect (inactivation) of tumour cells
over the target volume. Suitable downward correction of the depth-dose profile
of the “physical dose” is required to achieve a uniform depth distribution of
biological effect, as represented in Fig. 2. This is due to the enhanced RBE in
the distal region, especially in the case of the carbon beam. The choice and
application of a suitable radiobiological model is therefore the key element of
any clinically applicable therapy planning system for ion radiotherapy.

\begin{figure}[h!]
 \centering
 \includegraphics[width=8cm,bb=0 0 757 838]{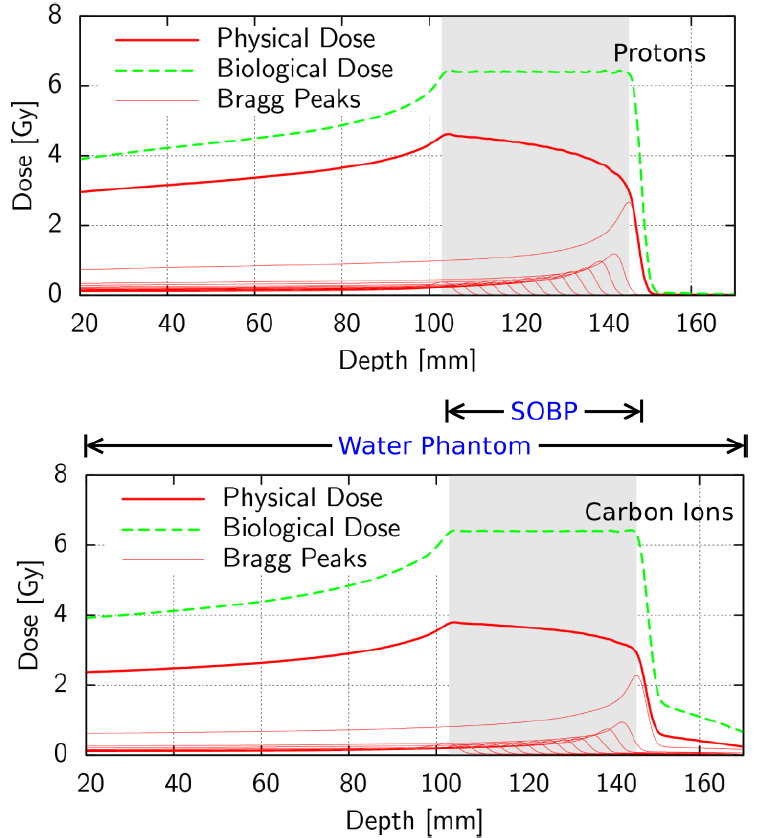}
 \caption{Illustration of depth distributions of “physical dose” (or of
depth-dose) and of “biological dose” (or of depth-cellular inactivation level)
in spread-out carbon and proton beams, in a water phantom. The spreading-out of
the Bragg peak is achieved by a suitable superposition of ion beams of different
initial energies and fluences. Reprinted from http://totlxl.to.infn.it/, Andrea
Attili, INFN-TPS project resources.}
 \label{fig:intro_carbon_proton_dose_survival}
\end{figure}

The requirements of the treatment planning system to be applied in ion
(especially carbon) beam radiotherapy are clearly much more complex than those
of the TPS applied in conventional radiotherapy. The beam transport component
must now also incorporate control of variation of the input beam energy,
required to spread out the Bragg peak region over the tumour volume.
Interactions of the beam ions with tissue lead to a complicated pattern since
the energetic carbon ions produce complex energy spectra of secondary and higher
generation ions and photons arising from nuclear fragmentation of lighter ions.
In inelastic collisions with target nuclei, carbon ions may change into lighter
fragments or may fragment nuclei of the medium, leading to production of
fragments of low energies, lighter than the original ion, traveling along the
direction of the original ion beam. Beam-produced fragments contribute to an
undesired dose in the tail region of the dose profile. A further requirement of
the beam transport component of the TPS is to calculate the locally deposited
dose as a superposition of the contributions of all these fragments, usually in
the form of a Monte Carlo calculation. An example of such a calculation for a
mono-energetic beam is shown in Fig. \ref{fig:intro_fragmentsC12}. 
In this representation, the contribution
to the dose due to lateral scattering of the primary beam is not included. Nor
is the additional complication to the ion beam transport calculation, due to the
need to calculate the spread-out dose-depth profile of the Bragg peak from a
suitable composition of ion beams of different initial energies . As a result,
the suitably optimised depth-dose distribution of the physical dose is expected,
as shown earlier in Fig. \ref{fig:intro_carbon_proton_dose_survival}.

\begin{figure}[h!]
 \centering
 \includegraphics[width=9cm,bb=0 0 131 74]{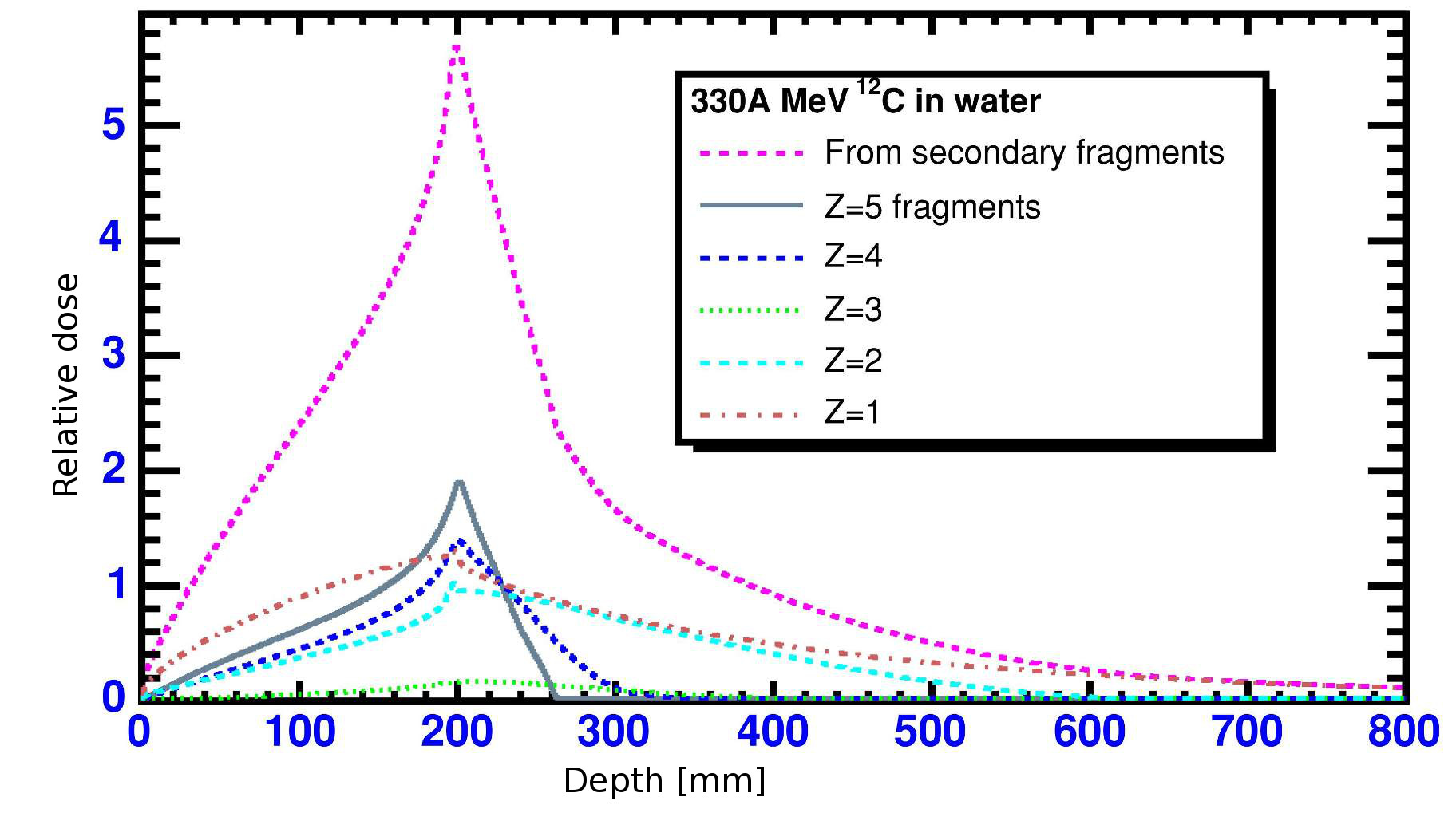}
 \caption{Monte Carlo-calculated contribution to the dose from secondary
fragments of a 330 MeV/amu carbon ion in water. The nominal range of this
monoenergetic carbon beam is 200 mm. Reprinted from Pshenichnov et al. \citep{Pshenichnov2005}}
 \label{fig:intro_fragmentsC12}
\end{figure}

Calculations of the depth-dose distributions of carbon ions can be verified by
measurements with ionization chambers, solid detectors or beam profilers.
However, cell survival (or inactivation), in carbon ion beams depends not only
on the physical dose, but also on the yield of lighter ion fragments and on
their energy spectra, which are much more difficult to determine experimentally.

The next element required in an ion beam TPS is the radiobiology component,
whereby, by using an appropriate radiobiological model, a reasonably uniform
distribution of the biological endpoint – survival or inactivation of cancer
cells in the target volume - can be achieved by suitable modification of the
physical depth-dose distribution. It is now generally accepted that in order to
incorporate the complex dependence of RBE of the different ions and on low-LET
radiation produced by the combination of ion beams of different initial
energies, it is necessary to apply the fluence approach, i.e. to base the
radiobiological calculations on the detailed knowledge of energy-fluence spectra
of all ions over the complete range of this initial ion beam combination. It is
essential to correctly calculate cell survival in the target region to ensure
that the treatment will be successful; moreover cell survival should be
estimated also outside the target region, to evaluate the survival of the
neighbouring healthy tissue cells the radiosensitivity of which may be different
from that of the tumour cells in the target volume. It is highly desirable to
seek analytically formulated radiobiology models to enable efficient
optimisation of the physical dose profile by suitable minimising computation
techniques.

An analytically simple and predictive model of ion RBE has been developed by
Robert Katz in the late 1960’s \citep{Butts1967}. The basic assumption of this track structure
model is the amorphous Radial Distribution of average Dose, $D(r)$, around the
path of a heavy ion. The average dose, due to delta-rays surrounding the ion
moving through the medium is assumed to be deposited in sensitive sites
representing radiosensitive elements of the cell, acting in a manner similar to
that after uniform irradiation of these sites by low-LET reference radiation.
The difference is in the highly non-uniform dependence (typically, $1/r^2$) of
average radial dose with radial distance from the ion’s path. The response of
the cellular system to a uniformly distributed average dose of reference
radiation is given here by the “m-target” formula (see Chapter 1) which, unlike
the linear-quadratic representation, gives a zero initial slope after low doses
of low-LET radiation. By folding the low-LET response of the cellular system
with the $D(r)$ distribution of delta-ray dose, a radial distribution of
activation probability is obtained, which, when integrated over all radii,
yields the activation probability cross-section per ion. While the shape of the
$D(r)$ is determined by the speed and the charge of the ion, the inactivation
cross section also depends on the radiosensitivity of the cell system, as
described by the m-target parameters of is response after doses of reference
radiation. Another important feature of the Katz model is in its application of
two modes of inactivation, via “ion-kill” and “gamma-kill” components of cell
inactivation probability. The model also proposes an analytical formulation of
the response of a cellular system to a superposition of a mixed radiation field
composed of different ions and of low-LET radiation. In its analytical (or
scaled) version, based on a suitable approximation of results of numerical
integrations, the model applies four parameters to describe a given cellular
system and requires the knowledge of energy-fluence spectra of the ion
irradiation to calculate the survival of this cellular system after any
composition of ions and low-LET radiation. The Katz model has not yet been
implemented in clinical treatment planning systems for ion therapy, though its
applicability to ion radiotherapy has already been demonstrated \citep{roth1980heavy_ref74}.
 The
simplicity and proven ability to predict in-vitro cell survival after heavy ion
irradiation make the Katz model a promising candidate for application in ion
radiotherapy planning systems. However, additional verification studies on the
consistency of the analytical approximations used in this model are required
\citep{KorcylPhD_ref45}, as well as further development and verification of the
mixed-field calculation which may then be applied with a more precise physical
ion beam model. The optimization algorithms used to find a beam configuration
which may deliver uniform cellular in the volume of interest, also need to be
further developed.

Currently, the only radiobiological model fully incorporated into a clinical
treatment planning system for carbon ion radiotherapy is the Local Effect Model
(LEM) \citep{Scholz1994_ref34} developed at GSI Darmstadt, Germany. To treat their patients, the
Japanese groups use an approach based on their past clinical experience with
fast neutron radiotherapy, developed at NIRS, Chiba, Japan \citep{MKM_ref50}. 
In the LEM and
Japanese approaches ion fluence is applied and the general concept of the
amorphous Radial Distribution of Dose originally proposed by Katz, however
incorporating the linear-quadratic description of cellular survival dependences
after reference (photon) radiation. LEM relates the response (cellular survival)
after ion irradiation via photon dose-response, however with additional
assumptions concerning the spatial distribution of dose. The distinct feature of
LEM is the assumption that cell survival is related to the spatial distribution
of the local lethal events resulting from the dose deposited by delta-electrons
and ions. The GSI Darmstadt group led by Gerhard Kraft has developed four
subsequent versions of the LEM model. A version of the LEM is used clinically in
the carbon ion beam TPS at the Heidelberg HIT ion radiotherapy centre.

In the Japanese treatment planning system, elements of LEM are also applied. The
former NIRS experience from fast neutron radiotherapy is used to aid the
calculation of the “biological dose”.

Although the Japanese treatment planning system has been used in the treatment
of several thousand patients and the LEM-based German planning system – in a almost 
1500 patients, both systems are still deficient in several areas: there is a lack of
consistency with experimental data for ions lighter than carbon and quite
complex computer-intensive calculations are required. Since the TPS codes of the
LEM are now unavailable due to commercial limitations, it is difficult to obtain
sufficient information on the German therapy planning system to analyse its
performance more closely.  

\section*{Aim and scope of work}
The general aim of this work was to develop and test the basic algorithms of a
kernel of a future therapy planning system (TPS) for carbon ion radiotherapy,
using in its \textit{radiobiology component} the cellular track structure model of Katz
and applying as its \textit{physical component} a realistic Monte Carlo-generated data
base describing transport in water of carbon beams of various initial energies.
Using this data set it should be possible to simulate the formation of the
spread-out Bragg peak structure and to evaluate, at all beam depths, the
energy-fluence spectra of the primary beam ions and of all generations of
secondary ions, as required by the Katz model.

It was decided that the \textit{libamtrack computer code library} would be used as the
resource for all the computer codes that had to be developed for the purposes of
this work. The libamtrack library had been co-developed earlier by the author,
in collaboration with Steffen Greilich and other colleagues at the DKFZ and
Aarhus research centres, as an open-source research tool, freely available to
all users. Codes of the libamtrack library have already been applied in
calculations of the response of alanine \citep{Hermann-libamtrack_ref41} 
and aluminium oxide \citep{Klein20111607_ref42} detector
response and in radiobiological modelling of cell survival \citep{grzanka2011application_ref43}.

The \textit{physical component} to be used in this work was a Monte Carlo
(SHIELD-HIT) generated  data base describing transport in water of carbon beams
of various initial energies, developed by Pablo Botas and available to
the author. A suitable averaging algorithm to generate the energy-fluence
spectra of all ions in the beam (primary and secondary) at the required depths
would need to be developed. Next, a method for modelling the depth-dose profile
in the spread-out Bragg peak would also need to be devised.

The Cellular Track Structure model developed by Katz \citep{Butts1967}, \citep{KatzParamCollection_ref47},
\citep{KatzTracks_ref48}, \citep{roth1980heavy_ref74} would be
applied as the \textit{radiobiological component} of the TPS kernel. Use of this model
would provide the author with an alternative to the Local Effect Model (LEM)
developed by Scholz and Kraft \citep{Scholz1994_ref34}, \citep{Scholz_ref62}
 and used in the carbon ion therapy
planning system (TPS) applied clinically at GSI Darmstadt and at the HIT
facility at Heidelberg. A method of calculating a suitable adjustment of the
depth-dose profile in order to obtain, at a given level of cell survival, a flat
survival vs. depth dependence over a given depth range, would need to be found.
The author had earlier collaborated with Marta Korcyl \citep{KorcylPhD_ref45} on developing
elements of the Katz model \citep{grzanka2011application_ref43}.
 
Results of a radiobiology experiment where Chinese Hamster Ovary (CHO) cells
were irradiated by a set of carbon beams  to verify the Spread Out Bragg Peak
(SOBP) calculations and the LEM-based TPS approach, are available \citep{mitaroff1998biological_ref66}.
 This offered the possibility of verifying the author’s method of calculating the
SOBP, based on the carbon beam data set and of the results of cell survival
calculations using the Katz model, against results of this experiment.

Developing a Katz model-based kernel of a carbon ion TPS would offer the
possibility of comparing this approach with systems based on other
radiobiological models and to suggest future directions for further work in this
area.
\end{wstep}

\chapter{The physics and radiobiology of track structure}
In this chapter the basic physical and radiobiology concepts required in track
structure modelling of ion beam radiotherapy are introduced. 
In the physics part, stopping power, Linear Energy Transfer, dose, fluence
and the Bragg peak, related to the passage of an ion through the medium
(typically water), are discussed. In the radiobiology part, description of
cellular survival after doses of reference radiation ($\gamma$-rays) by
linear-quadratic or m-target formulae, and Relative Biological Effectiveness are
introduced, followed by a brief presentation of the Local Effect Model (LEM). 

\section{Particle Track Physics}

Energetic ions are able to ionise the medium through which they pass. This may
occur \emph{directly} (by charged particles, such as electrons or ions) or \emph{indirectly}
(by neutrons or photons). Neutral particles, through their interaction with
orbital electrons or atomic nuclei of the medium, produce charged particles
which then ionise the medium in direct processes. We will focus here mainly on
ionisation and excitation processes due to the passage of energetic carbon ions
through water.

As ionising particles travel through the medium they interact with atoms or
higher structures of the medium (such as molecules), producing ionisations and
excitations. Ions of energies ranging between a few to a few hundred MeV
interact mainly by Coulomb interaction with electrons of the outer shells of
atoms of the medium. For ions and media relevant to radiotherapy, direct nuclear
interactions with nuclei of the atoms of the medium can be neglected, as they do
not contribute much to the total dose nor do they affect the range of the
primary ions. However, nuclear interactions are responsible for fragmentation of
the primary beam ions. This process cannot be neglected, as secondary and higher
generations of secondary fragments contribute to the total dose deposited in the
medium and to the effective range of the radiotherapy beam.

\begin{figure}[h!]
 \centering
 \includegraphics[width=0.8\textwidth]{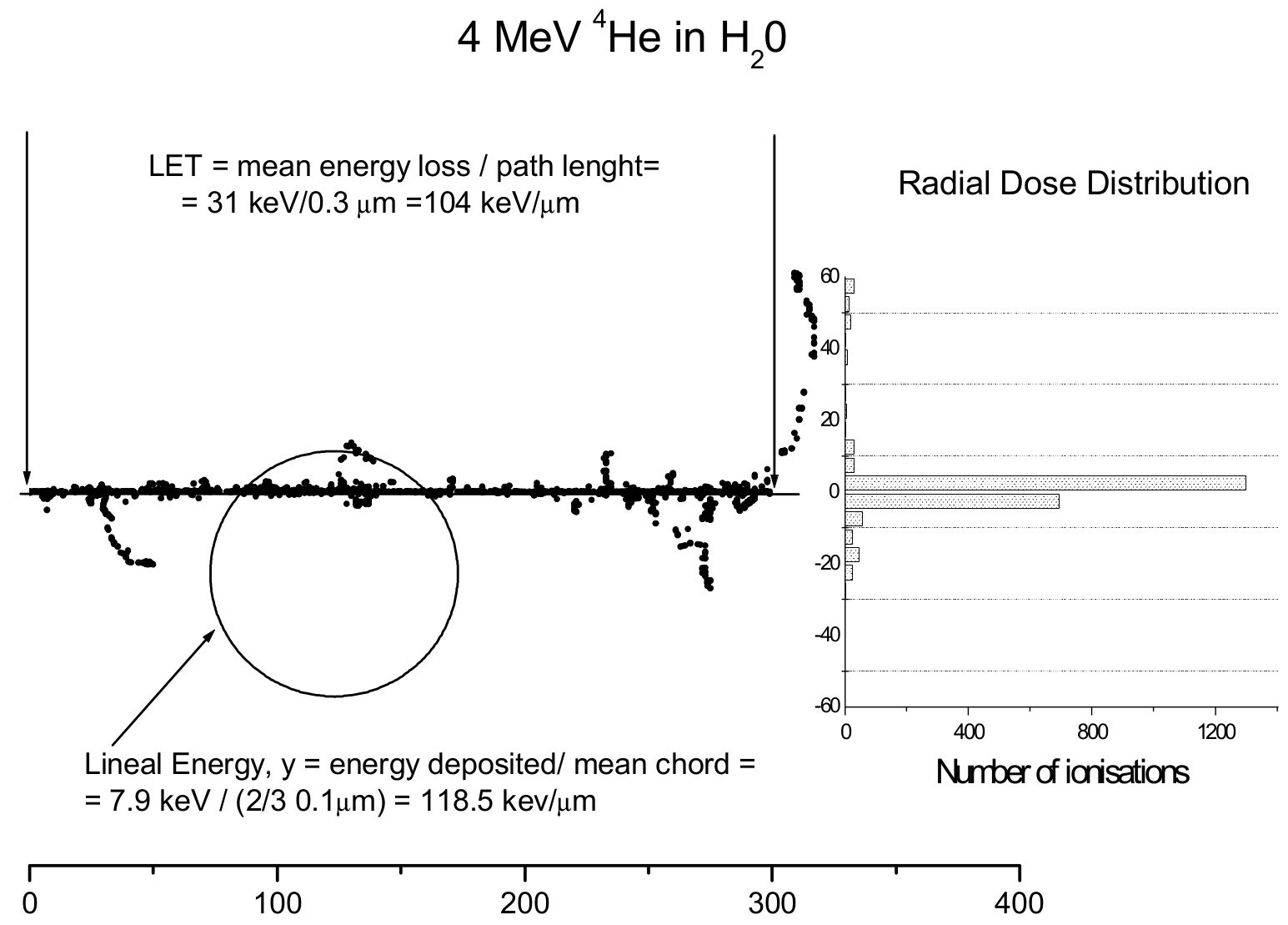}
 \caption{A two-dimensional representation of a 3-D track-segment of a 4-MeV
$\alpha$-particle in H$_2$O (water vapour normalised to density of 1 g/cm$^3$), simulated
using the MOCA-14 Monte Carlo track structure code \citep{paretzke1987radiation_ref79}.  Dots
denote the positions of individual ionisations.  The value of Linear Energy
Transfer, LET (Eq. \ref{eq:ch1_let}) is calculated as the mean energy imparted to water by
the 4 MeV $^4$He ions (31 keV) divided by the length of the illustrated path length
(0.3 $\mu$m).  The insert on the right shows the radial histogram of ionisations,
corresponding to the radial dose distribution , $D(r)$. Figure adapted from
Goodhead \citep{goodhead1987relationship_ref80}, and reprinted from Olko \citep{Olko2002_ref69}.}
 \label{fig:ch1_track}
\end{figure}

The principal concept of an \emph{ion track} is illustrated in Figure \ref{fig:ch1_track}. 
The illustrated passage of an $\alpha$-particle of energy 4 MeV over a distance of 300
nm of water will not change much its energy over that distance. The trail (or
the sequence of coordinates) of excitation and ionisation events which the ion
creates along its path as it passes through the medium, is called a \textit{track} of
this ion. A \textit{track segment} is the part of an ion track where the energy $E$ of the
ion changes only by an incremental value $dE$ (i.e. is practically unchanged). The
gradual loss of the ion's energy as it transverses the medium can then be
represented by a sequence of track segments with gradually decreasing energies,
by $dE$, as applied, e.g., in the \textit{Continuous Slowing Down Approximation} (CSDA).

An energetic ion will knock out electrons from the atoms of the medium. The
angular distribution of these electrons exhibits a maximum at about 90 degrees to the
ion’s path. These electrons may have sufficient energies to travel noticeable
distances from the ion and also to further ionise. Such electrons are called
\textit{delta-electrons} ($\delta$-electrons) or \textit{delta-rays} ($\delta$-rays). Ionisations due to $\delta$-rays
form a cloud of excitations and ionisations around the ion’s path, very dense at
small radii and of decreasing density at larger radial distances from the ion’s
path. Thus, the ion track appears to be composed of excitation and ionisation
events occurring along the path of the ion and due to the passage of
delta-rays. 

The maximum energy, $\omega_{\max}$ , transferred to the $\delta$-electron by an 
ion of speed $\beta =v/c$  in an elastic collision is given by

\begin{equation}
 \omega_{\max}  = 2 m_e c^2 \beta^2 \gamma^2
\label{eq:ch1_omegamax}
\end{equation}

where $m_e$ is rest mass of the electron and $\gamma = \frac{1}{\sqrt{ 1 - \beta^2}}$

As the energy of the primary ion gradually decreases, so does the maximum energy
of the emitted $\delta$-rays, which limits the radial “thickness” of an ion track at
the end of the ion's range. This effect is called track “thindown”.

\subsection{Stopping power and Linear Energy Transfer}

The linear rate of energy loss, $dE$, by a charged particle to atomic electrons of
the medium per unit path length of the particle, $dl$, $-dE/dl$, (commonly in units
of MeV/cm or keV/$\mu$m) is called the stopping power of the medium for the
particle. It reflects the energy lost by the particle to the medium it
transverses. According to the physical process in which energy is lost and
transferred to the medium, electronic and nuclear stopping power are
distinguished, as defined more precisely in ICRU Report 73 \citep{icru73stopping}.

\begin{figure}[h!]
 \centering
 \includegraphics[width=0.8\textwidth]{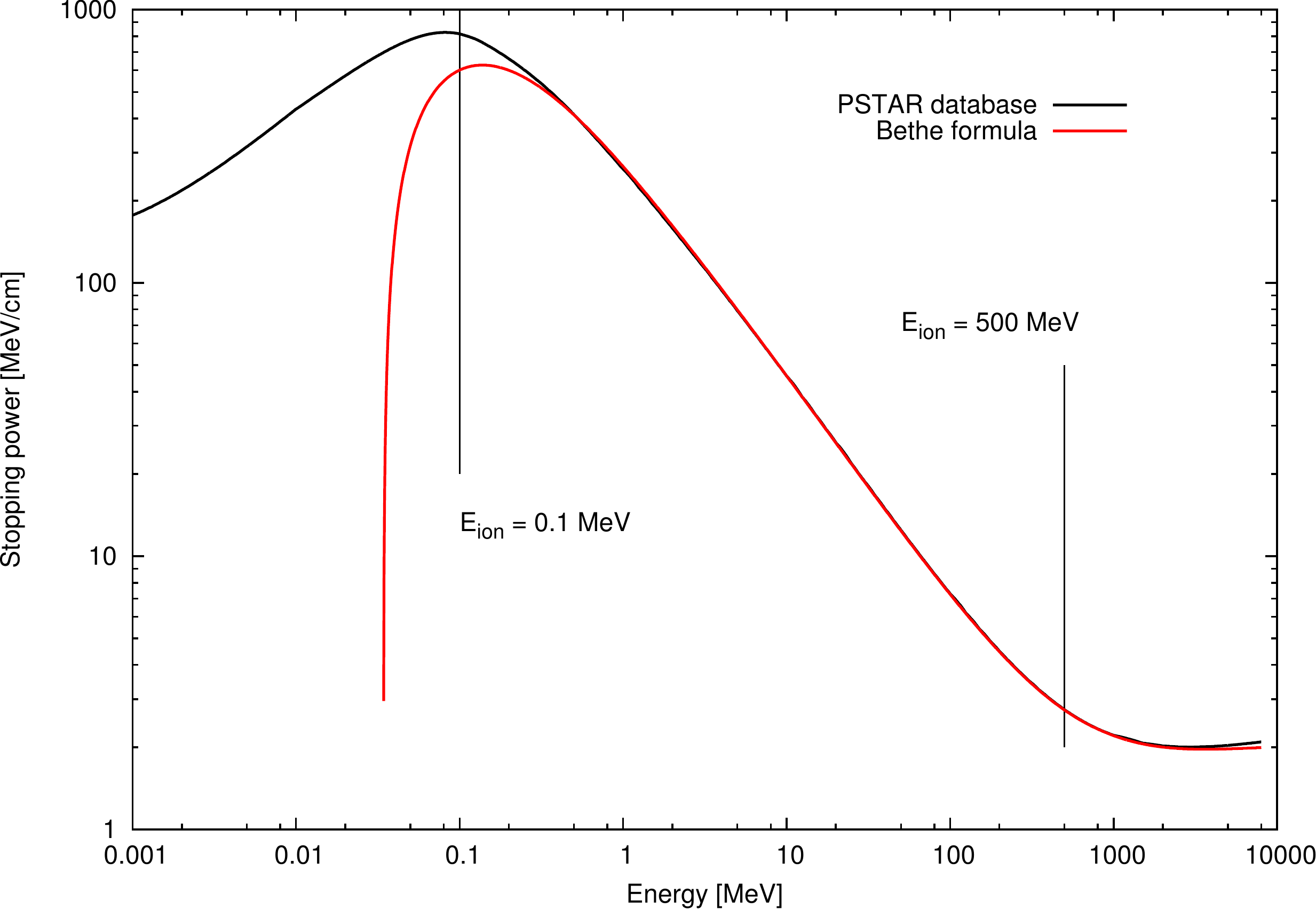}
 \caption{Stopping power of protons in liquid water based on PSTAR
database \citep{NISTESTAR} and on the Bethe formula \ref{eq:ch1_bethe}, 
implemented in the libamtrack library. The range of ion energies (0.1 – 500 MeV) 
relevant to ion radiotherapy is shown by vertical lines}
\label{fig:ch1_stop_power}
\end{figure}

The Bethe formula \citep{bethe1932}, derived by applying quantum mechanics to describe the
stopping power of a heavy charged particle, is as follows: 

\begin{equation}
- \frac{dE}{dl}
= \frac{4\pi}{m_e c^2}
\frac{n z^2}{\beta^2}
  \left(\frac{e^2}{4 \pi \varepsilon_0}\right)^2 
   \left(\frac{1}{2} \ln{2m_ec^2 \beta^2\gamma^2 T_{\rm max}\over I^2} - \beta^2  - \frac{\delta(\beta\gamma)}{2}\right)
\label{eq:ch1_bethe}
\end{equation}

where $n$ is the electron density of the medium, $e$ is the elementary electric
charge, $\varepsilon_0$ is the permittivity of vacuum, $\delta(\beta\gamma)$
is a density effect correction and:

\begin{equation}
 T_{\rm max} = \frac{2 m_e c^2 \beta^2 \gamma^2}{ 1 + 2 \gamma \frac{m_e}{M} + \frac{m_e^2}{M^2}}
\end{equation}

which reduces to $\omega_{\max}$  (equation \ref{eq:ch1_omegamax}) 
if  $m_e \ll M$ (here $M$ is the mass of the incident particle and $m_e$ the mass of the electron). 
For ions of energies
relevant to radiotherapy (0.1 - 500 MeV/amu), the effect of the density effect
correction may be neglected. The stopping power of protons in liquid water
calculated using the Bethe formula and that extracted from the PSTAR database are
shown in Fig. \ref{fig:ch1_stop_power}.

The PSTAR database published by NIST \citep{NISTESTAR} combines experimentally evaluated
values of stopping power of protons in a several media. As can be seen from the
plot of proton stopping power calculated in water using Bethe’s formula and from
the PSTAR database, at energies below 1 MeV results of calculations using
Bethe’s formula deviate from experimental values. Bethe’s formula is valid down
to proton relative velocity $\beta \approx 0.05$. Below this value there is no reliable
theory, so one may use experimental data fits \citep{andersenZiegler_ref29} to this formula \citep{Passage_ref28}. The
PSTAR database contains values of proton stopping power published in the
ICRU-49 report \citep{icru49protons_ref30} (at low energies, below 0.5 MeV) combined with the
Bethe-Bloch formula (at energies above 0.5 MeV). Other sources of stopping power
data are the ICRU-73 report \citep{icru73stopping} (stopping power of ions heavier than helium),
and tables published by Janni \citep{Janni_ref25}.

The PSTAR database was chosen by the author as the source of stopping power for
calculations, due to its completeness and to comply with ICRU-49
recommendations \citep{icru49protons_ref30}. Moreover, as the Bethe formula does not agree
with experimental data for ions of energies below 1MeV, range estimates
calculated with CSDA (Continuous Slowing Down Approximation) and the Bethe-Bloch
formula will not agree with those calculated using data from the PSTAR database.

To distinguish between the energy transferred from a charged particle to the
medium and the energy actually absorbed by the medium (i.e. absorbed dose),
Linear Energy Transfer (LET), is defined as the average energy locally imparted
to the medium by a charged particle, $dE_i$ traversing a distance $dl$ in the medium
(to denote LET, $L$ will be used in this work):

\begin{equation}
 L = \frac{dE_i}{dl}
 \label{eq:ch1_let}
\end{equation}

Not all energy lost by a particle is transferred to the medium (e.g. radiative
losses, or bremsstrahlung) and some energy lost by the projectile at one
location may be imparted to the medium elsewhere (e.g. carried by long range
delta electrons or neutrons). To focus only on the energy deposited in the
medium in the vicinity of the particle’s track, the concept of \textit{restricted linear
electronic stopping power}, LET$_{\Delta}$ , is introduced, also referred to as restricted
linear energy transfer (LET$_{\Delta}$) which denotes the energy loss $dE_{\Delta}$ due to
electronic collisions minus the kinetic energies of delta electrons of energy
larger than $\Delta$, per unit path length $dl$ (again, here we shall use $L_{\Delta}$ to represent
LET$_{\Delta}$ ): 

\begin{equation}
 L_{\Delta} = \frac{dE_{\Delta}}{dl}
 \label{eq:ch1_letdelta}
\end{equation}

In this work we shall assume that over the energy range of carbon (and of
lighter ions) relevant to radiotherapy, LET of these ions does not significantly
differ from their stopping power. We will use both concepts interchangeably
(i.e., if in the definition of $LET_{\Delta}$ all the energy deposited in the medium is
accounted for, LET = LET$_{\infty}$ = dE$_i$/dl ). In what follows, $L$ will denote LET$_{\infty}$,
i.e. unrestricted linear energy transfer.

For ions of energies below a few MeV, partial ``screening'' of the ion charge (or
``electron pickup'') occurs, which effectively diminishes its charge. Following
other authors the ``effective charge'', calculated using the formula of Barkas \citep{barkas1963nuclear}
will be used:

\begin{equation}
 z^{\star} = Z \left( 1 - \exp\left(-125 \beta Z^{-\frac{2}{3}}\right) \right)
\label{eq:ch1_zeff}
\end{equation}

where $Z$ the is atomic number of the ion.

One may calculate the linear energy transfer of an ion of charge $Z$ and energy $E$
(in MeV/amu) from that of the proton of the same energy by the following
relationship involving the ratio of the effective charges of both ions, $z^\star$: 

\begin{equation}
 \frac{L_{1}}{L_{2}} = \left( \frac{z^\star_1}{z^\star_2} \right)^2
 \label{eq:ch1_zeff_let}
\end{equation}

As we shall see later in this Chapter, the biological effectiveness of energetic
ions strongly depends on their LET, generally increasing as LET increases. For
this reason, energetic ions are often termed ``high-LET'' radiation while sparsely
ionising photons (X-rays or $\gamma$-rays) are termed ``low-LET'' radiation. However,
while LET is indeed related to track structure, LET alone is not a good
predictor of the biological effects of irradiation by different types of ions of
different energies, if track structure considerations are not included.

\subsection{Dose and fluence}

Particle fluence and dose are basic quantities describing beam intensity and
energy deposited by ions of the beam in the medium. ICRU Report 85 \citep{icru85units}
defines particle fluence $F$ as the ratio of number $dN$ of particles incident on a
sphere with cross-sectional area $dA$:

\begin{equation}
 F = \frac{dN}{dA}
 \label{eq:ch1_fluence}
\end{equation}

In this work, particle fluence $F$ will be interpreted as the number $N$ of beam
particles (all travelling in parallel along a given direction) incident on
planar area $A$, perpendicular to the beam direction, i.e.  $F = N/A$.  This
interpretation is close to the concept of vector fluence, defined in ICRU Report
85 \citep{icru85units}.

Absorbed dose $D$ is defined as the mean energy $dE$ delivered by ionizing
radiation to a small but finite volume of matter of mass $dm$.

\begin{equation}
 D = \frac{dE}{dm}
 \label{eq:ch1_dose}
\end{equation}

The unit of absorbed dose in the SI system is the gray (Gy) which is equal to
one joule per one kilogram. Considering our interpretation of fluence, both
quantities are related by the following formula:

\begin{equation}
D = \frac{1}{\rho}  F  L 
\label{eq:ch1_dose_let}
\end{equation}

where $\rho$ is the density of medium. More details on fundamental quantities and
definitions can be found in ICRU Report 85 \citep{icru85units}.

\subsection{Scattering and energy straggling}

The Bethe formula assumes that the primary beam is composed of monoenergetic
ions, on their entrance to the medium and at all depths.  In reality, the energy
loss of a beam of charged particles fluctuates around a mean value. This effect,
observed particularly in the passage of particles through thin absorbers, is
called energy straggling. The probability distribution of energy loss is
described by the Landau-Vavilov formula \citep{Landau1944_ref32}, \citep{Vavilov1957_ref33}, 
while for thick absorbers the mean energy loss is closer to values obtained with Bethe’s formula.

In addition, the ion undergoes frequent collisions with electrons of the medium,
changing its direction by relatively small angles. This results in a broader
spectrum of ion energies with increasing depth. Ion scattering can be described
by the scattering power $T$ (or mass scattering power, $T/\rho$) defined as the mean
square angle d$\theta^2$ of scattering per unit length of the medium traversed, $dz$. 
The process of multiple scattering is described by Moliere’s theory \citep{Moliere_ref31}.

The processes of energy loss straggling and of multiple scattering lead to a
gradual broadening of the primary ion energy spectrum with increasing depth in
the absorbing medium.

\subsection{The Bragg peak}

Linear energy transfer of ions in a given medium depends on their energy, as
seen in Fig \ref{fig:ch1_stop_power}. In the case
of an ion beam there is much less scatter by outer-shell electrons of the
medium (due to the high ratio of the ion and electron masses). Scattering of
ions in the beam occurs rather via the electrostatic field of the nuclei of the
medium and depends on their impact parameters. Since the dimensions of the
atomic nuclei are much smaller than those of the atomic volumes occupied by
their electron structures, ion-nucleus scattering does not contribute
significantly to all possible processes in which ions interact with the medium.
Thus, the depth-dose profile of the ion beam consists of a low and flat dose
region at lower depths, and then of a distinct peak at a depth close to the end
of their range - the Bragg peak. Finally a ''tail'' appears which consists mainly
of the dose delivered by fragmentation products which, arising at some depths,
will travel further than the primary ions. The characteristic depth-dose
dependence of an ion beam is thus explained by the low value of LET of ions at
their initially high energy while the occurrence of the Bragg peak follows from
the rapid increase of stopping power at low ion energies, as the ions slow down
in the medium to below a few MeV/amu. In ion radiotherapy proton beams of
initial energies between 60 MeV and 250 MeV and carbon ion beams of initial
energies between 150 MeV/amu to 400 MeV/amu are applied.  Heavier ions of higher
initial energies and higher charges and of higher ranges and stopping powers are
not considered to be suitable for radiotherapy.

\begin{figure}[h!]
 \centering
 \includegraphics[width=0.8\textwidth]{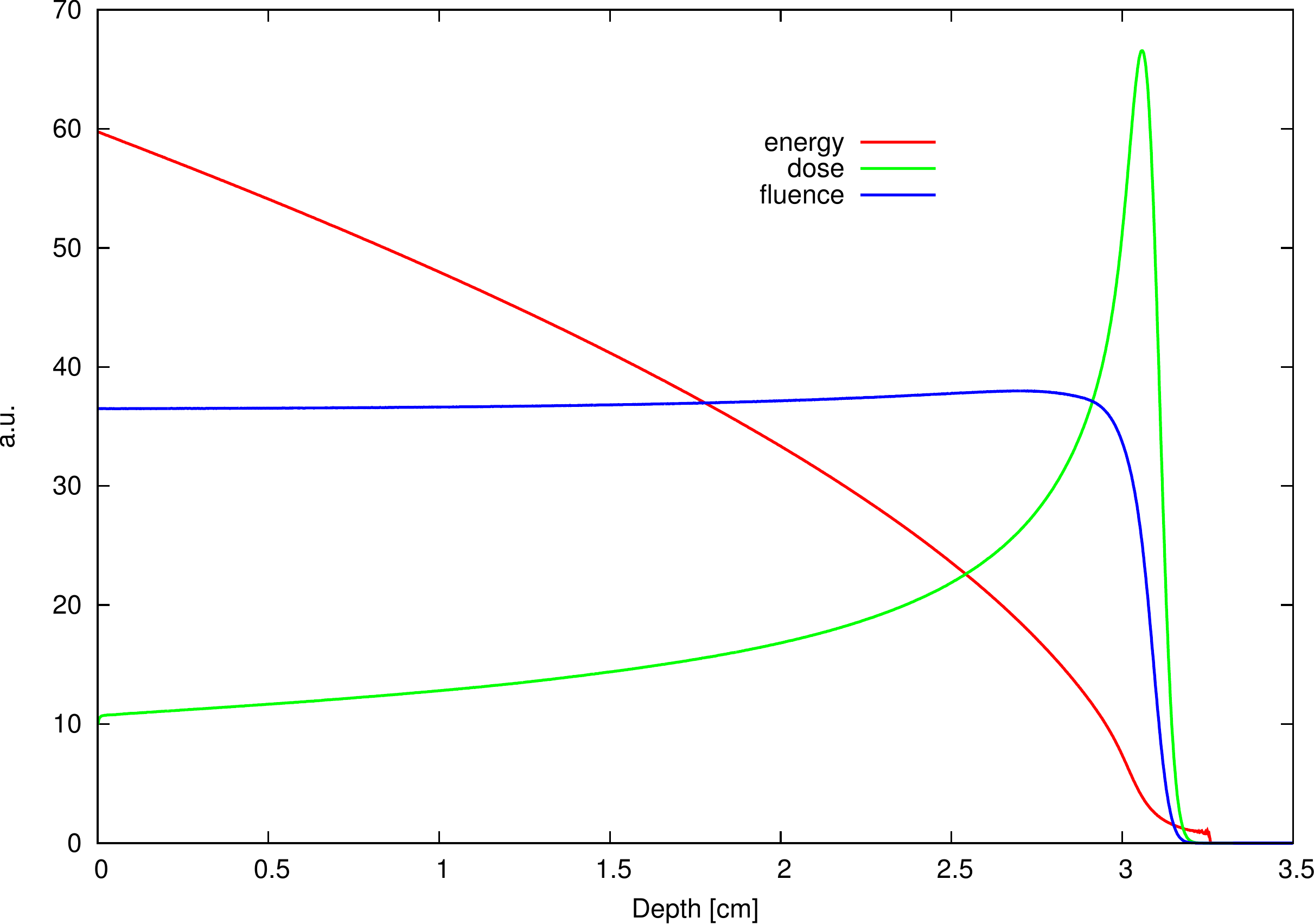}
 \caption{Depth-dose dependence, mean energy vs. depth, and fluence vs. depth of
a pristine (non-modulated) proton beam. The entrance values (at 0 cm depth) 
are: energy:  $E$ = 59.75 MeV; dose: $D$ = 1.727 Gy; and fluence: $F$ = $10^9$ 
ions/cm$^2$. Calculations made by the author, using the Geant4 package and the
libamtrack library. }
 \label{fig:ch1_bragg}
\end{figure}

For illustration, in Fig. \ref{fig:ch1_bragg} results are shown of the author’s calculations of
the depth-dose dependence, mean energy of ions in the beam vs. depth, and beam
fluence vs. depth, of a pristine (non-modulated) proton beam of initial energy
about 60 MeV. This represents the proton beam used for ocular radiotherapy at the
Institute of Nuclear Physics of the Polish Academy of Sciences (IFJ PAN) in
Krakow. As no fragmentation of protons occurs in this case, the general trends
are easier to follow than in the case of a carbon radiotherapy beam, which is
discussed in Chapter 4 and where the respective depth dependences are shown in
Figs. 4.1 - 4.3.

\section{Radiobiology and biophysical models of cellular survival}

The mechanisms of interaction of ionizing radiation with living matter which
radiobiology is concerned with, have been studied for several decades, but at
the level of whole organisms, such as man, they are still largely unknown. At
the cellular and sub-cellular organisation levels, of interest in this work is
cell survival as the biological endpoint relevant to radiotherapy, where the
objective is to inactivate, or kill (i.e., to stop proliferation of) the tumour
cells by doses of ionising radiation while maintaining the normal life processes
in neighbouring normal tissues. The mammalian cell is the basic autonomous
component of any mammalian organism, such as man. An eukaryotic cell consists of
the cell nucleus containing DNA as the basic cell replication matrix, and
cytoplasm with organelles. The cytoplasm and organelles are considered to be
rather insensitive to doses of ionising radiation. It is generally considered
that ionising radiation affects (via production of biologically active free
radicals) mainly the cell nucleus, leading to changes in the amino-acid sequence
of the DNA double helix. It is possible to verify experimentally that ionising
radiation may damage one of the two DNA strands the other strand remaining
intact (a single strand break, SSB), or cause a double strand break (DSB) which
is believed to be primarily responsible for cell death. Some of the
radiation-induced changes (or deletions) in the DNA, especially DSB, may not be
repaired by the many repair processes available to the cell. As a consequence,
the cell may lose its capability to proliferate (i.e. will not survive) or may
develop mutations, some of which may turn out to be lethal or cause cancer in
later generations of daughter cells. According to these concepts, cell damage
after a dose of ionising radiation may be classified as being lethal (i.e.
leading directly to cell inactivation), or sub-lethal , i.e. repairable shortly
after irradiation, unless additional sub-lethal damage occurs which may
eventually lead to cell death or proliferation of mutated cells.

A more extensive presentation of topics related to radiobiology can be found in
the textbook of Hall and Giaccia \citep{hall2006radiobiology}.

\subsection{Cell survival and Relative Biological Effectiveness}

\begin{figure}[h!]
 \centering
 \includegraphics[width=0.8\textwidth]{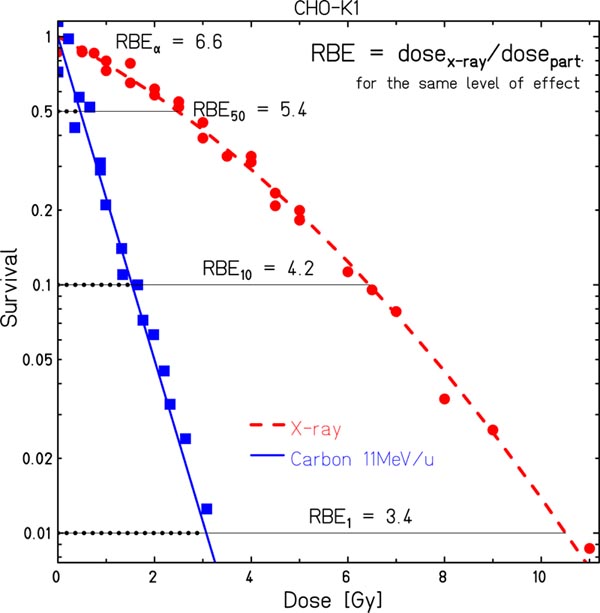}
 \caption{Survival of CHO cells irradiated with 1 MeV/u carbon ions or 250 kVp
X-rays (reprinted from \citep{Weyrather_ref51}). 
For the carbon data, the best fitted value is $\alpha$ = 1.387 Gy$^{-1}$. 
For the X-ray data points, the best-fitted parameters of the LQ formula are: 
$\alpha$ = 0.227 Gy$^{-1}$, $\beta$ =  0.017 Gy$^{-2}$ (linear fit) or 
$\alpha$ = 0.224 Gy$^{-1}$ and, $\beta$ = 0.0185 Gy$^{-2}$ (fit to logarithms of data points). 
The best fitted parameters of the m-target formula to the X-ray data points are:  
m=2.31 and $D_0$=1.69 Gy 
(from a linear fit to a larger set of data, see also Fig. \ref{fig:ch3_fit} and chapter \ref{seq:ch3_fitting}).}
 \label{fig:ch1_survival}
\end{figure}

A frequently studied biological endpoint applied in biophysical modelling is
cell survival \textit{in vitro} after irradiation with different types of ionising
radiation. The experimental technique relies on culturing a known number of
cells in glass or plastic vessels (Petri dishes, hence the term \textit{in vitro} -in
glass), irradiating them and counting the number of cells which have survived
and are able to multiply into cell colonies on the Petri dishes, after a fixed
number of cell multiplication cycles.  Cells that do not proliferate following
their irradiation are not able to form such colonies. Cellular survival, $S$, is
defined as the fraction of irradiated cells that are able to proliferate (i.e.
to form colonies) after irradiation (the probability of cell inactivation or
cell killing is then $1-S$). The relationship between cell survival and absorbed
dose $D$ is usually presented on a semi-logarithmic survival plot (linear scale
for dose and logarithmic for survival). An example of a survival plot is shown
in Fig. \ref{fig:ch1_survival}. Apparent is the difference between cell survival after doses of
reference radiation (X-rays) and after doses of 1 MeV/amu carbon ions, where the
probability of cell killing after a dose of this  high-LET radiation is much
higher than that after the same dose of the low-LET reference radiation.

Two types of expressions are most frequently used to represent cellular survival
after doses of reference radiation: the linear-quadratic and the $m$-target (or,
more precisely,the 1-hit $m$-target) formulae.

The Linear-Quadratic (LQ) \citep{douglas1976} formula is given by the equation:

\begin{equation}
 S(D) = \exp( - \alpha D - \beta D^2 )
 \label{eq:ch1_lq}
\end{equation}

where $\alpha$ (in units of 1/Gy) and $\beta$ (in units of 1/Gy$^2$) are constants best-fitted
to the experimental data points.

In a biological interpretation of these constants, $\alpha$ is believed to represent
the probability, per unit dose, of creating double strand breaks (DSB) after the
passage of a single track (e.g., the passage of a single ion), while $\beta$ could
represent the probability of DSB via two independent tracks or ionising events.
From the LQ formula a characteristic parameter $\alpha$/$\beta$ can be derived. It is equal
to the dose at which both components (linear and quadratic) are equal. In the
example shown in Fig. 1.4 cellular survival after X-rays is characterised by
both linear and quadratic components, while survival after carbon ions is purely
exponential, interpreted as resulting from each ion track passage through the
cell nucleus causing a lethal DSB.

The multi-target formula \citep{mtarget_ref54} is given by the equation:

\begin{equation}
 S(D) = 1 - \left( 1 - e^{- \frac{D}{D_0}} \right)^m
 \label{eq:ch1_mtarget}
\end{equation}

where $D_0$ (in units of Gy) and $m$ are constants best fitted to the experimental
data points.
The biological interpretation of this formula is based on the assumption that
each cell nucleus contains one or more ($m$) ``1-hit targets''. A single, or more,
``hits'' (i.e. energy deposition events) to each such target will inactivate it,
leading to cell inactivation after all m targets in the cell nucleus have been
``hit''. Simple biological systems, such as enzymes or viruses are well described
by the 1-hit, 1-target model (one or more ``hits'' to this target will inactivate
the enzyme or virus) while mammalian cells are considered to have $m$ such 1-hit
targets.  Curvature of the survival curve indicates that $m$ is larger than 1
while an exponential dependence indicates that $m$ = 1 (i.e., a ``1-hit, 1-target``
system). In 1-hit systems, $D_0$ represents the characteristic dose (or
radiosensitivity of the system), at which survival is equal to exp(-1), i.e. to
about 0.37. Non-integer values of m can be fitted to experimental data points.
This is interpreted as being an average over a large number of cells, each of
which has a different (integer) number of targets. In the case of $m = 1$
(1-target, 1-hit system) formula \ref{eq:ch1_mtarget} reduces to the purely exponential form
$S(D) = \exp(-D/D_0)$ which is the same as that of the LQ model (eq. \ref{eq:ch1_lq}) with $\beta =
0$ and $\alpha = 1/D_0$.

Relative Biological Effectiveness (RBE) has been introduced to quantify the
difference between survival curves after doses of reference radiation (usually
Co-60 $\gamma$-rays or 250 kVp X-rays) and doses of the tested radiation (e.g. of
high-LET radiation, such as heavy ions).  At a pre-selected level of survival
(or of another biological endpoint), RBE is defined as a ratio of the dose of
reference radiation and the dose of the tested radiation, required to reach that
level of survival (or of another biological endpoint):

\begin{equation}
{\rm RBE} |_S = \frac{D_{\rm ref}}{D_{\rm test}}
\label{eq:ch1_rbe} 
\end{equation}

where $S$ is the selected level of survival (or of another biological endpoint),
$D_{\rm ref}$ is the dose of reference radiation and $D_{\rm test}$ is the dose of the tested
radiation. As may be seen in Fig. \ref{fig:ch1_survival}, the value of RBE depends on the level of
survival chosen. In general, it depends in a complicated manner on many factors,
such as level of survival, dose rate, type of ion, its energy, its LET, etc.

\subsection{Biophysical models of cellular survival}

\subsubsection{Track Structure Theory (the Katz Model)}

Track Structure Theory, developed by Katz and co-workers around 1970, is based
on the concept of radial dose distribution. The model describes and predicts
cellular survival after irradiation with mixed beams of ions and photons and is
able to predict RBE dependences. Its principal assumption is that the energy
deposition within the ion track is entirely described by the radial dose
distribution due to $\delta$-rays and that at a given local dose, the same response is
observed after photons and $\delta$-rays surrounding the ion track. Thus, knowing the
photon dose response of the cellular system and using a suitable description of
track structure, it is possible to compute cell survival after ion irradiation. 
Several radiobiological models have since then been based on the concept of
radial dose distribution first introduced by Katz. A detailed description of
Katz’s cellular Track Structure Theory will be given in Chapter 3.

\subsubsection{The Local Effect Model}

The Local Effect Model (LEM) was developed around 1990 by Scholz and Kraft \citep{Scholz1994_ref34}
at GSI, Darmstadt as a tool for the then newly constructed experimental heavy
ion radiotherapy facility. The main goal of LEM was to efficiently calculate
cell survival and other biological endpoints in mixed heavy ion fields.
The response (survival) of a particular cell line after photon irradiation, $S_{\gamma}$
is parameterized by the generalized linear-quadratic equation:

\begin{equation}
S_{\gamma} (D) =  \begin{cases}
e^{-\alpha D - \beta D^2} \quad & 0 \leq D < D_t \\
e^{\beta D_t^2 - (\alpha + 2 \beta D_t ) D} \quad & D \geq D_t
\end{cases}
\label{eq:linquad}
\end{equation}

where $D_t$ is the dose at which a transition between the quadratic and linear
parts occurs. Following the track segment approach it is assumed that the
irradiated volume is thin enough to be able to perform two-dimensional
calculations using $x$ and $y$ coordinates to describe the spatial distribution of
local dose, $D(x,y)$. After a given set of energy deposition events $D(x,y)$ (due
to photon or ion irradiation) the number of lethal events, $N(x,y)$ is defined
as:

\begin{equation}
 N(x, y) = - \ln(S_{\gamma}( D(x,y) ))
\label{eq:ch1_lemN}
\end{equation}

\begin{figure}[h!]
 \centering
 \includegraphics[width=0.6\textwidth]{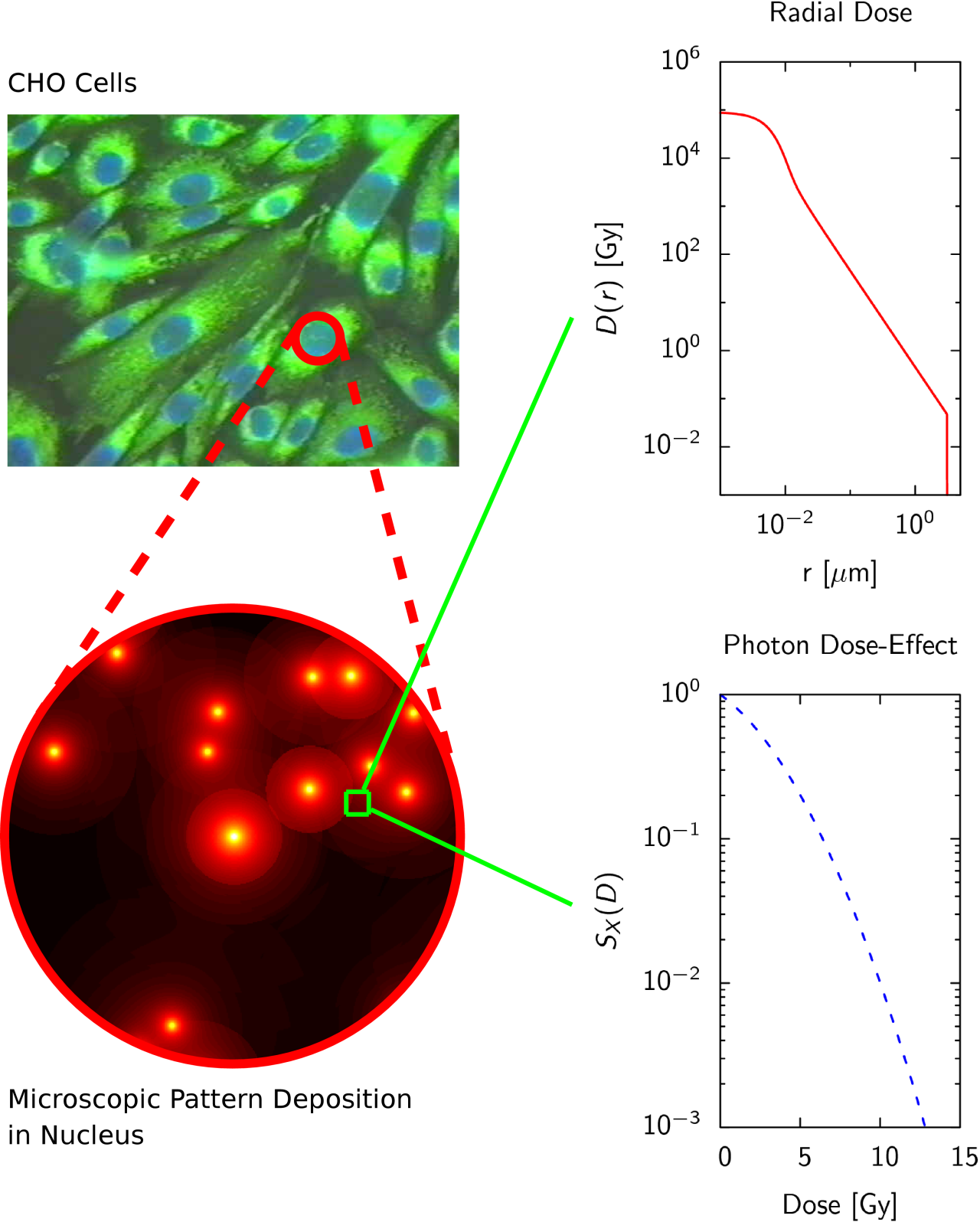}
 \caption{Basic assumptions of the LEM model, reproduced from Attili \citep{attiliLEM_ref55}}
 \label{fig:LEMpattern}
\end{figure}
                           
Finally, cell survival, $S_{\rm ion}$, after ion irradiation is calculated as  

\begin{equation}
S_{\text{ion}} = e^{-N_{\text{av}}}
\end{equation}

where $N_{\rm av}$ is the average number of lethal events in the cell nucleus within the
volume $V_{\rm nucl}$, defined as: 

\begin{equation}
 N_{\text{av}} = \frac{1}{V_{\text{nucl}}} \iint_{V_{\text{nucl}}} N(x,y) dx dy
\end{equation}

In the LEM several free parameters are applied: the response (cellular survival
endpoint) after reference radiation is described by $\alpha$, $\beta$, and $D_t$; $a_0$ is a radial
dose distribution parameter and $V_{\rm nucl}$ characterises the volume relevant to the
survival endpoint. Authors of the LEM model claim that it has only two free
parameters: $\alpha$ and $\beta$, which are determined from the known shape of the survival
curve after reference radiation. Other parameters relate to the description of
track structure, i.e. radial dose distribution ($a_0$) and to the description of
the biological endpoint ($D_t$ and $V_{\rm nucl}$). Some of these parameters are adjusted
in order for the model-calculated survival curves after ion irradiation
represent, as best possible, those measured experimentally. The original LEM,
the basic principles of which are described above, has since gone through
several further improvements, namely:

\begin{itemize}
 \item LEM I - original version of the model \citep{Scholz1994_ref34}, also including an
approximate version in which $\beta$ can be rapidly calculated from $\alpha$. 
 \item LEM II – addition of the effect of clustered damage in the DNA \citep{Elsasser2007b_ref35}
 \item LEM III – addition of an ion energy-dependent value of $a_0$ to compensate
for a systematic deviation in RBE predictions \citep{Elsasser2008_ref36}
 \item LEM IV – addition of a relationship involving the distribution of DSB in
the characteristic volume \citep{Elsasser2010_ref37}
\end{itemize}

\subsubsection{The Japanese approach to radiotherapy planning}
 
The Japanese group at the National Institute of Radiological Sciences (NIRS) in
Chiba has had an experience of several decades in clinical fast neutron
radiotherapy. On the basis of this experience an approach was created to deliver
carbon ion treatment, implemented within the HIMAC project. Instead of
radiobiological modelling of cell survival distributions, the following
procedure is applied: First, depth dose and LET distributions of pristine Bragg
peaks are evaluated and dose-averaged LET are calculated as a function of depth.
Next, a calculation of survival of the representative HSG (Human Salivary Gland)
cell line is performed. Survival is estimated on the basis of the
linear-quadratic formula (eq. \ref{eq:ch1_lq}) fitted to experimental cellular survival
data. Finally, the ''biological dose`` is calculated for HSG cells over the Spread
Out Bragg peak, assuming the LQ formula with coefficients averaged as follows:

\begin{equation}
 S_{\rm SOBP}(D) = \exp( - \alpha_{\rm ave} D - \beta_{\rm ave} D^2)
\end{equation}

Where:

\begin{equation}
\alpha_{\rm ave} = \sum_i f_i \alpha_i\qquad
\sqrt{\beta_{\rm ave}} = \sum_i f_i \sqrt{\beta_i} 
\end{equation}

And $f_i = D_i / \sum_i D_i$ is the fraction of the dose in the $i$-th component of the
SOBP.

Knowing the “biological dose” for HSG cells, the “physical dose” profile is
rescaled by a single scaling factor, termed “clinical RBE” by the authors of
this approach. The value of this factor is estimated on the assumption that the
therapeutic effect of a carbon ion beam of LET=80 keV/$\mu$m is clinically
equivalent to the effect of a fast neutron beam of the same average LET value. 
The ''clinical RBE'' of fast neutrons was found to be
equal to 3.0, thus corresponding scaling factors can also be established for
carbon ion beams of average LET values around 80 keV/$\mu$m. Finally, the calculated
individual “physical” depth-dose profile for each patient is rescaled using the
established scaling factor and applied in the patient’s treatment plan. This
scaling factor may additionally be corrected by retrospective analysis of
clinical results of treating selected tumour sites in selected groups of
patients. 

\mgrclosechapter

\chapter{Evaluation of elements of track structure models}
The Radial Dose Distribution, $D(r)$, is the central element of amorphous track
structure models, and particularly of Katz’s theory of Cellular Track Structure.
Apart from Katz’s derivation of $D(r)$, several other formulations have been
published. The author of this thesis has implemented some of these formulations
and supplementary elements in the open source libamtrack library of computer
codes, as a convenient platform for model calculations. Four formulations of
$D(r)$ are presented and discussed in this Chapter. The author discusses the
congruence of the selected $D(r)$ formulae with experimental data and the relation
between the radially integrated $D(r)$ formulae and LET of the ion. The $D(r)$
formula found to be most applicable in later track structure calculations is the
formula of Zhang with ionisation potential $I = 0$, thus $\theta = 0$ in Zhang's formula
shown in Table \ref{tab:ch2_rddmodels}.

\section{The libamtrack software library}

The libamtrack library is a volunteer scientific project \citep{libamtrack_ref40} to create an
open-access collection of routines, databases and functions, allowing
calculations to be performed of the response of biological and physical
detectors after doses of heavy charged particles.  The libamtrack library was
initiated by Steffen Greilich and is supported by several collaborating
scientists applying amorphous track structure track calculations to radiobiology
and detector physics. The present libamtrack library of codes allows
calculations of several physical elements of track interactions, such as basic
kinematics, stopping powers, detector response, and elements of track structure
modelling.  Subroutines of the library can be downloaded, edited, modified and
incorporated in other software. Together with the library codes, implemented in
ANSI C language, a set of wrapping methods is provided, making it possible to
use these codes in other computing languages (such as Python or Java), and in
numerical simulation tools (such as R or Matlab). Some of the library functions
were used to create a web interface (called libamtrack WebGUI, available under
\verb"http://webgui.libamtrack.dkfz.org/test"), where users connected to the Internet
can perform basic calculations using their web browsers. The libamtrack WebGUI was based on the 
work of Christian Kolb, but its capabilities being extended by the author of the thesis.
\cite{Kolb}

The libamtrack library
has been applied in calculations of the response of alanine \citep{Hermann-libamtrack_ref41}
 and aluminium oxide \citep{Klein20111607_ref42} detectors and in 
radiobiological modelling of cellular survival \citep{grzanka2011application_ref43}.

The libamtrack library consists of the following modules:
\begin{itemize}
 \item implementation of the “scaled” and “integrated” versions of the Katz model 
 \item grid summation method (similar to the first version of LEM)
 \item evaluation of compound Poissonian processes using the successive convolution
algorithm \citep{greilich2013efficient_ref91}
 \item basic radiobiological formulae (linear-quadratic, multi-target, etc.) to
describe cellular survival, 
 \item operations on energy-fluence databases stored in SPC format
 \item radial dose distribution formulae and their derivatives (integrated $D(r)$,
averaged $D(r)$)
 \item stopping power data
 \item electron and ion energy range data
 \item radiation absorber data (target medium, density, electron density, etc.)
 \item particle (projectile) data
 \item operations on histograms
 \item numerical routines
 \item physical routines
 \item physical and radiological constants
\end{itemize}

Within this thesis project, the author developed a set of analytical functions
fitted to electron energy-range data (Table \ref{tab:ch2_ermodels}) and radial dose distributions
around energetic ions (Table \ref{tab:ch2_rddmodels}), implementing them in C language as functions
in the libamtrack library module. The libamtrack library was also applied by the
author to implement different algorithms of $D(r)$ to the Katz model, and to
tabularize Monte Carlo beam data (originally calculated by Pablo Botas using the
SHIELD-HIT10A code) to be handled by optimization algorithms required to
suitably adjust the composition of the carbon ion beams to obtain optimal
depth-dose or survival vs. depth profiles.

The operating manual for the libamtrack library of computer codes is available
at the \verb"http://libamtrack.dkfz.de" webpage. Examples of calculations using radial
dose distribution formulae and of energy transfer from ions to delta-electrons,
are given in Appendix C. 

\section{Electron energy-range relationships}

\begin{figure}[!h]
 \centering
 \includegraphics[width=0.8\textwidth]{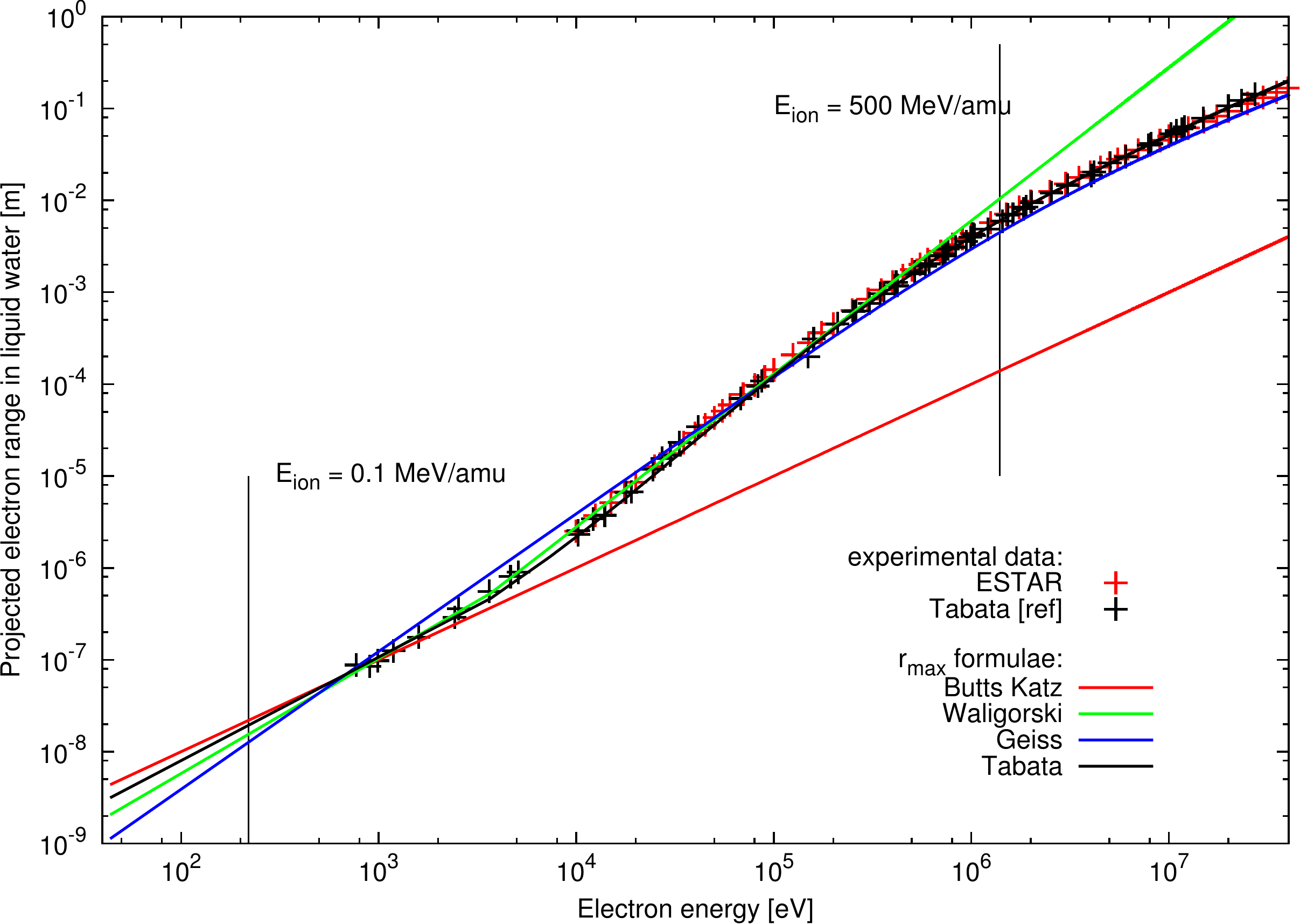}
 \caption{Projected electron ranges vs. initial electron energy in liquid water.
Experimental data are from Tabata \citep{Tabata1972} (measurements in different materials,
density-rescaled to liquid water). The ion energy range useful in ion beam
radiotherapy is indicated by vertical lines. Experimental data and the
respective electron energy-range relationships have been implemented in the
libamtrack library.}
 \label{fig:ch2_er}
\end{figure}

To determine extrapolated ranges of electrons simple analytical functions can be
applied to fit experimental data. The first semi-empirical linear or power-law
formulae describing the electron energy- range relationship were developed in
the early 1960s \citep{Butts1967}. These formulae could then be applied to drive the first
formulations of the radial dose distribution due to delta-rays \citep{Butts1967}, 
\citep{Waligorski1986_ref44}, \citep{Zhang1985}.
Later, to comply with improved experimental data, more complex energy-range
formulae were presented in the literature. An experimental data set collected by
Tabata (figure \ref{fig:ch2_er}), is available for electron energies from 900 eV to about 25
MeV. These data were determined from extrapolated ranges of electrons and from transmission and projected-range straggling curves,
density-scaled to water from measurements in different materials, according to
Eq. \ref{eq:ch2_rmax_scaling}. The formula of Tabata \citep{Tabata1972} currently offers the best representation of
experimental data over the broadest range of energies.

Electron energy-range formulae from the libamtrack library (see Appendix A) have
been applied to calculate projected delta-electron ranges, for electron energies
50 eV - 50 MeV.  Over the energy range of ions relevant for ion radiotherapy
(roughly 0.1 - 500 MeV/amu, corresponding to the ejected delta-electron energy
range of 220 keV - 1.4 MeV), the parameter fits of Tabata, Waligórski and Geiss all
represent experimental data with sufficient accuracy. 

To apply the absorber density scaling of electron range, the simple relation is
used: 

\begin{equation}
 \frac{R_{e1}}{R_{e2}} = \frac{\rho_1}{\rho_2}
 \label{eq:ch2_rmax_scaling}
\end{equation}

where $R_{e1}$ and $R_{e2}$ are the ranges of the two particles and $\rho_1$ and $\rho_2$ are the
respective densities of media they travel through.

Under the realistic assumption that delta-electrons are emitted at an angle of
90 degrees to the ion path, the maximum energy of these delta-electrons determines
their maximum range and the extent of the radial dose distribution. In Table \ref{tab:ch2_ermodels}
the energy-range formulae used by different authors are listed, together with
constants applied in these formulae. The units in which the electron or ion
energy, denoted by $\omega$ or $E$, are given are keV or MeV/amu, respectively. For more
details concerning these formulae, Appendix A should be consulted.

\begin{table}[!h]
\begin{tabular}{m{0.2\textwidth}p{0.4\textwidth}m{0.3\textwidth}}
\hline
\textbf{Name} & \textbf{Expression} & \textbf{Reference} \\
\hline
\begin{center}Butts and Katz\end{center}&
$r_{\max}=10^{-6} \text{cm}\cdot \frac{\omega}{\text{keV}}$
&\citep{Butts1967}\\

\begin{center}Walig\'orski\end{center}&
$r_{\max} =6\cdot 10^{-6} \text{cm} \cdot
\left(\frac{\omega}{\text{keV}}\right)^\alpha$
&\citep{Waligorski1986_ref44}\\

\begin{center}Geiss\end{center}&
$r_{\max}=4\cdot 10^{-5} \text{cm} \cdot
\left(\frac{E}{\text{MeV}}\right)^{1.5}$
&\citep{Geiss1997}\\

\begin{center}Scholz\end{center}&
$r_{\max}=5\cdot 10^{-6} \text{cm} \cdot
\left(\frac{E}{\text{MeV}}\right)^{1.7}$
&\citep{Scholz2001a}\\

\begin{center}Tabata\end{center}&
$r_{\max}= f(a_1,a_2,\ldots,a_7, \omega) $
&\citep{Tabata1972}\\
\hline
\end{tabular}
\caption{Energy-range formulae used to calculate the maximum range of
delta-electrons, $r_{\max}$, implemented in the libamtrack library. For more details
concerning the exponent $\alpha$ \citep{Waligorski1986_ref44}, and the function $f$ \citep{Tabata1972}, 
Appendix A should be consulted. 
The units in which the electron or ion energy, denoted by $\omega$ or $E$ are
given, are keV or MeV, respectively.}
\label{tab:ch2_ermodels}
\end{table}

\section{Radial Dose Distribution formulae}

Consider an energetic ion travelling through the medium along a straight line
(ion scattering is neglected here). The relative speed of the ion is $\beta$ , its
effective charge is $z^\star$ and the electron density in the medium is $N$. Assume that
delta-electrons, of mass $m_e$ and maximum energy given by eq. \ref{eq:ch1_omegamax}, will then be
ejected at straight angles due to ion-orbital electron Coulomb interactions, the
ion gradually slowing down as it loses its initial energy (a track segment is
considered here). The delta-electron energy spectrum may be given by
Rutherford’s formula \citep{Passage_ref28}: 

\begin{equation}
  \frac{dn}{d \omega} = \frac{2 \pi N e^4 {z^{\star}}^2}{m_e c^2 \beta^2} \frac{1}{(\omega+I)^2}
 \label{eq:ch2_rutherford}
\end{equation}

where $dn$ is the number of ejected delta-electrons of energies between $\omega$ and
$\omega+d\omega$.

Low energy delta-electrons and the ion itself will produce most of their
ionisation and excitation events close to the ion’s path. Per unit mass, the
average number of these events will decrease and no such events will occur
beyond the maximum range of delta-electrons. The spatial distribution of average
dose deposited in the medium by such events occurring around the path of the
heavy ion is called the Radial Dose Distribution, $D(r)$. Axial (cylindrical)
symmetry of the $D(r)$ can be assumed.

The first distributions of radial dose derived analytically by Butts and Katz
\citep{Butts1967} were soon confirmed by Monte Carlo track structure calculations of Paretzke
\citep{paretzke1973comparison_ref81}, which were able to numerically simulate, step-by-step, all
interactions of primary and secondary particles. The present track structure
calculations performed with PARTRAC (\citep{friedland1998partrac_ref82}), 
MOCA \citep{paretzke1974moca_ref83}, Fluka \citep{Fluka_ref27}, or Geant4 \citep{Geant4_ref26} 
codes are all consistent and in good agreement with
experiment, although their precision is not satisfactory at low ion energies
\citep{hauptner2006spatial_ref53}.

In the analytical derivation of the amorphous radial distribution of average
dose around the path of an energetic ion, the following assumptions were
initially made:
\begin{itemize}
\item the ion moves in the medium along a straight path (ion scattering is ignored)
\item only ionisations due to delta-electrons contribute to the average dose
\item every ionisation contributes to the dose with the same average energy, $\omega$
\item there is axial symmetry of average energy deposition around the ion’s path.
\end{itemize}

Following later studies, two more postulates were added:
\begin{itemize}
\item the contribution of excitations to the radial dose should also be included
\item on radial integration, the $D(r)$ should yield the LET of the ion in the medium
\end{itemize}

The Radial Dose Distribution, $D(r$), may be derived from eq. \ref{eq:ch2_rutherford} 
or from other
assumptions by calculating the average dose in a cylindrical shell of thickness
$dr$ located at radial distances $r$, $r+dr$ from the ion’s path. Depending on the
$D(r)$ formulations developed by different authors, the maximum dose where most of
the energy transfer events occur, will appear around a selected cut-off radius
(usually, $5 \cdot 10^{-11}$ or 10$^{-10}$ m), or be constant up to a selected radius. Next the
radial dose rapidly decreases, approximately as as $r^{-2}$ , and reaches zero at the
maximum range of the delta-rays, as calculated from the electron energy-range
relationships applied by their authors. Some of the published $D(r)$ formulae
implemented in the libamtrack library are listed in Table \ref{tab:ch2_rddmodels}.

\begin{table}[!h]
\begin{tabular}{m{0.15\textwidth}m{0.49\textwidth}m{0.27\textwidth}}

\hline
\textbf{Name} & \textbf{Expression} & \textbf{Reference} \\
\hline

\begin{center}Katz\end{center}&
$D(r)=C_1 \frac{z^{\star 2}}{\beta^2}\frac{1}{r}
\left(\frac{1}{r}-\frac{1}{r_{\max}}\right)$&\citep{Zhang1985}\\

\begin{center}Zhang\end{center}&
$D(r)=C_1 \frac{z^{\star 2}}{\beta^2}\frac{1}{\alpha}\frac{1}{r}
\frac{1}{r+\theta(I)}\left(1-\frac{r+\theta(I)}{r_{\max}+\theta(I)}\right)^{\alpha^{-1}}$&
\citep{Zhang1985}\\

\begin{center}Gei{\ss}\end{center}&
$D(r)=\begin{cases}C_2&\text{if $0<r<a_0$,}\\
\frac{C_2}{r^2}&\text{if } a_0\le r\le
r_{\max}\end{cases}$&\citep{Geiss1998}\\

\begin{center}Cucinotta\end{center}&
$D(r)=C_1 \frac{z^{\star 2}}{\beta^2} f(r) \frac{1}{r^2}
+ C_3 \frac{\exp(-\frac{r}{2d})}{r^2}$&\citep{Cucinotta1997}\\

\hline
\end{tabular}
\caption{Selected formulae describing the radial dose distribution,
implemented in the libamtrack library. For further details, see Appendix A.}
\label{tab:ch2_rddmodels}
\end{table}

Katz derived his first $D(r)$ formula \citep{Butts1967} analytically from Rutherford’s formula
(eq. \ref{eq:ch2_rutherford}) assuming a linear delta-electron energy range relationship (denoted as
Butts and Katz in Table \ref{tab:ch2_ermodels}).

Zhang later adapted this formula by accommodating a power-law electron energy
range relationship (denoted as Waligórski in Table \ref{tab:ch2_ermodels}). Zhang’s formula was
further improved by Waligórski \citep{Waligorski1986_ref44} who incorporated a correction effective at
small radii in order for the radial integral of $D(r)$ to yield the correct value
of LET of the ion (see Fig. \ref{fig:ch2_rdd}, right panel). For further details, see Appendix
A. For  a thorough review of the consecutive developments of $D(r)$ formulae by
Katz and co-workers, see \citep{KorcylPhD_ref45}. In further calculations presented in
Chapter 5 of this thesis the formula of Zhang, modified by setting the value of
ionisation potential $I = 0$ (see Table \ref{tab:ch2_rddmodels}) will be used.

Geiss in his derivation of the $D(r)$ used another expression for the electron
e\-ner\-gy-ran\-ge relationship (see Appendix A), postulated a $\frac{1}{r^2}$ dependence of
average radial dose with distance from the ion’s path at radii exceeding $a_0$ (a
free parameter) and a constant radial dose at smaller radii, the value of which
is calculated via the ion’s LET, assuring that when integrated radially, $D(r)$
should yield this value of LET \citep{Geiss1998}. For further details, see Appendix A.

Cucinotta proposed a two-component radial dose distribution formula \citep{Cucinotta1997}, the
first component describing the contribution to the average dose from ionisations
and the second - from excitations. In the first component, the electron energy
range formula of Tabata (see Table \ref{tab:ch2_ermodels}) is used and an angular distribution of
delta-rays applied which is more complex that that given by Rutherford’s formula
(eq. \ref{eq:ch2_rutherford}) and the assumption of delta-ray emission at 90 degrees. 
The contribution of the
second component due to excitations is significant at small radial distances.
The constant $C_3$ applied in this formula (see
Table \ref{tab:ch2_rddmodels}) is calculated by making the radially integrated sum of both
components equal to the value of ion’s LET.  This makes this $D(r)$ formula less
convenient in massive calculations, because radial integration of both its
components is required to calculate the above constants each time this formula
is needed. For further details, the original paper of Cucinotta et al. \citep{Cucinotta1997},
\citep{KorcylPhD_ref45} and Appendix A should be consulted. 

\section{Comparison of D(r) formulae with experiment}

While particle tracks, due to the size of silver bromide grains, can readily be
observed in nuclear emulsion \citep{Katz1972b}, it is quite difficult to measure experimentally the
distribution of energy (or local dose) deposited around the path of an energetic
ion. The main difficulty is in the small scale of this phenomenon, where maximum
ranges of delta-electrons in solid media, or water, are typically of the order
of  micrometres or less. However, radial dose distributions around a few ion
species have been measured in air, water vapour or tissue-equivalent gas, mainly
by Wingate, Baum and Varma \citep{katz1992radial_ref84} using ionisation chambers 
to evaluate the absorbed dose (or exposure) at various distances by controlling gas pressure.

\begin{figure}[p]
 \centering
 \includegraphics[width=0.8\textwidth]{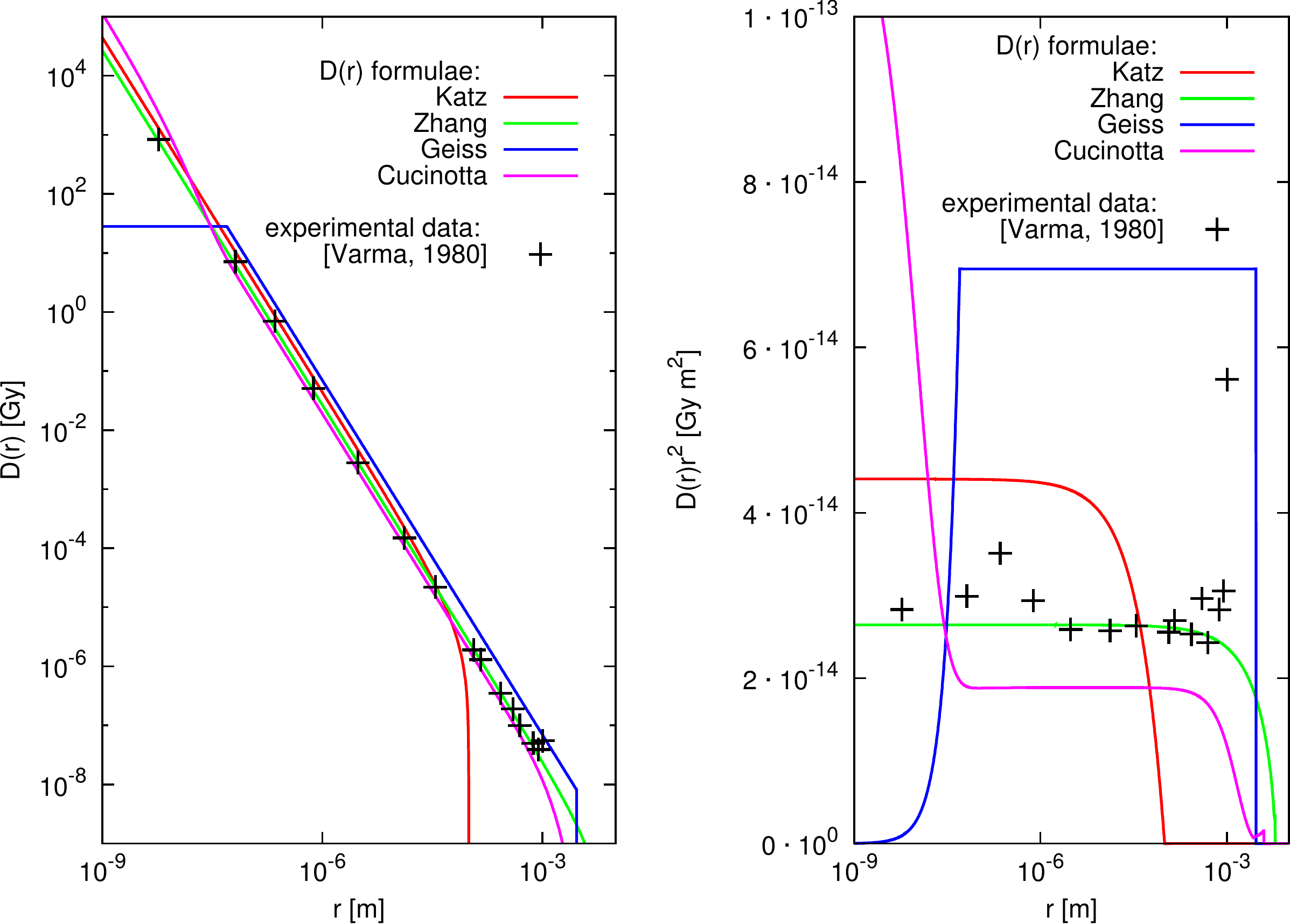}
 \caption{Comparison of experimentally measured radial distribution of dose
around  377 MeV/amu neon ions in tissue-equivalent gas \citep{Varma1980Neon} and
$D(r)$ calculated using the formulae of Katz, Zhang, Geiss and Cucinotta.
Experimental uncertainties were not provided by Varma et. al. The right-hand
panel shows the same data and calculations, but on a semi-logarithmic plot where
data points and results of versus of $D(r)$ calculations are multiplied by their
respective $r^2$ values. }
 \label{fig:ch2_rdd}
\end{figure}

In Fig. \ref{fig:ch2_rdd} the radial dose distribution around 377 MeV/amu neon ions, measured
by Varma et al. \citep{Varma1980Neon} in tissue-equivalent gas is compared with calculations 
using the above-discussed $D(r)$ formulae
of Katz, Zhang, Geiss and Cucinotta. This comparison is rather of a qualitative
nature, as Varma's experimental data points did not include any assessment of
their uncertainty. Disagreement of the original Katz’s formulation of $D(r)$ at
the low-dose range stems from his over-simplified electron energy-range
relationship, invalid at higher delta-ray energies (see Fig. \ref{fig:ch2_er}). The remaining
$D(r)$ formulations show much better agreement, except for the formulation of
Geiss (over the central dose region). For ease of comparison, the same data and
results of $D(r)$ calculations are also shown in a plot where the linear ordinate
is in terms of $D(r) r^2$ (the $r^{-2}$ dependence of data points and of $D(r)$
calculations is then represented as a horizontal line). The “bump” in the
experimental data around radial distances about $10^{-7}$ m was accounted for by the
$D(r)$ formulation of Waligórski et al. \citep{Waligorski1986_ref44} (not shown in Fig. 2.2). 
Monte Carlo simulations are able to represent the available experimental data to within an
order of magnitude \citep{Waligorski1986_ref44}. Zhang’s formulation appears to
offer the best agreement with experimental data.

\section{Radial integration of D(r) and LET}

According to the above-listed assumptions and postulates, since all the energy
deposited around the ion’s path is to be transferred by the delta rays,
integration of the radial distribution of dose over all radii should yield the
stopping power (or LET) of the ion, for a given track segment. For this reason,
the correction of $D(r)$ by Waligórski et al. \citep{Waligorski1986_ref44} and the later developed $D(r)$
formulations by Geiss \citep{Geiss1998} and by Cucinotta et al. \citep{Cucinotta1997} were designed to
comply with this postulate.

Linear Energy Transfer $L$, by definition, is equal to the energy $dE$, deposited
in the medium along the length $dz$ of a track segment, divided by $dz$: 

\begin{equation}
L = \frac{dE}{dz}
\end{equation}

$L$ is related to the average dose $\frac{dE}{dm}$ in track segment by the following
relationship:

\begin{equation}
 L = \frac{dE}{dz} = 
\frac{dE (\pi r_{\max}^2 - \pi r_{\min}^2 ) \rho}{dz (\pi r_{\max}^2 - \pi r_{\min}^2) \rho} =
\frac{dE}{dm} (\pi r_{\max}^2 - \pi r_{\min}^2 ) \rho
\label{eq:ch2_letnorm_part1}
\end{equation}

where $\rho$ is the density of the medium, and $r_{\max}$ and $r_{\min}$ are the maximum range
of the delta rays and cutoff radius, respectively. 
 
The average dose $\frac{dE}{dm}$ can be also calculated by radially integrating $D(r)$:

\begin{equation}
\frac{dE}{dm} = 
\frac{1}{\pi r_{\max}^2 - \pi r_{\min}^2} \int_{r_{\min}}^{r_{\max}} 2 \pi r D(r) dr
\label{eq:ch2_letnorm_part2}
\end{equation}

From eq. \ref{eq:ch2_letnorm_part1} and eq. \ref{eq:ch2_letnorm_part2} the following relation 
between Linear Energy Transfer $L$ and Radial Dose Distribution $D(r)$ holds for an ion track segment:

\begin{equation}
L =  \left(\frac{1}{\pi r_{\max}^2 - \pi r_{\min}^2} \int_{r_{\min}}^{r_{\max}} 2 \pi r D(r) dr\right)
(\pi r_{\max}^2 - \pi r_{\min}^2 ) \rho
\end{equation}

Therefore:

\begin{equation}
L =  2 \pi  \rho \int_{r_{\min}}^{r_{\max}} r D(r) dr
\label{eq:ch2_rddnormdef}
\end{equation}

\begin{figure}[p!]
 \centering
 \includegraphics[width=0.8\textwidth]{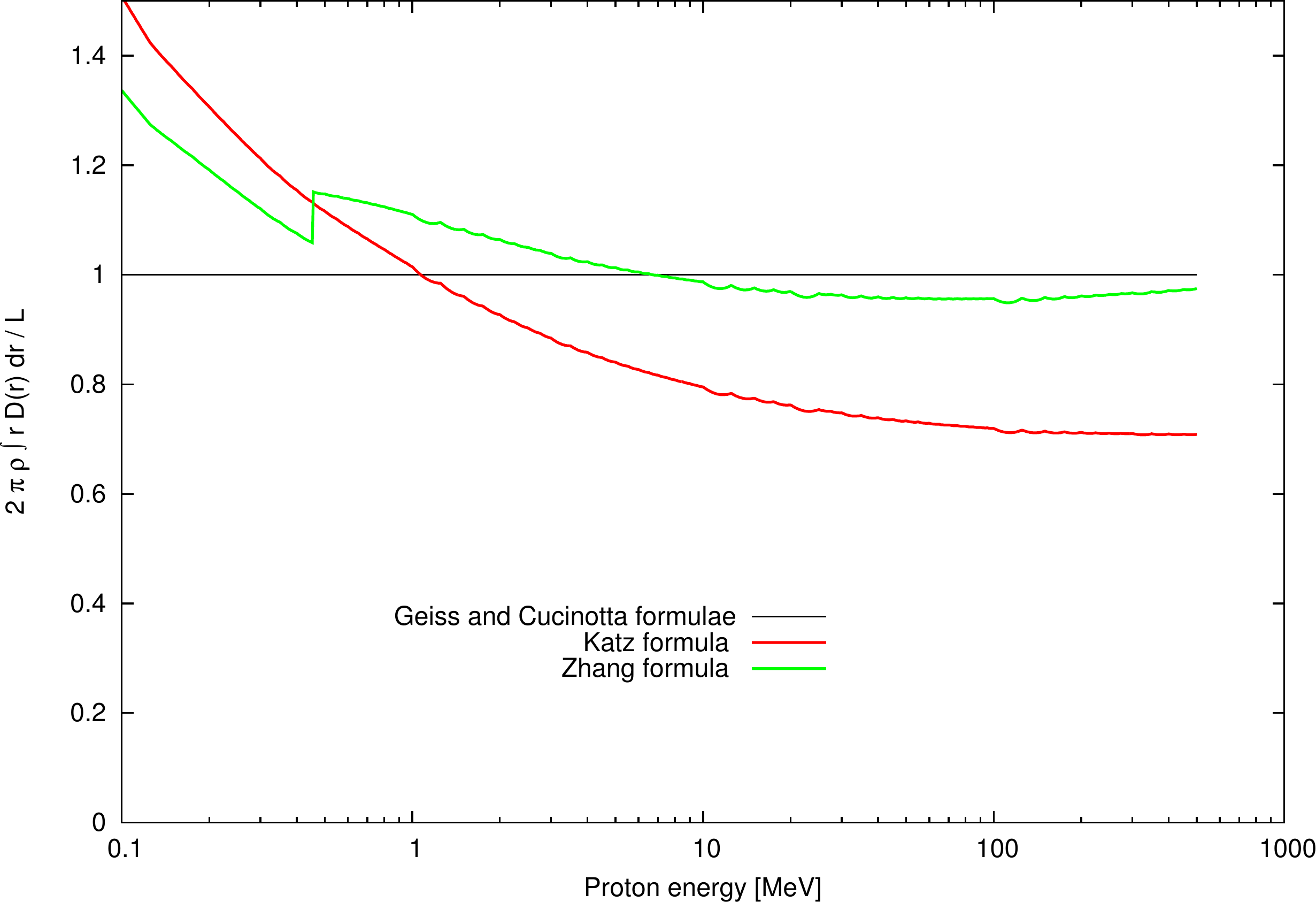}
 \includegraphics[width=0.8\textwidth]{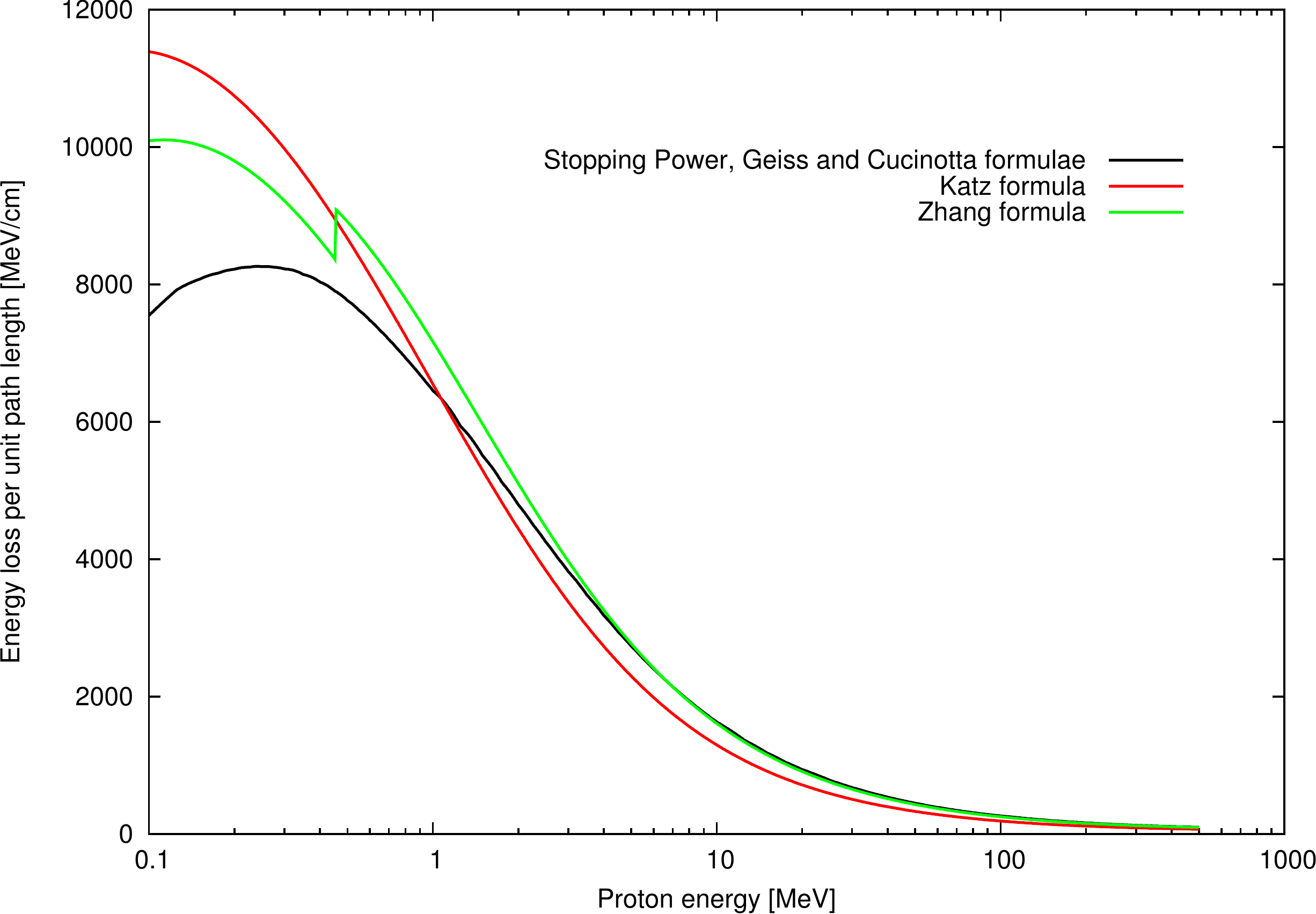}
 \caption{ Upper plot: Ratio of the radially integrated D(r) formulae of Katz
and of Zhang ($\theta$ = 0) and the values of LET versus proton energy The formulae of
Geiss and Cucinotta on radial integration return the values of LET by
definition.  Lower plot: Proton LET (i.e. the result of radial integration of
the formulae of Geiss and Cucinotta) and results of radial integration of the
$D(r)$ formulae of Katz and of Zhang ($I = 0$) vs. proton energy. As the LET values,
the proton stopping power values from the PSTAR database \citep{NISTESTAR} are used. Author’s
calculation based on the libamtrack library.}
 \label{fig:ch2_rddnorm}
\end{figure}

\section{Conclusions}

On the basis of Fig. \ref{fig:ch2_er} and its analysis, the electron energy-range formulae of
Geiss, Waligórski and Tabata are found to represent the measured data well over
the energy range of interest to ion radiotherapy, i.e. up to ion energy 400
MeV/amu.

Complexity of implementation and calculation time are important factors when
considering the application of any $D(r)$ formulation in massive calculations
required in treatment planning systems. For these reasons, the most accurate
$D(r)$ formula of Cucinotta et al. which appears to best fulfil the requirements
of such a $D(r)$ formula, was discarded, as evaluation of the constants in this
formula requires multiple integrations every time this formula is needed in the
TPS calculations or in fits to radiobiology data. Therefore, in further work the
delta electron energy-range function listed here as Waligórski’s was implemented
in Zhang’s D(r) formula, however modified by neglecting the value of ionisation
potential (i.e., by assuming $\theta$ = 0 in that formula). Zhang’s formula satisfies
the postulated requirements well enough and with its simple analytical form, can
be conveniently applied in the complex calculations presented in further
chapters of this thesis. 

\clearpage\null
\mgrclosechapter

\chapter{The Katz model of cellular survival}
The Cellular Track Structure model (the Katz model) is presented in detail and
discussed. Following Korcyl’s review of this model where model elements were
recalculated and the scaling approximations originally introduced by Katz were
generally confirmed, the analytically simple set of basic formulae of Katz’s
model will be termed here the \textit{scaled version} of the model. The author’s results
of model calculations involving integration of its elements (thus termed the
\textit{integrated version} of Katz’s model in what follows) suggest the possibility of
further developing the Katz model. The author next justifies the use of the
scaled version of the Katz model in further calculations and uses it to best fit
model parameters from published data of in vitro survival of CHO (Chinese
Hamster Ovary) cells after X-ray and ion irradiation. These cellular parameters
and the principle of performing model calculations of cellular survival after
mixed-field irradiation applied in Katz’s model will be applied in later parts
of this thesis.

\section{Principles of the Katz model}
\label{sec:ch3_principles}

Robert Katz introduced his Track Structure Theory (TST) model of RBE around
1960, basing it on the m-target formula to describe cellular survival after
doses of reference radiation. In that aspect, the Katz model differs from the
LEM where the linear-quadratic formula is applied. Katz’s TST
phenomenological analytical model is aimed at calculating the survival of
biological cells and the response of physical detectors after ion irradiation.
The model originates from the early works of Robert Katz and co-workers from
Lincoln University, Nebraska \citep{katz1971cellular_ref58} \footnote{Robert Katz 
and co-workers publications are freely available online on 
\verb|http://digitalcommons.unl.edu/physicskatz/| by Digital Commons platform.}.
The general assumption of this model is that
the response of physical or biological detectors after ion irradiation can be
calculated from the distribution of energy deposition (or dose) around the ion
track by scaling their response from their dose response after reference
$\gamma$-rays.

Katz postulated that the probability of cellular survival after a dose $D$ of
radiation, $S(D)$, is a product of two probabilities, or modes of  cell
inactivation:

\begin{equation}
 S(D) = \Pi_i(D) \cdot \Pi_{\gamma}(D)
 \label{eq:ch3_katz_survival}
\end{equation}

Katz termed the first term of this product, $\Pi_i(D)$, ion-kill probability (since
the cell is either “killed” or survives, the probability of “killing” a cell is
$1- S(D)$). $\Pi_i(D)$ assumes 1-hit probability of cell inactivation by direct passage
of an ion:

\begin{equation}
 \Pi_i(D) = \exp( -\sigma F ) = \exp\left( -\frac{\sigma \rho}{L} D \right)
\label{eq:ch3_ionkill}
\end{equation}

The inactivation cross section, $\sigma$, is calculated for a single ion. Here, $F$ is
ion fluence, related to ion dose $D$ and ion LET, $L$, and density of the medium,
rho, via eq. \ref{eq:ch1_dose_let}. To evaluate $\sigma$, knowledge is required of the averaged (or
extended target) radial dose distribution (see par. 3.2 below).
In the second term in eq. \ref{eq:ch3_katz_survival}, named gamma-kill probability 
(again, kill $K_\gamma(D) = 1- S_{\gamma}(D)$ ), 
Katz assumes that the response of the cell system after doses of
reference radiation, and also after delta-rays from overlapping ion tracks, is
described by the multi-target formula:

\begin{equation}
 \Pi_{\gamma}(D) = 1 - \left( 1 -\exp\left( - \frac{(1-p)D}{D_0} \right)\right)^m
\label{eq:ch3_gammakill}
\end{equation}

Here, $D_0$ is the radiosensitivity of the cell system and  $(1-p)$ is the fraction
of the ion dose, $D= \frac{1}{\rho} F L$, involved in gamma-kill mode. To calculate
$\Pi_{\gamma}(D)$, values of the model parameters $m$ and $D_0$ are needed, and knowledge of
the value of $p$. The factor $p$, which may assume values in the range 0-1 and
appears in the “ion-kill” and “gamma-kill” expressions, is the “mixing
parameter” of the Katz model, which enables smooth transition between the purely
exponential cell survival after ion irradiation and the m-target form if the
system is irradiated only by $\gamma$-rays. In the case of irradiation by a dose $D$ of
reference $\gamma$-radiation only, $p \approx 0$ and the expression in brackets 
in eq. \ref{eq:ch3_gammakill} becomes the usual m-target formula, eq. \ref{eq:ch1_mtarget}, 
while the “ion-kill” probability, eq. \ref{eq:ch3_ionkill} then becomes unity. 
On the other hand, in the case of ion
irradiation only, $p \approx 1$, hence the “gamma-kill” probability becomes unity and
only the “ion-kill” term remains in eq. \ref{eq:ch3_katz_survival}. 
The value of the model’s “mixing
parameter” $p$ may be interpreted as the degree of overlap of delta-ray “clouds”
of neighbouring ions irradiating a cellular system, which strongly depends on
the radial extension and ion charge-dependent dose values of the radial dose
distributions of these ions. The value of $p$ may either be evaluated by numerical
calculations in the integrated version of Katz’s model (see par. \ref{sec:ch3_integrated}) or
calculated analytically from a formula in the scaled version of the model (see
par. \ref{sec:ch3_scaled}). The case of mixed radiation (i.e. cell irradiation by a mixture of
ions of different charges and energies), of particular interest in this thesis,
is discussed in paragraph \ref{sec:ch3_mixed} of this chapter.

Calculations in the Katz model are performed in two steps. First, the
inactivation cross section of a single cell (target) by a single ion, $\sigma$, is
calculated.  Then, cellular survival is calculated by combining the inactivation
cross section with ion fluences and doses resulting from the applied ion beams,
where $\Pi_i$ and $\Pi_{\gamma}$ both contribute.  

\section{The extended target radial dose distribution}

The Katz model assumes that the spatial distribution of dose delivered by X-rays
or $\gamma$-rays to the entire target volume is homogenous. It is also assumes that
the spatial distribution of dose delivered by ions and surrounding
delta-electrons is given by the radial dose distribution.

The analytical expressions of $D(r)$ discussed in Chapter 2 which describe the
radial dose absorbed in an infinitesimally small volume at distance $r$ from the
path of the ion are called \textit{point-target} distributions. By averaging the
\textit{point-target} dose distribution function $D_p$  over a small
cylindrical target of radius $a_0$ and length of a track segment $dz$ (representing
the cell nucleus or some volume contained within the cell, of size typically of
the order of a few micrometres), another, averaged over target function, $D_e$, 
is obtained, which is called the \textit{extended target} or \textit{average dose}
distribution: 

\begin{equation}
 D_e(r) = \frac{1}{S_r} \iint_{S_r} D_p(x,y) dx dy
 \label{eq:ch3_exttarget}
\end{equation}

where $S_r$ is the target area, $S_r = \pi a_0^2$  , located at a distance $r$ from the path
of the ion.

As long as the track segment assumption is valid, we can use averaging over surface,
instead over volume averaging.

Assuming track-segment irradiation ($dE/dx = \rm const$, track length $dz$), the integral
in eq. \ref{eq:ch3_exttarget} can be reduced to a two-dimensional intersection of the target
volume. Furthermore, by changing the Cartesian coordinate system to polar
coordinates, eq. \ref{eq:ch3_exttarget} can be rewritten as follows:

\begin{equation}
 D_e(r) = \frac{1}{\pi a_0^2} \int_{t_{\rm lower}}^{a_0+r} D_p(t) \Phi(a_0,r,t) dt
 \label{eq:ch3_exttarget_polar}
\end{equation}

where $\Phi$ is the length of an arc of radius $t$ inside the target (a circle) of
radius $a_0$ at the distance $r$ from the ion track. More details on calculating
extended target distributions of radial dose are given in Appendix B.

\section{Probability of inactivation --- the m-target formula}
In the Katz model it is assumed that the target volume consists of several
sensitive sub-targets. Each sub-target can change its state on being inactivated
due to an energy deposition event from ionising radiation (in other words, by
the target being hit).

The probability $P(n)$ that a single target or sub-target will be hit exactly n
times is given by the Poisson distribution:

\begin{equation}
P(n) = f^n \frac{e^{-f}}{n!}
\end{equation}

where $f$ is the average number of hits. If $D_0$ is the dose after which, on
average, each subtarget receives one hit, then the average number of hits, $f$,
can be related to dose by:

\begin{equation}
 f = \frac{D}{D_0}
\end{equation}

The probability that a sub-target will receive one or more hits equals:

\begin{equation}
 P( n \geq 1 ) = 1 - P( n = 0) = 1 - f^0 \frac{e^{-f}}{0!} = 1 - e^{-f} = 1 - e^{-\frac{D}{D_0}}
\end{equation}

If the target contains only a single 1-hit sub-target, the probability of the
target being inactivated is equal to $P(n \geq 1) = 1 - e^{-D/D_0}$. In case of an
m-target configuration, where the target consists (on average) of $m$ 1-hit
sub-targets, the probability of target activation is given by the following
general expression:

\begin{equation}
P = (1 - e^{-D/D_0})^m 
\label{eq:ch3_katz_probability_general}
\end{equation}

This is the single-hit multi-target formula representing cellular survival after
reference radiation, eq. \ref{eq:ch1_mtarget}, where survival is defined as the average number
of targets not activated. In the Katz model it is assumed that cellular survival
$S$ after a dose $D$ of reference radiation (X-rays or $\gamma$-rays) is described by
this m-target expression:

\begin{equation}
S(D) = 1 - P = 1 - (1 - e^{-D/D_0})^m
\label{eq:ch3_katz_mtarget}
\end{equation}

Two parameters of the m-target expression, eq. \ref{eq:ch3_katz_mtarget}: the radiosensitivity (or
characteristic dose) $D_0$ and the number of 1-hit sub-targets in the target, $m$,
can be obtained by fitting this equation to experimental data (e.g., cell
survival curves after doses of reference radiation). Typically, for mammalian
cells, $D_0$ is of the order of a few Gy and $m$ is a number (not necessarily
integer) ranging between 1 and 5.

One of the basic assumptions of amorphous track structure models is that the
effect of a dose $D$ non-homogenously distributed and deposited by delta-rays in a
small target of size $a_0$, is the same, per dose unit, as that of a dose $D$ of
reference radiation distributed homogenously. As already discussed, in the Katz
model, survival curves after doses of reference gamma radiation are described ex
definitione by the m-target formula, eq. \ref{eq:ch3_katz_mtarget}, or eq. \ref{eq:ch1_mtarget}.

\section{Inactivation cross-section}
\label{sec:ch3_inactivation}
The probability of the target being “killed” or inactivated by an ion passing at
a distance $t$ from the target can be calculated from eq. \ref{eq:ch3_katz_probability_general}, 
assuming a t-dependent distribution of extended target radial dose distribution, $D_e(t)$
where $t$ is the radial distance between the ion path and the centre of the
target: 

\begin{equation}
P(t) = (1 - e^{ -D_e(t)/D0})^m
\end{equation}

The \textit{single-event inactivation cross section} is defined as an average probability
$P(t)$ over all distances $t$ available to the delta rays surrounding the ion, i.e.
up to $r_{\max}+a_0$ :

\begin{equation}
 \sigma = \int_0^{r_{\max}+a_0} P(t) 2 \pi t dt
 \label{eq:ch3_sigma_general}
\end{equation}

In eq. \ref{eq:ch3_sigma_general} $t$ can is interpreted as the ion’s impact parameter, 
while $r$ in the average radial dose distribution formula, $D_e(r)$, is the radial distance between
the centre of the extended target and the ion’s path. Equation \ref{eq:ch3_sigma_general} also
obtains if the centre of the extended target is considered to be displaced from
the ion’s path by a radial distance $t$ , over the range of $t$ from $0$ to $r_{\max}+a_0$.

The unit of inactivation cross-section is $cm^2$ , i.e. that of area. Its value
strongly depends on the ion’s energy (which affects the extent of $D_e(t)$ via the
maximum range of delta-rays, $r_{\max}$ ), and on the ion’s charge (which affects the
dose values in $D_e(t)$, via $z^{\star 2}/\beta^2$ of the ion).

\begin{figure}[p!]
 \centering
\includegraphics[width=0.8\textwidth]{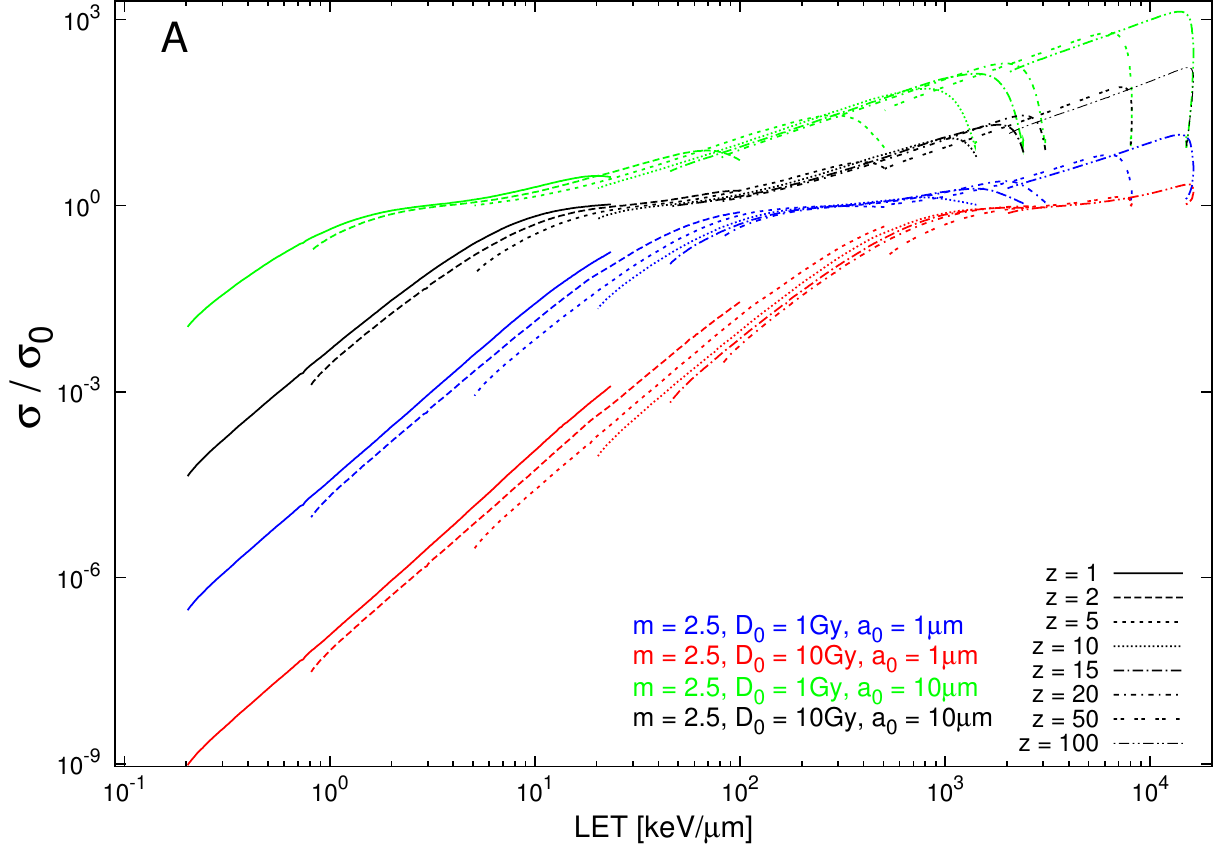}
\includegraphics[width=0.8\textwidth]{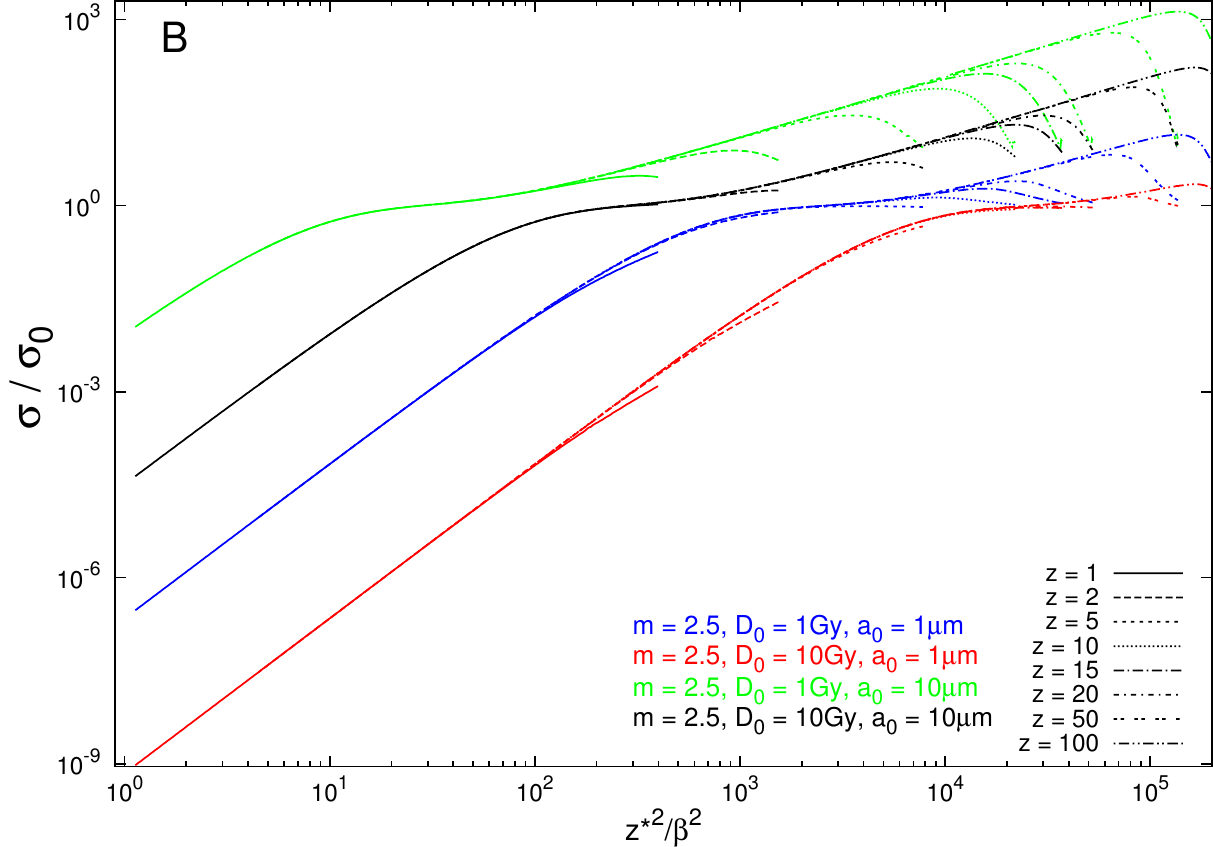}
 \caption{Inactivation cross-sections normalized to their plateau values ($\sigma_0$),
plotted as a function of LET (panel A) or $z^{\star 2}/\beta^2$ (panel B) for $m$=2.5, 
$D_0$ = 1~Gy or 10~Gy and $a_0$ = 1~$\mu$m or 10~$\mu$m. Each curve was plotted for ions with $Z$ ranging
from 1 to 100 and for relative ion speed $\beta$, from $\beta$= 0.05 to $\beta$=0.99. The radial
dose distribution function of Zhang was used in these calculations. Figure
reprinted from Korcyl \citep{KorcylPhD_ref45}}
 \label{fig:ch3_inact_plateau}
\end{figure}

In Fig. \ref{fig:ch3_inact_plateau} are shown examples of systematic calculations of extended target
cross sections for sensitive targets of different dimensions, in cellular
systems of different radiosensitivity, versus ion LET (panel A) or $z^{\star 2}/\beta^2$ (panel
B) for ions of different charges and different energies (as given by their
relative speeds, $\beta$) \citep{KorcylPhD_ref45}. 
In these calculations Zhang’s $D(r)$ formula ($I=0$) was
used to calculate the respective extended target radial distributions of dose,
$D_e(r)$. The calculated cross-section values have all been normalised to their
plateau values, represented by $\sigma_0$ . Here, the scaling properties of the Katz
model become evident: when plotted versus $z^{\star 2}/\beta^2$ (Fig. \ref{fig:ch3_inact_plateau} 
panel B), rather than versus ion LET (Fig. \ref{fig:ch3_inact_plateau}, panel A), 
the activation cross sections for
targets of widely differing sizes, calculated for ions of widely different
charges and energies, appear to follow singular trends (“universal curves”),
dependent on the target size and cell radiosensitivity, and only branching apart
at their highest values (we note that in this example the $m$-target
parameter remains the same, $m=2.5$, in all calculations).

As may be observed in  panel B of Fig. \ref{fig:ch3_inact_plateau}, the general trend of the
dependence of the single-particle activation cross sections, versus $z^{\star 2}/\beta^2$ is as
follows: with increasing $z^{\star 2}/\beta^2$ the cross sections first rapidly rise through
several orders of magnitude, then saturate at some plateau values (represented
by the $\sigma_0$ Katz model parameter) and next further increase to their maximum
values and finally rapidly decrease, forming characteristic “hooks”.
For each family of inactivation cross-section dependences on $z^{\star 2}/\beta^2$ (as
determined by parameters $m$, $D_0$ and $a_0$) one may find such a value $\kappa$ on the
$z^{\star 2}/\beta^2$ abscissa at which:
\begin{itemize}
\item if $z^{\star 2}/\beta^2 < \kappa$ - inactivation cross-sections increase,
\item if $z^{\star 2}/\beta^2 \approx \kappa$ - inactivation cross-sections saturate at around 
$\sigma(z^{\star 2}/\beta^2) = \sigma_0$,
\item if $z^{\star 2}/\beta^2 > \kappa$ - inactivation cross-sections increase above $\sigma_0$ and “hooks”
occur.
\end{itemize}

The first region, where $z^{\star 2}/\beta^2 < \kappa$, is called \textit{grain-count regime}, while the
last one, $z^{\star 2}/\beta^2 > \kappa$ is called the \textit{track-width regime}. These names, often
used in the publications of Robert Katz, originate from his studies of ion
tracks in nuclear emulsion: in tracks of ions of high energy (i.e., of low LET
and low $z^{\star 2}/\beta^2$) ionizations appear mostly along the ion paths as sparsely
distributed exposed grains. As the ion gradually slows down (i.e., as its LET
and $z^{\star 2}/\beta^2$ increase) denser ionisations make the track continuous and then
thicker and broader (hence the “track width” regime). Close to the end of the
ion track (where LET and $z^{\star 2}/\beta^2$ are highest) the track appears to be very dense
and then gradually thins down before it finally ends. Track “thin-down” effect
is represented by the “hooks” in the cross-section calculations of Fig. \ref{fig:ch3_inact_plateau}. 
It is worth noting that the track thin down occurs over the region where LET is
highest (i.e. over the Bragg peak region). Consequently, over the thindown
region most of the dose may be “wasted”, i.e. may not fully contribute to the
biological effect, due to the limited range of the radial dose distribution.

\section{The ion-kill and gamma-kill modes of inactivation}

As mentioned earlier in par. \ref{sec:ch3_principles} ion-kill and gamma-kill modes of cell
inactivation are introduced in Katz’s model. The ion kill mode is related to
passage of a single ion through the cell nucleus which results directly in cell
death, i.e. a purely exponential survival curve. The probability of inactivation
in the ion-kill mode $\Pi_i$ is defined as:

\begin{equation}
 \Pi_i = \exp( - \sigma F)
 \label{eq:ch3_ionkill_mode}
\end{equation}

where $F$ is the ion fluence and $\sigma$ is the single-ion inactivation cross
section.

In the case of ion irradiation, cell death in the gamma-kill mode is related to
the overlap of delta rays emitted from different ions. In this case accumulation
of effects from several delta-rays emitted from different ions is needed to
produce a lethal effect. The gamma-kill mode of inactivation is represented by a
shouldered survival curve (due to fact that typically, $m$>1). The probability of
inactivation in the gamma-kill mode is given by a formula based on the m-target
model:

\begin{equation}
\Pi_{\gamma} = 1 - (1 - e^{-D_{\gamma}/D_0})^m
 \label{eq:ch3_gammakill_mode}
\end{equation}

where $D_0$ and $m$ are m-target model parameters and $D_{\gamma}$ is a fraction $P_{\gamma}$
of the ion dose $D$, delivered in gamma-kill mode:

\begin{equation}
D_{\gamma} = P_{\gamma}  D
\end{equation}

In Katz's model, the fraction $P_{\gamma}$ of the ion dose delivered in the
gamma-kill mode is assumed to be given by the following formula:

\begin{equation}
 P_{\gamma} = \begin{cases} 1 - \sigma / \sigma_0 &\text{if } \sigma \leq \sigma_0\\
0&\text{elsewhere}\end{cases}
\end{equation}

where $\sigma_0$ is the saturation value of the inactivation cross-section $\sigma$. 
In the case of a monoenergetic beam of ions of one type, survival in the Katz model
is defined as the product of gamma-kill and ion-kill components:

\begin{equation}
 S = \Pi_i \Pi_{\gamma}
 \label{eq:ch3_survival_general}
\end{equation}

These two modes of inactivation are not mutually exclusive. The total ion dose $D$
always contributes to the ion-kill mode and at the same time, a fraction of this
total dose, $D_{\gamma}$ contributes to the gamma- kill mode. 
Gamma kill dominates in the grain-count regime and ion-kill dominates in the
track-width regime.

\section{Algorithm of the integrated version of the Katz model}
\label{sec:ch3_integrated}
In this thesis a distinction is made between the \textit{integrated} version of the Katz
model (free parameters: $m$, $D_0$, $\sigma_0$ and $a_0$) and the \textit{scaled} version of the Katz
model (free parameters: $m$, $D_0$, $\sigma_0$ and $\kappa$). Here, the integrated version
is discussed, while the scaled version is discussed in the next paragraph.

The single-particle inactivation cross-section can be calculated using formula
of Section \ref{sec:ch3_inactivation}, equation \ref{eq:ch3_sigma_general}. 
This leads to a model with four
free parameters:  $m$, $D_0$, $\sigma_0$ and $a_0$. Once the single-particle inactivation
cross section is known, calculation of cell survival for different ion doses (or
fluences) can be readily performed, using the ion-kill and gamma-kill mode
formulae \ref{eq:ch3_ionkill_mode}-\ref{eq:ch3_survival_general}.

The calculation algorithm consists of the following steps:

\begin{itemize}
 \item[Step A] --- calculation of inactivation cross section:
\begin{align}
 D_e(r) &= 1 / \pi a_0^2 \int_{r_{\min}}^{a_0+r_z} D(r) \Phi( a0, r, t ) t dt \\
 P(t) &= (1 - e^{ -D_e(t)/D_0})^m \\
 \sigma &= \int_0^{r_{\max}+a_0} P(t) 2 \pi t dt
\end{align}
 \item[Step B] --- calculation of ion- and gamma-kill modes:
\begin{align}
 P_\gamma &= \begin{cases} 1 - \sigma / \sigma_0 &\text{if } \sigma \leq \sigma_0\\
0&\text{elsewhere}\end{cases} \\
 D_{\gamma} &= P_{\gamma}  D \\
 \Pi_\gamma &= 1 - (1 - e^{-D_\gamma/D_0})^m\\
 \Pi_i &= e^{- \sigma F} \\
 S &= \Pi_\gamma \Pi_i
\end{align}
\end{itemize}

This calculation becomes more involved if cellular survival needs to be
calculated after irradiation by different ions species of different energies, in
which case double integration is required to calculate each single-particle
inactivation cross section. This is a time consuming procedure. Evaluation of
extended target dose from equation \ref{eq:ch3_exttarget_polar} 
requires calculation of the integral of
function $D(t) \Phi(t) t$ which cannot be done analytically. The result of this
calculation is substituted into equation \ref{eq:ch3_katz_probability_general} 
and again integrated in equation \ref{eq:ch3_sigma_general} to calculate the 
inactivation cross-section.

While Katz performed several such calculations to establish the scaling
principles of his analytic Cellular Track Structure Theory (termed in this
thesis the scaled version of his model), to apply the above approach to best-fit
model parameters to survival data was impractical, as 30-40 years ago when numerical
integration was  a much more time-consuming task than it is today. The novel
approach in this thesis is to investigate the possible introduction of an
“integrated version of the Katz model” in modelling cellular survival.

One of the advantages of this integrated version of the Katz model is that it
offers the possibility of applying any radial dose distribution function to
describe ionisation distributions around the ion’s path. In particular, in the
integrated version of the Katz model it would be possible to include Geiss’s
radial dose distribution function and perhaps the linear-quadratic description
of cellular survival (both are used in the LEM). It would then be interesting to
investigate differences between LEM and Katz’s model approaches if the same
radial dose distribution function is used in the calculations, and perhaps also
if two different representations of cellular survival curves are applied.

\section{Algorithm of the scaled version of the Katz model}
\label{sec:ch3_scaled}
The Cellular Track Structure Theory which Robert Katz \citep{katz1971cellular_ref58} originally proposed, 
consists of a set of simple analytical formulae in which the scaling properties
of his track structure approach are exploited. The possibility of applying
$z^{\star 2}/\kappa\beta^2$ as a scaling factor in calculating inactivation cross sections is
demonstrated in Fig. \ref{fig:ch3_inact_plateau}. A key result from Katz’s numerical calculations
involving multiple integrations (called here the integrated version and
performed in a manner described in the preceding paragraph), was the observation
that $z^{\star 2}/\kappa\beta^2$ could be used to scale the results of such
calculations. The possibility of such scaling is closely related to choice of
the point target radial dose distribution formula, $D(r)$, and to the choice of
the m-target formulation in describing cellular survival after doses of
reference radiation.  A careful re-analysis of the scaling properties of Katz’s
model equations was performed by Marta Korcyl \citep{KorcylPhD_ref45} and has been submitted for
publication \citep{Korcyl2013_ref85}

In Fig. \ref{fig:ch3_inact_plateau} inactivation cross sections calculated for different ion species of
different energies using Zhang’s radial dose distribution and plotted versus
$z^{\star 2}/\kappa\beta^2$ are seen to form families of overlapping curves. Similar behaviour is
also observed if Katz’s radial dose distribution function is used in such
calculations, but not if Cucinotta’s formula is applied \citep{KorcylPhD_ref45}. If we neglect the
region of “hooks” in Fig. \ref{fig:ch3_inact_plateau}, then for each family of curves of the same colour
a “universal” dependence can be drawn, as a scaled approximate representation of
inactivation cross-section for ions various values of $Z$ and $\beta$.

Robert Katz proposed the following formula for the “universal
curve” $g(m,x) = (1 - \exp(-x))^m$ , which would be a valid approximation of
inactivation cross-section in the grain-count regime:

\begin{equation}
\frac{\sigma}{\sigma_0} = g(m, z^{\star 2}/\kappa\beta^2) = (1 - \exp( - z^{\star 2}/\kappa\beta^2))^m      
\label{eq:ch3_universal_curve}
\end{equation}

In the track-width regime there is no simple “universal” formula, but a high
order polynomial $f( m, z^{\star 2}/\kappa\beta^2)$ can be fitted to the curves
calculated using eq \ref{eq:ch3_sigma_general}.

As values of $z^{\star 2}/\kappa\beta^2$ increase, the value of $g(m, z^{\star 2}/\kappa\beta^2)$
converges to unity. The $g$ function will not reach exactly unity, thus $0.98$ is
taken as the value at which saturation is achieved. At this value smooth
transition to approximation of the cross-section by polynomial $f$ is possible.

The presently used formulae of the original Katz model (here termed as the
scaled version of this model) are the following

For $( 1 - \exp(- z^{\star 2}/\kappa\beta^2 )^m < 0.98$ :

\begin{equation}
\sigma = \sigma_0 ( 1 - \exp(- z^{\star 2}/\kappa\beta^2) )^m 
\end{equation}

and elsewhere:

\begin{equation}
\sigma = \sigma_0 f(m, z^{\star 2}/\kappa\beta^2)
\end{equation}

where f is a polynomial approximation of the inactivation cross section.

The algorithm of the scaled Katz model calculation consists of the following
steps:

\begin{itemize}
 \item[Step A] --- calculation of inactivation cross section:
\begin{align}
 \sigma &= \begin{cases} \sigma_0 ( 1 - \exp(- z^{\star 2}/\kappa\beta^2) )^m  &\text{if } ( 1 - \exp(- z^{\star 2}/\kappa\beta^2 )^m < 0.98\\
\sigma_0 f(m, z^{\star 2}/\kappa\beta^2) &\text{elsewhere}\end{cases}
\end{align}
 \item[Step B] --- calculation of ion- and gamma-kill modes:
\begin{align}
 P_\gamma &= \begin{cases} 1 - \sigma / \sigma_0 &\text{if } \sigma \leq \sigma_0\\
0&\text{elsewhere}\end{cases} \\
 D_{\gamma} &= P_{\gamma}  D \\
 \Pi_\gamma &= 1 - (1 - e^{-D_\gamma/D_0})^m\\
 \Pi_i &= e^{- \sigma F} \\
 S &= \Pi_\gamma \Pi_i
\end{align}
\end{itemize}

\section{Comparison between Scaled and Integrated versions of the Katz Model}
As presented above, a distinction was made in this thesis between the integrated
version of the Katz model, where the free model parameters are: $m$, $D_0$, $\sigma_0$
and $a_0$,  and the scaled version of the Katz model, where the free parameters
are: $m$, $D_0$, $\sigma_0$ and $\kappa$. It should be stressed that when fitting either of
these versions of the Katz model to e.g., experimental data points on measured
survival curves of cells in vitro, the values of best fitted parameters using
either approach may differ.

Representations of the cross section in the scaled approach ignore the presence
of “hooks” , or track thindown over the final parts of ion tracks. Over this
region, where $z^{\star 2}/\kappa\beta^2 > 1$ and ions are in their Bragg peak region, values
of inactivation cross sections calculated using integrated and scaled versions
of the Katz model differ by up to an order of magnitude. However, satisfactory
agreement between results of these two versions over the grain count regime
($z^{\star 2}/\kappa\beta^2 < 1$) and the fact that the “hook” regions constitute only a
small fraction of the ion’s total range at ion energies of interest to
radiotherapy, justifies the use of the scaled (i.e. the original) version of the
Katz model. It is much simpler and faster to calculate in applications relevant
to ion radiobiology and radiotherapy planning.

\begin{figure}[h!]
 \centering
 \includegraphics[width=0.8\textwidth]{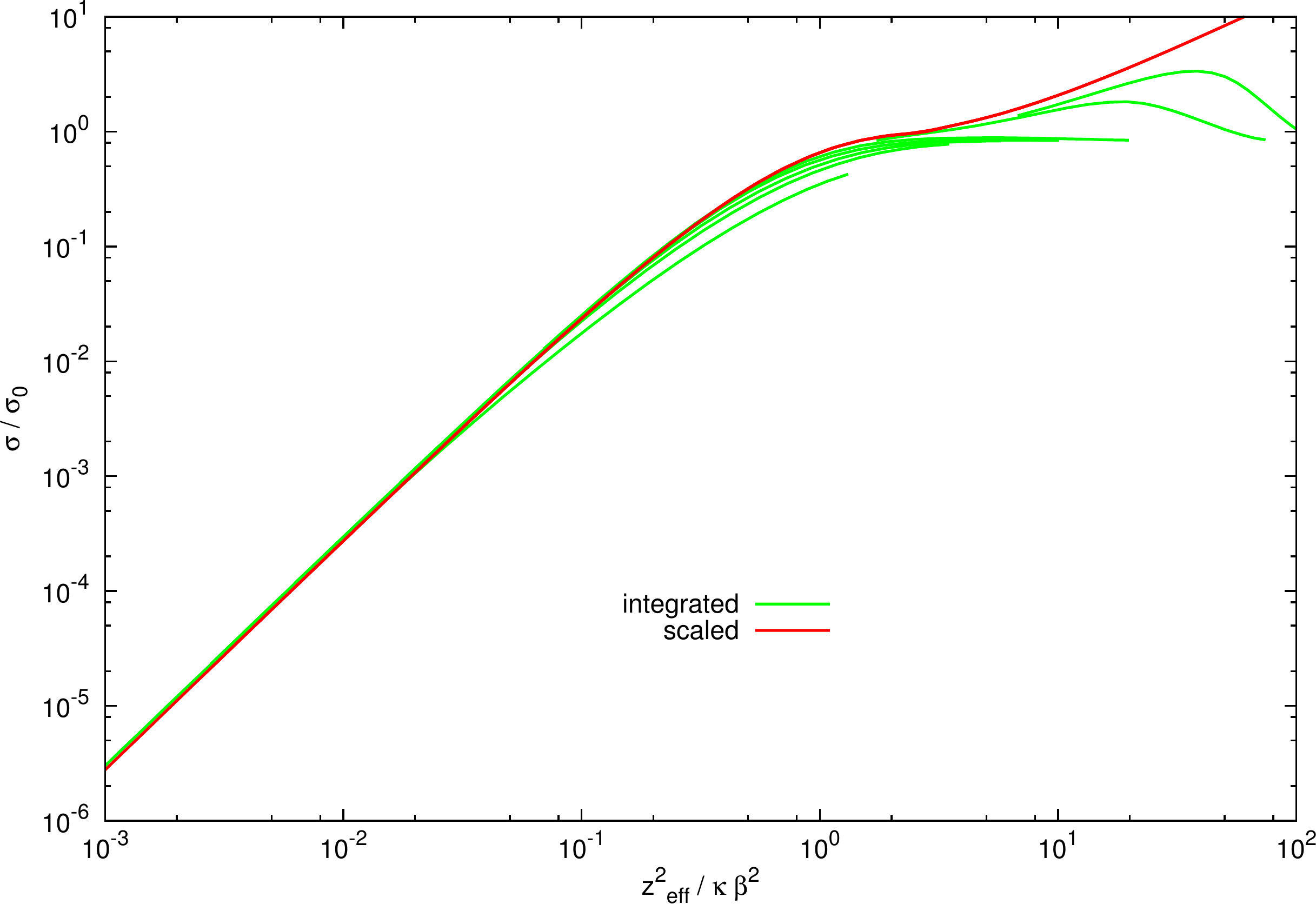}
 \caption{Inactivation cross-sections, normalised to $\sigma_0$ vs. $z^{\star 2}/\kappa\beta^2$, based on
the radial dose distribution function of Katz. Calculations were performed using
the integration (green lines) and scaled (red lines) versions of the Katz model,
for ions of Z=1,2,3,5,10,50 and 100 and energies 0.1 - 500 MeV/amu.  Katz model
parameters used: $m$ = 2, $D_0$ = 5 Gy, $a_0$ = 1$\mu$m, $\kappa$ = 1500, $\sigma_0$ = 1.2 $\pi a_0^2$}
 \label{fig:ch3_inactKatz}
\end{figure}

It appears that $z^{\star 2}/\beta^2$ scaling in Katz’s model may be applicable only to
selected radial dose distributions \citep{KorcylPhD_ref45}. This is illustrated in 
Fig. \ref{fig:ch3_inactKatz} where
results of scaled and integrated model calculations of inactivation cross
sections, obtained using the Katz $D(r)$ formula are compared. Calculations were
performed over a broad range of ion energies (0.1 - 500 MeV/amu) and ion species
($Z=1\ldots100$).

\begin{figure}[h!]
 \centering
 \includegraphics[width=0.8\textwidth]{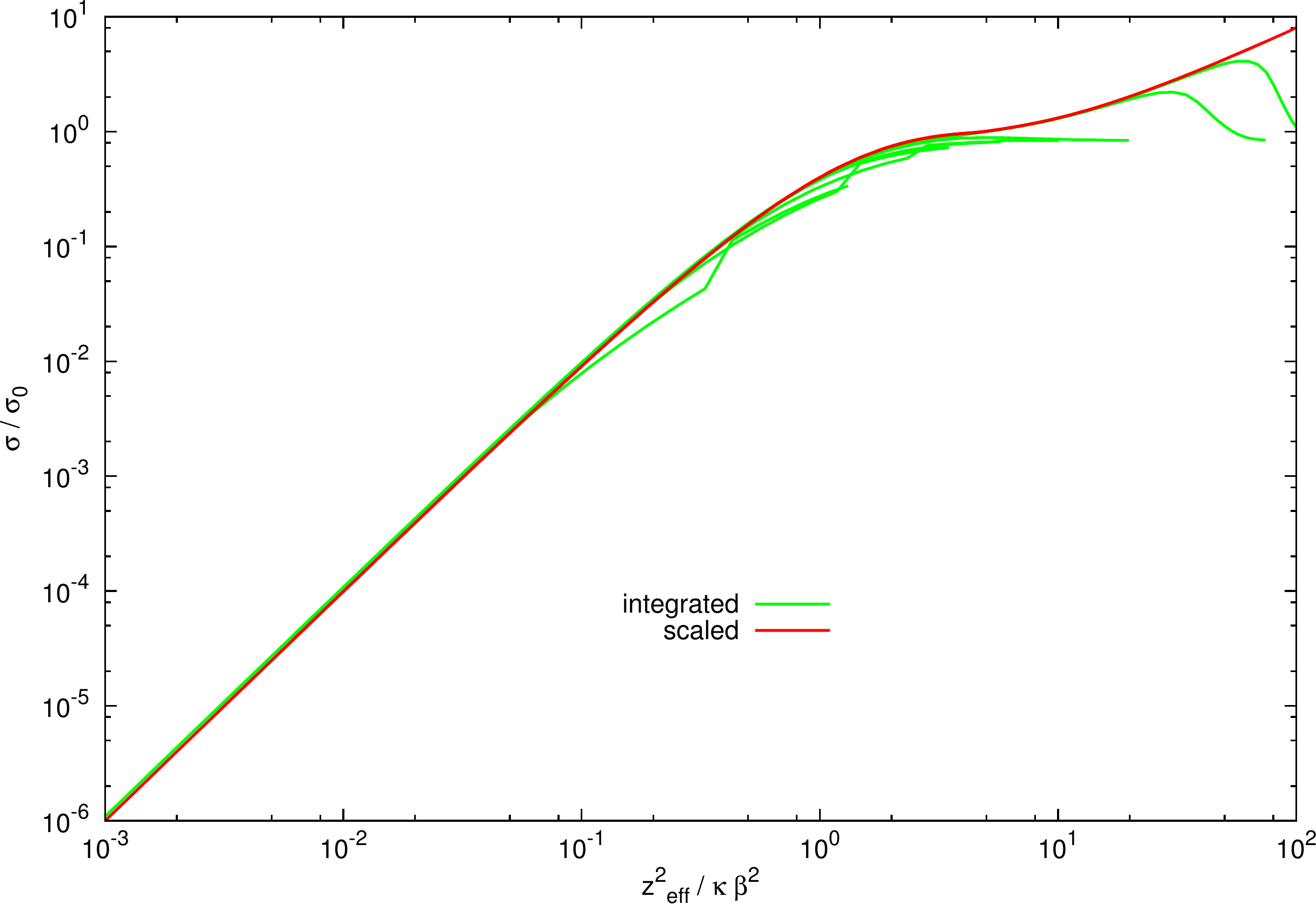}
 \caption{Inactivation cross-sections, normalised to $\sigma_0$ vs. $z^{\star 2}/\kappa\beta^2$, based on
the radial dose distribution function of Zhang. Calculations were performed
using the integration (green lines) and scaled (red lines) versions of the Katz
model, for ions of Z=1,2,3,5,10,50 and 100 and energies 0.1 - 500 MeV/amu.  Katz
model parameters used: $m$ = 2, $D_0$ = 5~Gy, $a_0$ = 1~$\mu$m, $\kappa$ = 2500, $\sigma_0$ = 1.2 $\pi a_0^2$}
 \label{fig:ch3_inactZhang}
\end{figure}

In Fig. \ref{fig:ch3_inactZhang} a similar comparison is made, but after calculations based on
Zhang’s radial dose distribution formula, demonstrating that better agreement
between results of either version of the model can be obtained using Zhang’s
$D(r)$ formulation.

The scaling elements of Katz’s model, such as the “universal curve”, eq. \ref{eq:ch3_universal_curve},
which is apparent, e.g., in Fig. \ref{fig:ch3_inactZhang}, could not be established for cross
sections calculated using radial dose distribution formulae of Geiss or of
Cucinotta, both of which satisfy the LET condition, eq. \ref{eq:ch2_rddnormdef}, by definition
\citep{KorcylPhD_ref45}. Thus, while they would not appear to be suitable for use in the scaled
version of Katz’s model, use could be made of either of these $D(r)$ formulae in
the integrated version.  It would be particularly interesting to apply Geiss’s
$D(r)$ formula in such calculations, since it is used in LEM. The integrated
version of Katz’s model could be applied in radiobiology modelling but it is too
computation-intensive to be used in radiotherapy planning. While, in principle,
the processing times of the nested numerical integrations could be significantly
reduced by special programming techniques, such as parallel code execution,
application of the integrated version of the Katz model in radiotherapy planning
would still be impractical.

\section{Fitting Katz model parameters from cell survival data}
\label{seq:ch3_fitting}

Values of cellular parameters of the Katz model were extracted from the
published set of cellular survival curves measured after irradiation of Chinese
Hamster Ovary (CHO) cells by carbon ion beams, data of Weyrather et al. \citep{Weyrather1999_ref24}.
Data points on the published survival curve plots \citep{Weyrather1999_ref24} were digitized and formed
into a data base from which values of four parameters of the Katz model (scaled
version) were found, such that model-predicted values of these data points were
best reproduced, as determined by the minimum value of $\chi^2$. Computer codes of
the libamtrack library were used for this purpose.

The implementation of Katz’s model in the libamtrack library was used in the
following configuration:
\begin{itemize}
 \item Waligorski delta electron range model
 \item Zhang radial dose distribution function
 \item inactivation cross section calculated with the approximation method
 \item stopping power tables based on PSTAR database
\end{itemize}

Fitting of the Katz model parameters to the experimental data was performed by
minimizing $\chi^2$ defined as the sum of squared differences between
model-predicted and experimental values of all data points (logarithm of
survival) obtained from irradiation of CHO cells by ion beams and by reference
radiation (250 kVp X-rays) :

\begin{equation}
 \chi^2 = \chi^2_{\rm ion} + \chi^2_{\rm Xrays}
\end{equation}

In the $\chi^2_{\rm ion}$ - ion beam component of the $\chi^2$ sum, survival $S_{\rm Katz,ion}$ was
calculated using Katz’s model (scaled version):

\begin{equation}
\chi^2_{\rm ion} = \sum_E \sum_D ( \ln( S_{\rm Katz,ion}(D,E) ) - \ln(S_{\rm experiment,ion}(D,E)) )^2 
\end{equation}

In the X-ray component of the $\chi^2$ sum, survival $S_{\rm model,Xrays}$ was calculated using
the m-target formula:

\begin{equation}
\chi^2_{\rm Xrays} = \sum_D ( \ln( S_{\rm model,Xrays}(D) ) - \ln(S_{\rm experiment,Xrays}(D)) )^2 
\end{equation}

where:

\begin{equation}
S_{\rm model,Xrays}(D) = 1 - (1-\exp(-D/D_0))^m
\end{equation}

Data published by Weyrather contain six survival curves, and in total n = 103 data
points.

The numerical algorithm of gradient minimization L-BFGS-B (implemented in python
scipy library, \citep{L-BFGS-B_ref46}) was used to minimize the $\chi^2$ function 
($\chi^2$ gradient was approximated numerically). The minimization algorithm was implemented by the
author in Python programming language, cell survival calculations were
implemented in the libamtrack library. As the starting point of L-BFGS-B, the
following parameters were selected:

\begin{equation}
m = 2 \quad D_0 = 5\ Gy \quad \sigma_0 = 1.42 \cdot 10^{-12}\ m^2 \quad \kappa = 1230 
\end{equation}

The minimization algorithm allows limits to be set on the parameter space. In
this case search for best fitted parameters was limited by following conditions:
\begin{eqnarray}
1 <& m &< 5 \nonumber\\
1.1\ Gy <& D_0 &< 3\ Gy \nonumber \\
10^{-13}\ m^2 <& \sigma_0 &< 10^{-9}\ m^2 \nonumber \\
200 <& \kappa &< 5000
\end{eqnarray}

As a result of the calculation, four best fitting parameters were found:

\begin{equation}
m = 2.31 \quad D_0 = 1.69\ Gy \quad \sigma_0 = 5.96 \cdot 10^{-11}\ m^2 \quad \kappa = 1692.8
\label{eq:ch3_bestfitted}
\end{equation}

Such parameter values lie well within the range of typical values of cellular
parameters of the Katz model published in the literature \citep{KatzParamCollection_ref47}.

The goodness of the fit, expressed as $\chi^2/(n-n_{\rm dof})$ is equal to 0.0061. Here
$n_{\rm dof}$ denotes the number of degrees of freedom which here is equal to the number
of parameters ($n_{\rm dof}$ = 4).

Substituting $p_1 = m$, $p_2 = D_0$, $p_3 = \sigma_0$, $p_4 = \kappa$, elements $C_{jk}$ of the Hesse
matrix of second derivatives of $\chi^2$ may be written as:

\begin{equation}
C_{jk} = \frac{\partial^2 \chi^2}{\partial p_j \partial p_k} ( p_1^{\rm opt}, p_2^{\rm opt}, p_3^{\rm opt}, p_4^{\rm opt}) \text{ where } j,k = [1\ldots4]   
\end{equation}

and $p_1^{\rm opt}, p_2^{\rm opt}, p_3^{\rm opt}, p_4^{\rm opt}$ are the parameters for which the $\chi^2$
function is at minimum.

Assuming Gaussian errors on the measured data, uncertainties of the fitted
parameters may be read from the correlation matrix diagonal $C^{-1}_{jj}$, estimated
from the inverse Hesse matrix: 

\begin{equation}
\sigma^2_j = \frac{\chi^2(p_1^{\rm opt}, p_2^{\rm opt}, p_3^{\rm opt}, p_4^{\rm opt})}{n-n_{\rm dof}} C^{-1}_{jj}
\end{equation}

The uncertainties of the best-fitted parameter values, calculated by numerical
evaluation of the correlation matrix are:

\begin{equation}
m^{\rm err} = 0.026 \quad D_{0}^{\rm err} = 0.016\ Gy \quad \sigma_{0}^{\rm err} = 9.45 \cdot 10^{-13}\ m^2 \quad \kappa^{\rm err} = 9.7
\label{eq:ch3_bestfitted_uncert}
\end{equation}

\begin{figure}[ht!]
 \centering
 \includegraphics[width=0.8\textwidth]{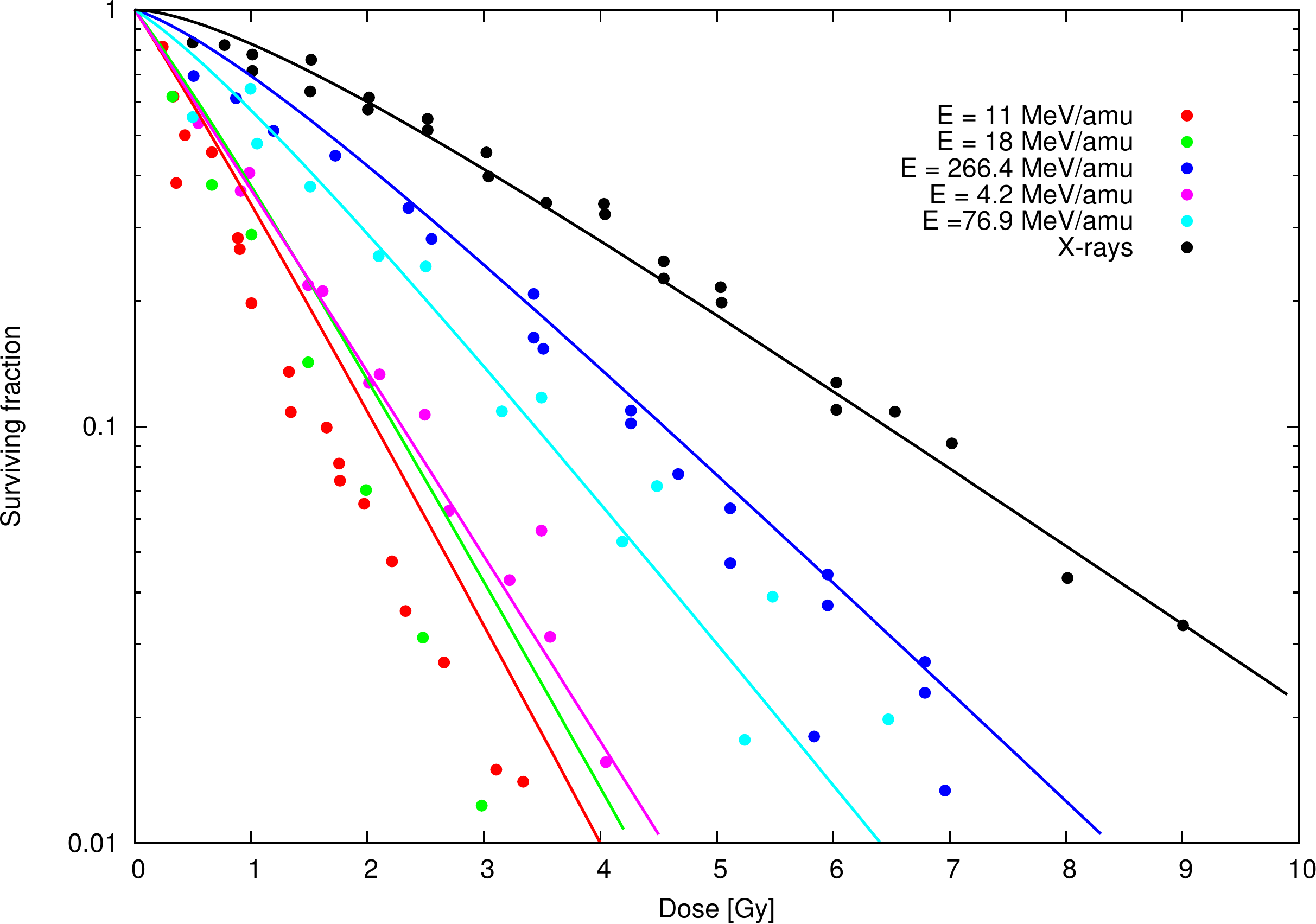}
 \caption{Katz model-predicted CHO cell survival curves (based on best-fitted
parameters) and experimentally measured data points for carbon beams (of
energies 4.2, 11, 18, 76.9 and 266.4 MeV/amu) and for 250 kVp X-rays. The
experimental data is from \citep{Weyrather1999_ref24}. 
The Katz model parameters are: $m = 2.31, \, D_0 = 1.69\ Gy, \, \sigma_0 = 5.96 \cdot 10^{-11} m^2, \, \kappa = 1692.8$
Calculations based on the libamtrack library.}
 \label{fig:ch3_fit}
\end{figure}

In Fig. \ref{fig:ch3_fit} comparison is shown between the Katz model-predicted survival curves
(based on best-fitted parameters) and experimentally measured data points, over
the dose range 0-10 Gy of carbon ion beams (of energies 4.2, 11, 18, 76.9 and
266.4 MeV/amu) and 250 kVp X-rays. Although model calculations reproduce the
experimental data quite well, for two survival curves (11 MeV and 18 MeV carbon
ions) cell survival is overestimated by these calculations.

\section{Mixed-field calculations}
\label{sec:ch3_mixed}
Katz model parameters were derived in the previous section under the assumption
that all irradiations were performed using monoenergetic carbon ions. In mixed
radiation fields, irradiation by a set of carbon ions of different energies and
by different ion species, due to carbon fragmentation, needs to be accounted
for. Track segment irradiation is assumed, i.e. that the energy and fluence of
each ion species are specified. According to the principles of the Katz model,
gamma kill and ion kill need to be calculated for each component and combined
together to yield the final cell survival. The following procedure of a
mixed-field calculation has been proposed by Katz \citep{katz1971cellular_ref58}:

Let us assume that the total dose $D$ is delivered as a sum of $N$ dose components
$D_i (i=1\ldots N)$ due to $N$ ions, each of charge $Z_i (i=1\ldots N)$, energy $E_i (i=1\ldots N)$.
 Let the respective fluence of each component be denoted as $F_i$.

Let $P_i$ be fraction of the dose $D_i$ delivered in gamma-kill mode by the $i$-th field
component, defined as:

\begin{equation}
 P_i = \begin{cases} \sigma_i / \sigma_0 &\text{if } \sigma_i \leq \sigma_0\\
1&\text{elsewhere}\end{cases}
\end{equation}

where $\sigma_i$ is the inactivation cross section for the $i$-th component.

The combined ion-kill mode survival is then equal to: 

\begin{equation}
\Pi_i = \exp\left( - \sigma_0 \sum_{i=1}^N P_i F_i \right)  = \prod_{i=1}^N \exp(-\sigma_0 P_i F_i)
\label{eq:ch3_mixed_ionkill}
\end{equation}

The combined gamma-kill mode survival is equal to:

\begin{equation}
\Pi_{\gamma} = \left(1 - \left(1 - \exp\left(- \frac{1}{D_0} \sum_{i=1}^N (1-P_i)D_i\right)\right)^m\right)
\label{eq:ch3_mixed_gammakill}
\end{equation}

Finally, the combined surviving fraction SF can be written as the product of
both modes 

\begin{equation}
 S = \Pi_i \Pi_{\gamma}
\end{equation}

In the ion kill mode the total surviving fraction is a product of ion kill
survivals of the components, according to equation \ref{eq:ch3_mixed_ionkill}. 
In the gamma kill mode the total surviving fraction is neither a product nor a sum of surviving
fractions of the components. The overall result is calculated by estimating the
contribution of each component to the “ion dose” and then mixed using the
non-linear m-target formula, as shown in equation \ref{eq:ch3_mixed_gammakill}.

\section{Conclusions}

The Katz model (here termed as its scaled version) is a fast and reliable method
to describe and predict biological cell survival in carbon ion beams. The model
has been continuously developed over the last 50 years \citep{KatzTracks_ref48}, \citep{KorcylPhD_ref45}. 
One of the
aims of this thesis was to demonstrate the features of the model (here - its
integrated version) if integration is applied explicitly, as compared to its
usual analytical form (scaled version). In the scaled version, track thindown,
as represented by “hooks” in the cross-section dependences at the highest ion
LET values, is neglected. This may lead to discrepancies for slow, stopping
ions. For ions of energies relevant in radiotherapy, the predictive capability
of the Katz model (its scaled version) is quite satisfactory. Due to its
analytic simplicity, it is extremely time-efficient in computation. The model
offers a well-specified procedure for calculating cellular survival in mixed ion
fields, provided that energy-fluence spectra are available for all the ion field
components.

The author of this thesis has implemented in the freely available open-source
libamtrack library both versions of the Katz model: Katz’s original formulation
(scaled version) and the integrated version.

The scaled version of Katz’s model, applying Zhang’s radial dose distribution
formula, was used by the author to find best-fitted values of model parameters
describing cell survival of CHO cells irradiated with carbon ions. Although the
overall agreement between model predictions and experiment is quite
satisfactory, model calculations overestimate the measured survival of these
cells after low energy (11 MeV/amu and 18 MeV/amu) carbon ions. This variant of
the Katz model will be applied in later calculations in Chapter 5 of this
thesis.

The best fitted values of the Katz model parameters will be applied in the
mixed-field calculations of survival of CHO cells in a realistic carbon beam
model. This beam model includes ion fragmentation and spread-out Bragg peak
(SOBP) configuration, as discussed in Chapter 4 of this thesis.

\mgrclosechapter

\chapter{Modelling the Transport of Carbon Beams}
A therapeutic carbon ion beam travelling through the patient’s tissues undergoes
a complicated pattern of interactions, including slowing down, scattering and
ion fragmentation.  In order to apply the Katz model to calculate survival of
cells of those tissues after their exposure to this carbon beam, prior knowledge
of fluence and energy spectra of primary and secondary particles is required at
all beam depths. The region of the spread out Bragg peak, which is adjusted to
match the tumour volume, is one where establishing such energy-fluence spectra
may be particularly difficult.

In this chapter an approach is developed and presented which allows
energy-fluence spectra of carbon ions and all fragments to be calculated at
different depths in water. The results of Monte-Carlo transport calculations of
the carbon beam in water performed by Pablo Botas using the SHIELD-HIT10A code
(Aarhus branch of the original development line) \citep{SHIELD-HIT_ref38}, \citep{Hansen-SHIELD_ref60} were organized in the form of a look-up
database attached to the libamtrack library. The author of this thesis developed
an interpolation algorithm to calculate fluence and energy values at
intermediate energies, and also a tool to calculate a linear combination of
pristine Bragg peaks which gives the desired depth-dose profile (i.e. a flat
depth-dose distribution over the Spread-Out Bragg Peak region). The developed
codes are now included in the libamtrack library.

\section{Monte-Carlo simulations and energy-fluence spectra.}

\subsection{Monte-Carlo simulations of carbon ion beams.}
The SHIELD hadron transport code performs Monte Carlo simulation of the
interaction of hadrons and atomic nuclei with complex extended targets. Its
medical version, SHIELD-HIT (Heavy Ion Therapy), is designed to simulate
interactions of therapeutic beams of protons and heavier ions with human tissues
over the energy range relevant
for radiotherapy. Models of nuclear reactions which describe various stages of
the inelastic hadron-nucleus and nucleus-nucleus interactions, developed mainly
by Demenyev and Sobolevsky at JINR (Dubna) and INR RAS (Moscow), are applied.
The models are grouped together in the MSDM-generator (Multi Stage Dynamical
Model) which allows simulation of a whole chain of nuclear reactions. Ionization
losses of charged hadrons and nuclear fragments in SHIELD-HIT10A are calculated
according to the Bethe-Bloch equation but include various models and data for
computation of mean ionization loss, fluctuations of the ionization loss and of
multiple Coulomb scattering.

Transport of 50-500 MeV/amu carbon ion beams in water was simulated using
SHIELD-HIT10A by Pablo Botas (DKFZ). Results of Monte-Carlo (MC) simulations
concerned primary beam attenuation, beam broadening with increasing depth and
production of secondary fragments. Fluences of the primary beam and its
fragments were scored over planes perpendicular to the ion paths and averaged.
Thus, a one-dimensional set of energy-fluence spectra versus depth was obtained
for a large number of carbon beams over a large range of initial energies. It is
in principle possible to extend the existing MC simulation to obtain results in
three-dimensional sets by increasing significantly the number of simulated
particles and using a three-dimensional scoring grid (or to assume rotational
symmetry and use only two-dimensions with cylindrical scoring volumes). 
SHIELD-HIT10A results were applied in this work as they reproduced the
experimental data well \citep{SHIELD-HIT_ref38} and because the code was suitable for producing
spectral data in binary form. Other Monte Carlo codes, such as Geant4, FLUKA and
PHITS were considered for this purpose, but their application would require more
effort in adapting their output to the existing libamtrack environment.

\begin{figure}[!p]
 \centering
 \includegraphics[width=0.8\textwidth]{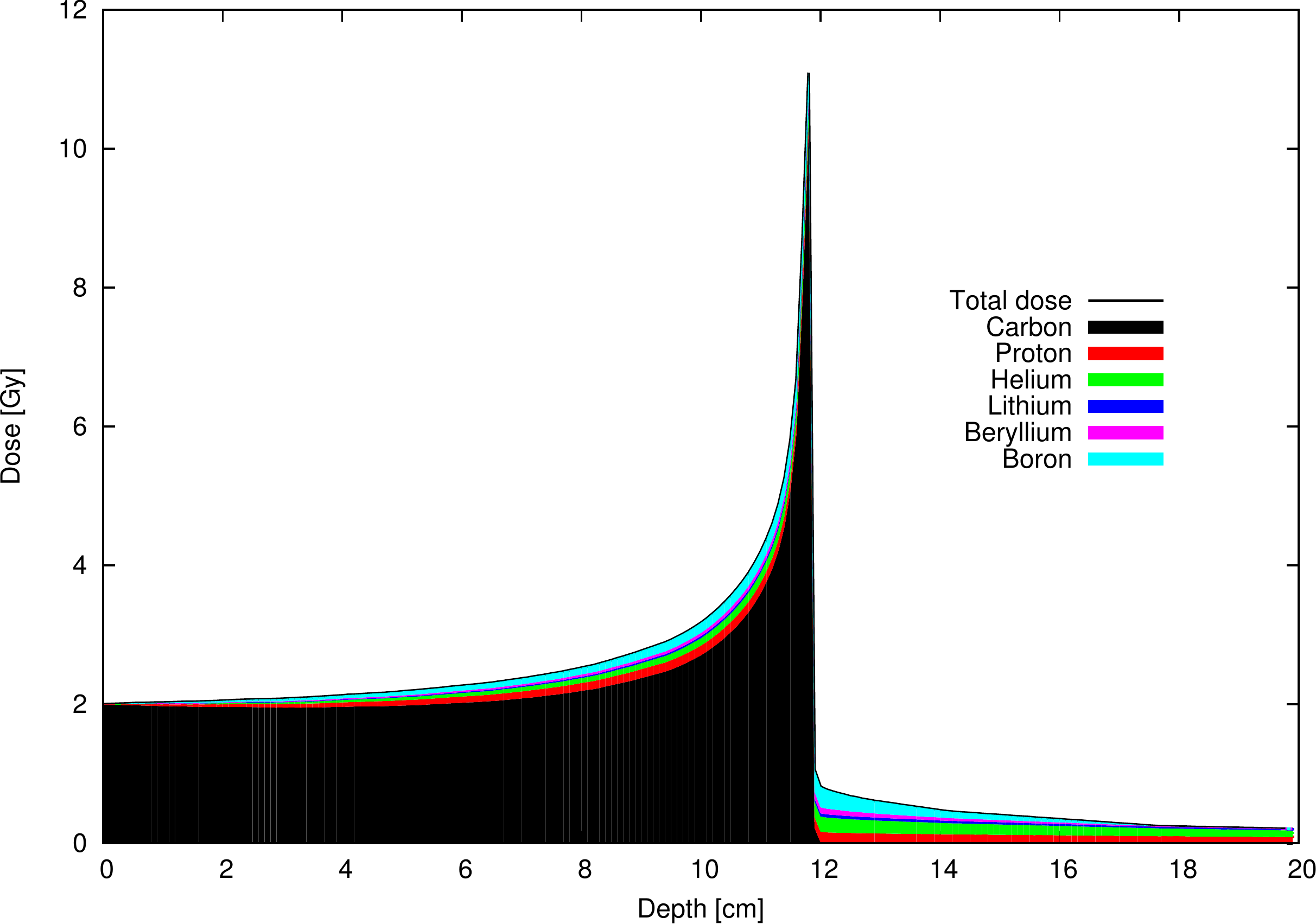}
 \includegraphics[width=0.8\textwidth]{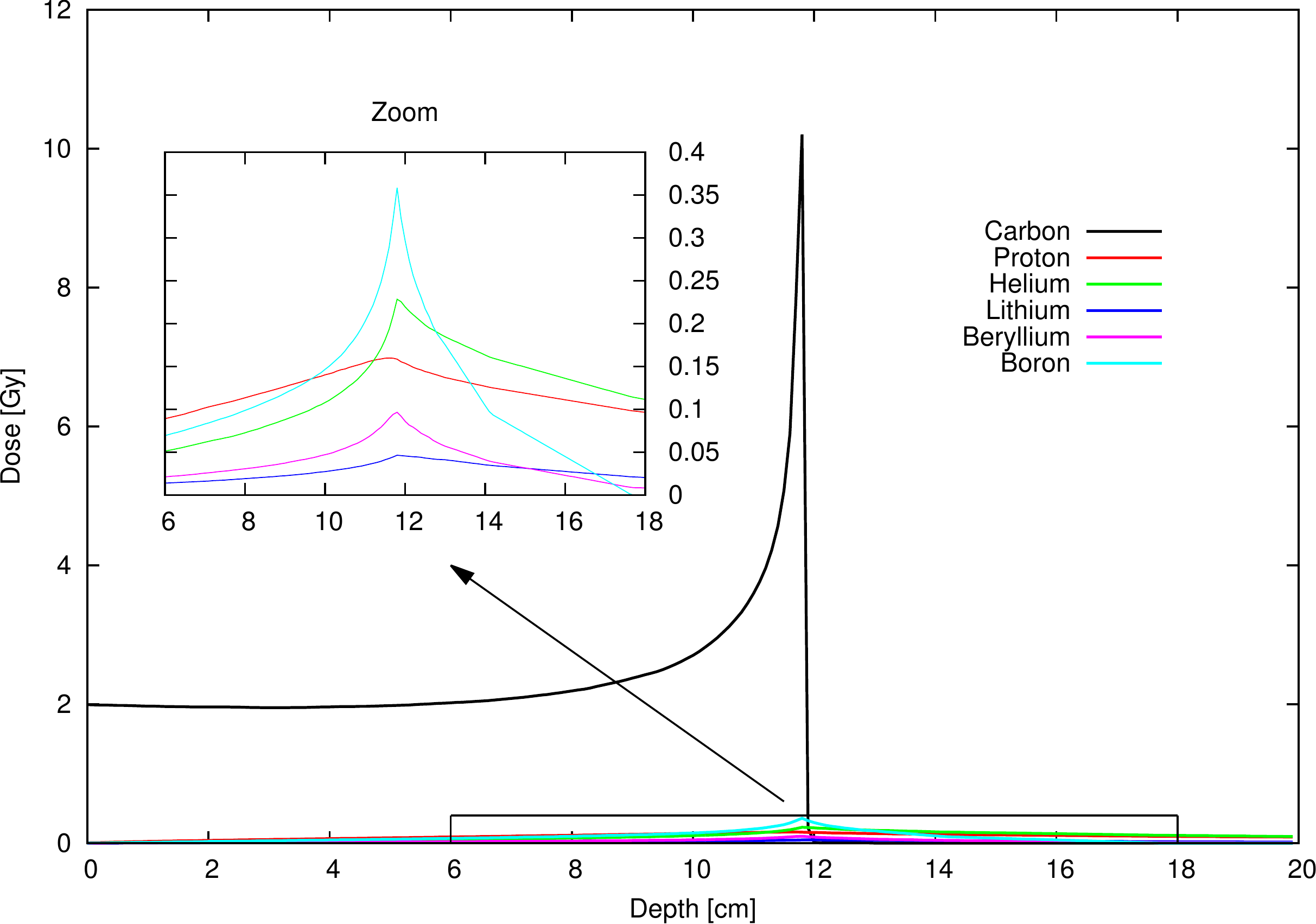}
  \caption{Dose vs depth of a carbon beam of initial energy 270 MeV/amu in
liquid water and of beam fragments. Data extracted from data sets available in
the libamtrack library, were generated using the SHIELD-HIT10A code \citep{SHIELD-HIT_ref38}. The
entrance channel dose is 2 Gy. Upper panel: cumulative representation. Lower
panel: individual depth-dose distributions of the primary beam and
of secondary ion species. In the lower panel the 6-18 cm depth region is
magnified to better visualise the individual dose contributions of secondary
particles. Data beyond the depth of 20 cm have not been plotted. }
 \label{fig:ch4_depth_dose}
\end{figure}

\begin{figure}[!p]
 \centering
 \includegraphics[width=0.8\textwidth]{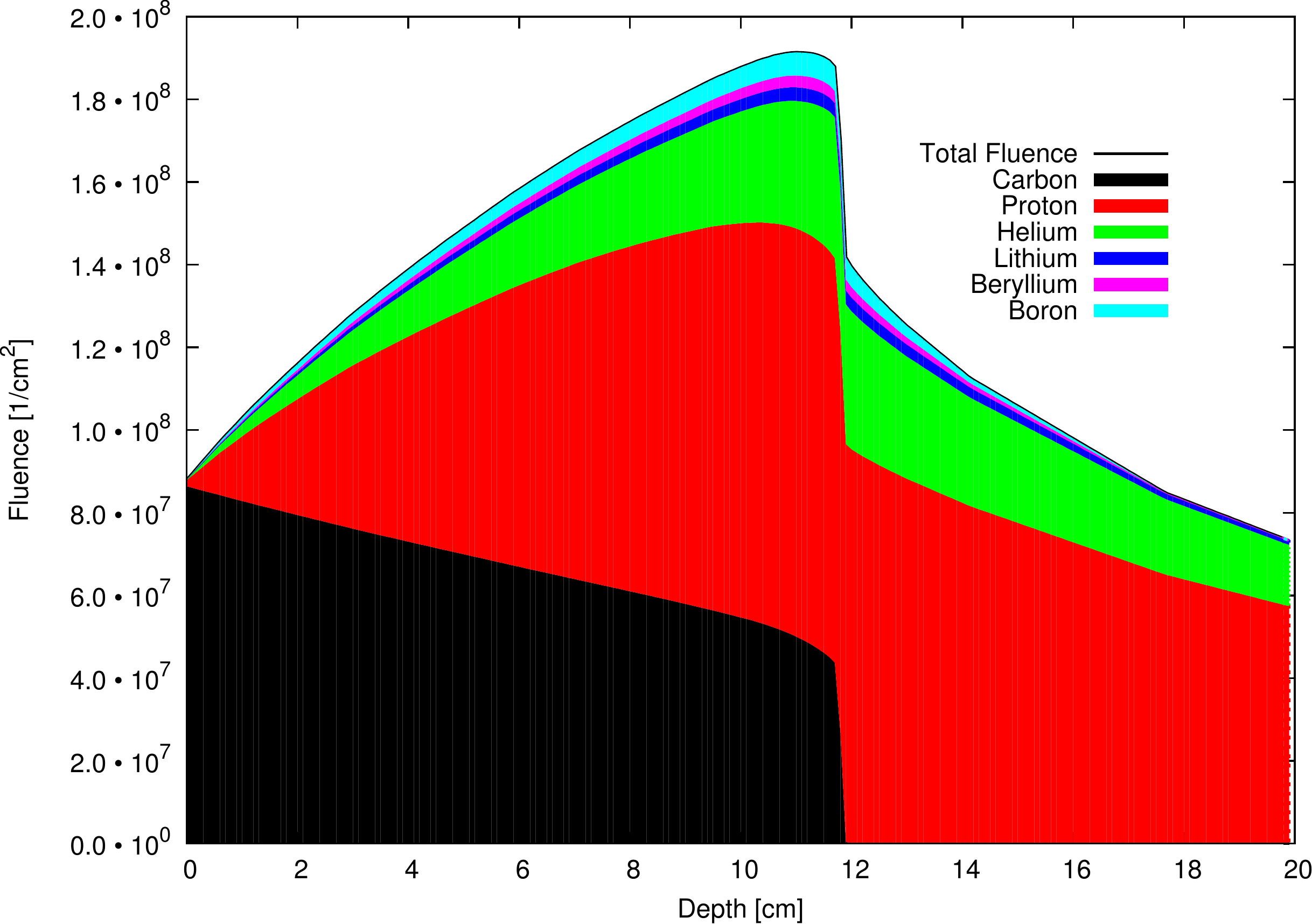}
 \includegraphics[width=0.8\textwidth]{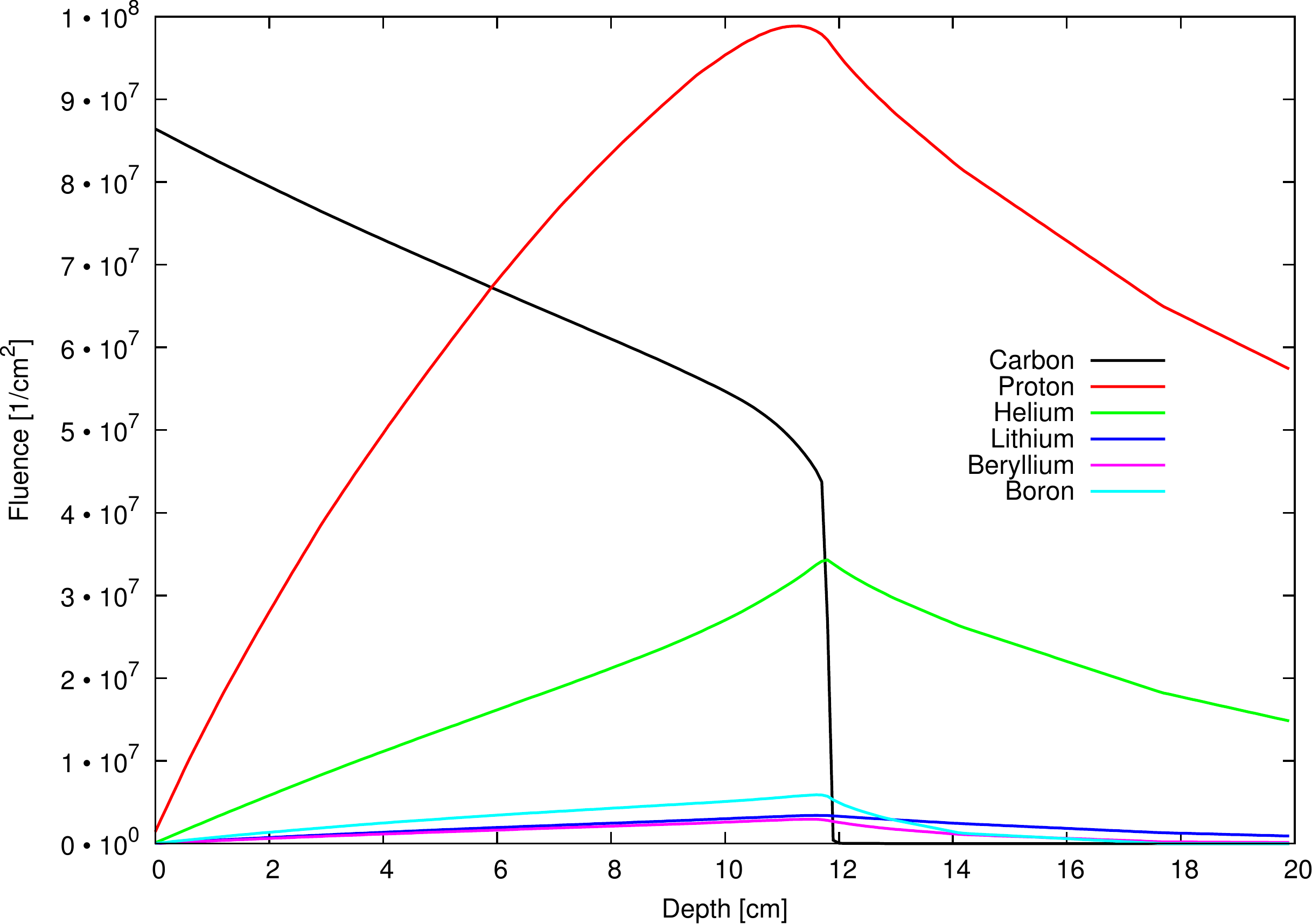}
 \caption{Fluence vs depth of a carbon beam of initial energy 270 MeV/amu in
liquid water and of beam fragments. Data extracted from data sets available in
the libamtrack library, were generated using the SHIELD-HIT10A code \citep{SHIELD-HIT_ref38}. The
primary beam fluence in the entrance channel was adjusted to represent the beam
entrance dose of 2 Gy, to conform with Fig. \ref{fig:ch4_depth_dose}. Upper panel: cumulative
representation. Lower panel: individual depth-fluence distributions of the
primary beam and of secondary ion species. Data beyond the depth of 20 cm have
not been plotted.}
 \label{fig:ch4_depth_fluence}
\end{figure}

\subsection{The energy-fluence spectra database in the libamtrack library --- SPC files}

A module for reading and writing energy-fluence spectra in binary format was
implemented in the libamtrack library. Binary SPC file format is used in the
TRiP98 treatment planning system and also in the Siemens “Syngo PT Planning”
carbon therapy planning system at HIT, Heidelberg. The author of this thesis
participated in implementing the SPC module in the libamtrack library.
In the libamtrack code library a sample set of SPC files is provided. Each file
contains spectra of carbon ions in water at various depths for a number of
initial beam energies. Files included in the libamtrack library cover the range
of initial carbon beam energies between 50 MeV/amu and 400 MeV/amu. 
A typical SPC file\footnote{\verb|http://bio.gsi.de/DOCS/TRiP98/DOCS/trip98fmtspc.html|} contains energy fluence spectra at ~50 depths, densely
covering the region of the Bragg peak and less densely over the entrance
channel. At each depth, spectra of up to 6 different ion species (from protons
to carbon ions) are provided. At each depth and for each ion species, the energy
histogram is divided into bins of the same width, containing fluences of
particles of corresponding energy.

\subsection{Algorithm for interpolation of the energy-spectrum data files.}
Energy-fluence spectra at an arbitrary depth in water $d$ for a carbon beam of
initial energy $E$ can be interpolated from the discrete SPC files using the
following algorithm. Interpolation is performed in the following steps: 
\begin{enumerate}
\item Over the initial beam energy, 
\item Over penetration depth 
\item Over energy in the energy-fluence spectrum.
\end{enumerate}

Let us calculate the fluence $F$ of particles (fragments) of charge $Z$, and
energy $E$, at depth $d$ in a beam of initial energy $E_{\rm in}$: $F( Z , E, d, E_{\rm in} )$.
Two SPC files are taken, with initial beam energies $E_{\rm in,min}$ and $E_{\rm in,max}$
closest to $E_{\rm in}$. 
The initial energy interpolation is performed as follows:

\begin{equation}
F(Z, E, d, E_{\rm in}) = (1 - \alpha) F(Z, E, d, E_{\rm in,min}) + \alpha F(Z, E, d, E_{\rm in,max})
\end{equation}

where $\alpha = (E_{\rm in} - E_{\rm in,min}) / (E_{\rm in,max} - E_{\rm in,min})$.
The next step is to interpolate between depths: two depths $d_{\min}$ and $d_{\max}$,
closest to $d$ are needed (assuming they are at the same depth binning as those
for $E_{\rm in,min}$ and $E_{\rm in,max})$:

\begin{eqnarray}
F(Z, E, d, E_{\rm in,min}) &=& (1 - \beta) F(Z, E, d_{\min}, E_{\rm in,min}) + \beta F(Z,E,d_{\max},E_{\rm in,min}) \nonumber \\
F(Z, E, d, E_{\rm in,max}) &=& (1 - \beta) F(Z, E, d_{\min}, E_{\rm in,max}) + \beta F(Z,E,d_{\max},E_{\rm in,max}) \nonumber
\end{eqnarray}

where $\beta = (d - d_{\min})(d_{\max} - d_{\min})$.

The next step is to interpolate between energies in energy-fluence spectra
(histograms): two energies $E_{\min}$ and $E_{\max}$, closest to $E$ are needed (assuming
there is the same energy binning for $d_{\min}$ and $d_{\max}$): 

\begin{eqnarray}
F(Z,E,d_{\max},E_{\rm in,min}) &=& (1- \gamma) F(Z,E_{\min},d_{\max},E_{\rm in,min}) + \gamma F(Z,E_{\max},d_{\max},E_{\rm in,min}) \nonumber\\
F(Z,E,d_{\max},E_{\rm in,max}) &=& (1- \gamma) F(Z,E_{\min},d_{\max},E_{\rm in,max}) + \gamma F(Z,E_{\max},d_{\max},E_{\rm in,max}) \nonumber\\
F(Z,E,d_{\min},E_{\rm in,min}) &=& (1- \gamma) F(Z,E_{\min},d_{\min},E_{\rm in,min}) + \gamma F(Z,E_{\max},d_{\min},E_{\rm in,min}) \nonumber \\
F(Z,E,d_{\min},E_{\rm in,max}) &=& (1- \gamma) F(Z,E_{\min},d_{\min},E_{\rm in,max}) + \gamma F(Z,E_{\max},d_{\min},E_{\rm in,max}) \nonumber
\end{eqnarray}

where $\gamma = (E-E_{\min})/(E_{\max}-E_{\min})$.

Energy-fluence spectra stored in SPC data files are normalized to entrance
fluence of 1 cm$^{-2}$, thus a method was provided to normalize spectra to the
fluence corresponding to a given entrance dose (i.e. 2 Gy). Methods to access
SPC data are provided in the \verb"AT_SPC" module in the libamtrack library.

\subsection{A sample calculation of energy-fluence spectra}

In Fig. \ref{fig:ch4_depth_fluence} fluences of the ion species generated by a  270 MeV/amu carbon beam
in water are shown as a function of depth and of ion energy. The range in water
of this carbon beam is about 12 cm. The fluence of carbon ions decreases from
its initial value ($8.8 \cdot 10^7 $ cm$^{-2}$, corresponding to 2 Gy) to zero at depths beyond
12 cm. Buildup of secondary fragments is observed at all depths. The
contribution to the total fluence of protons and helium ions is the highest. The
total particle fluence, understood as the fluence of carbon ions plus the
fluence of all fragments, at some depths exceeds the fluence of carbon ions in
the entrance channel (see Fig. \ref{fig:ch4_depth_fluence}). This is because in nuclear reactions a
single ion may produce several secondary fragments - ions of lower charges. 

\begin{figure}[!h]
 \centering
 \includegraphics[width=0.45\textwidth]{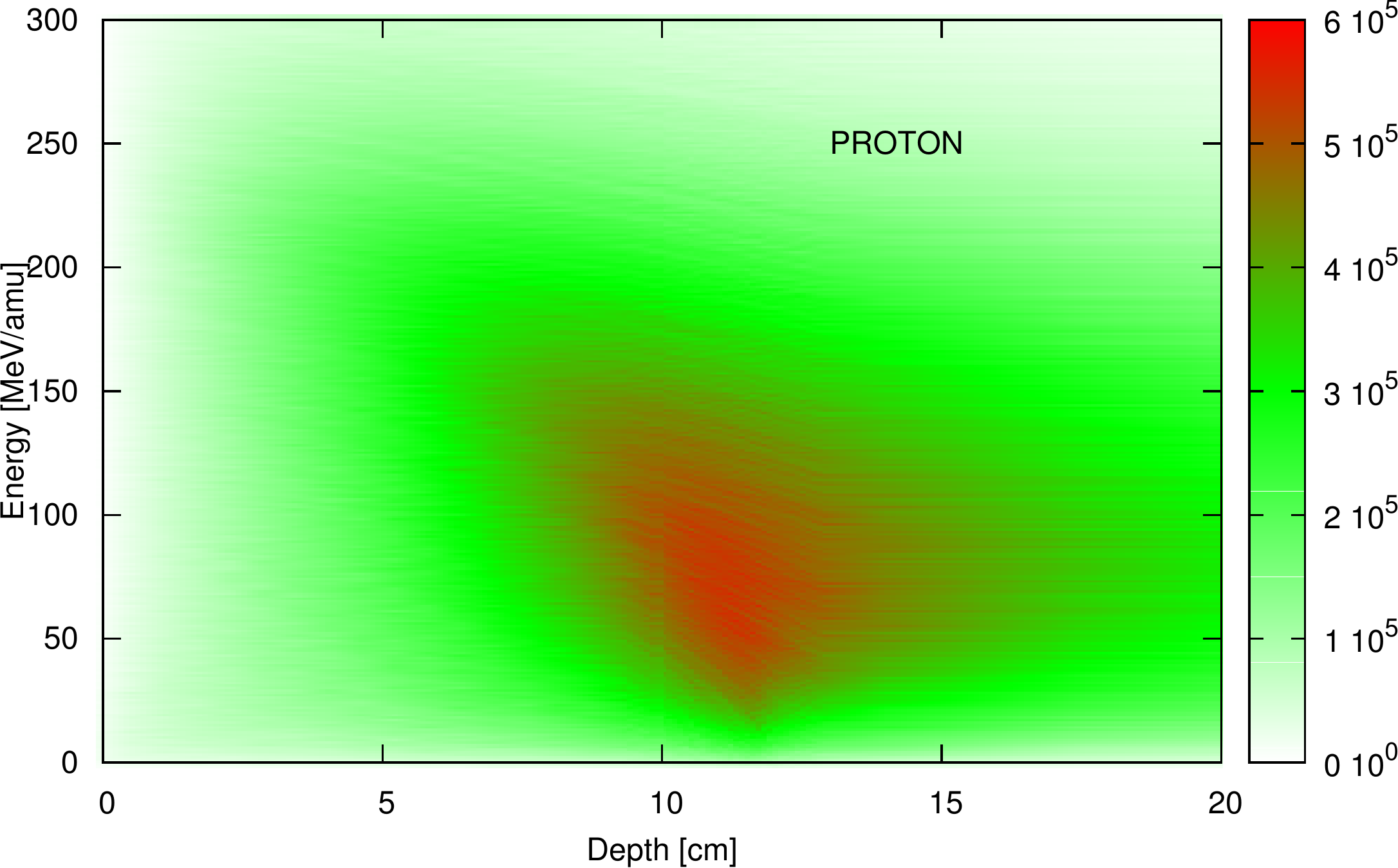}
 \includegraphics[width=0.45\textwidth]{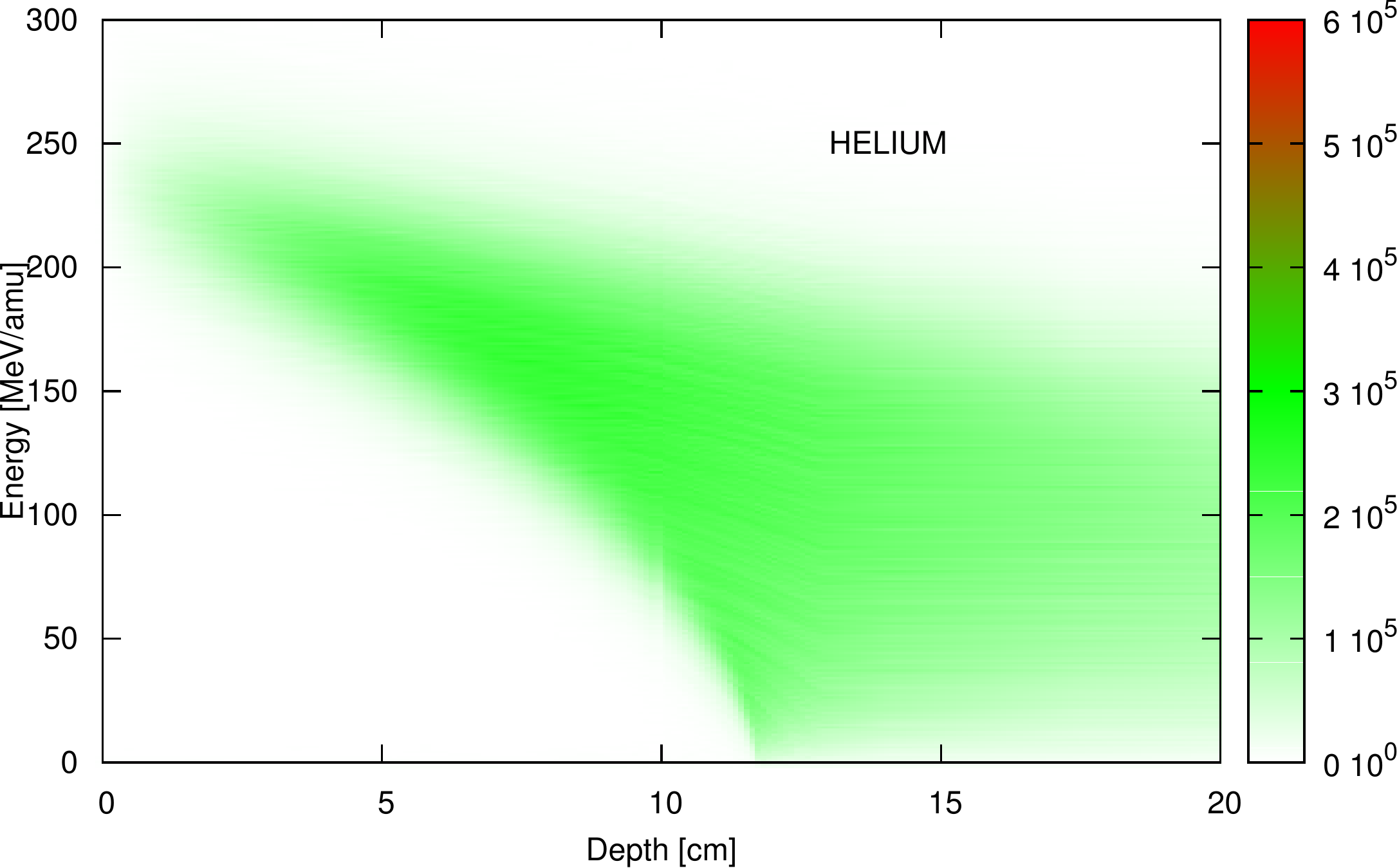}
 \includegraphics[width=0.45\textwidth]{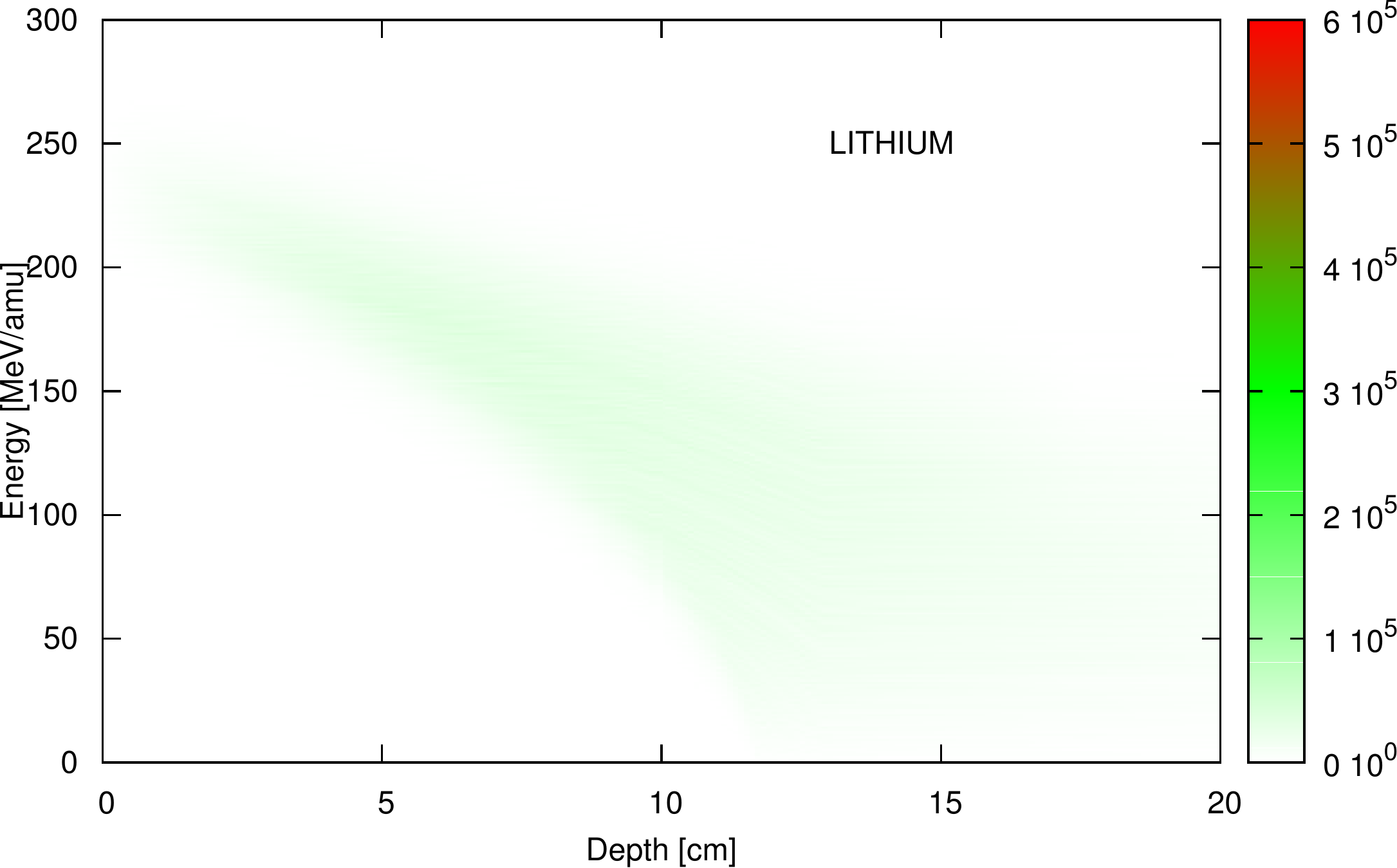}
 \includegraphics[width=0.45\textwidth]{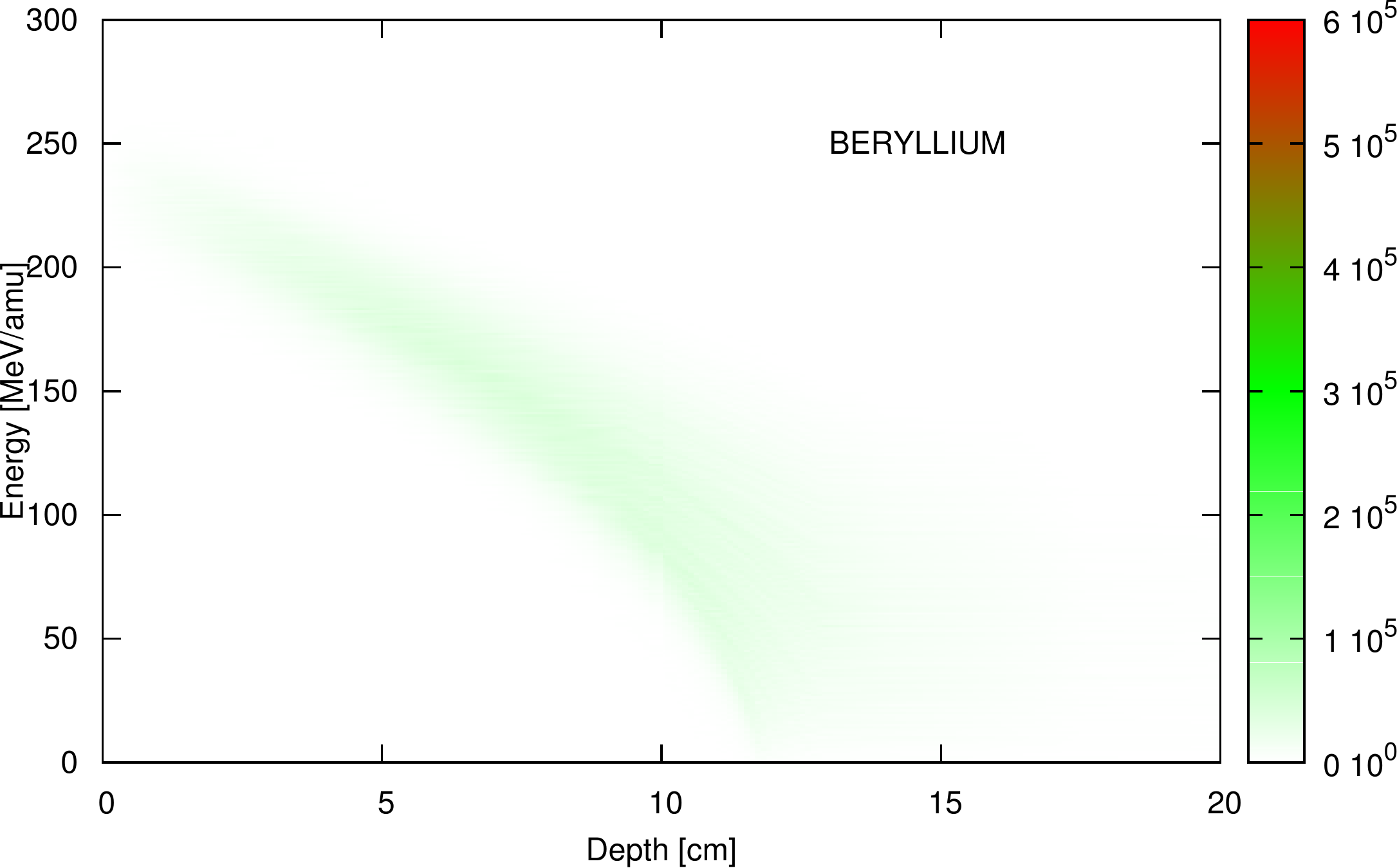}
 \includegraphics[width=0.45\textwidth]{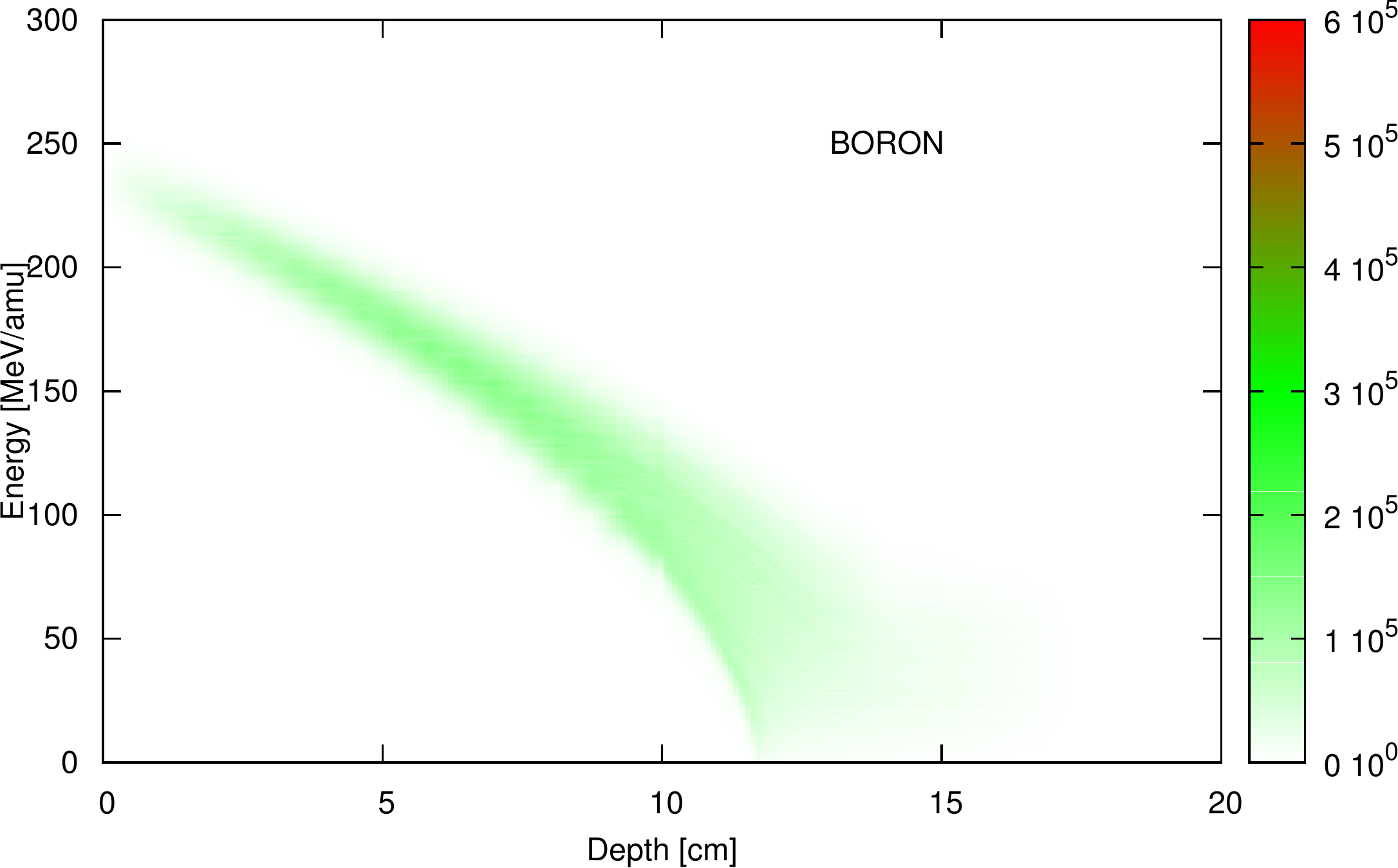}
 \includegraphics[width=0.45\textwidth]{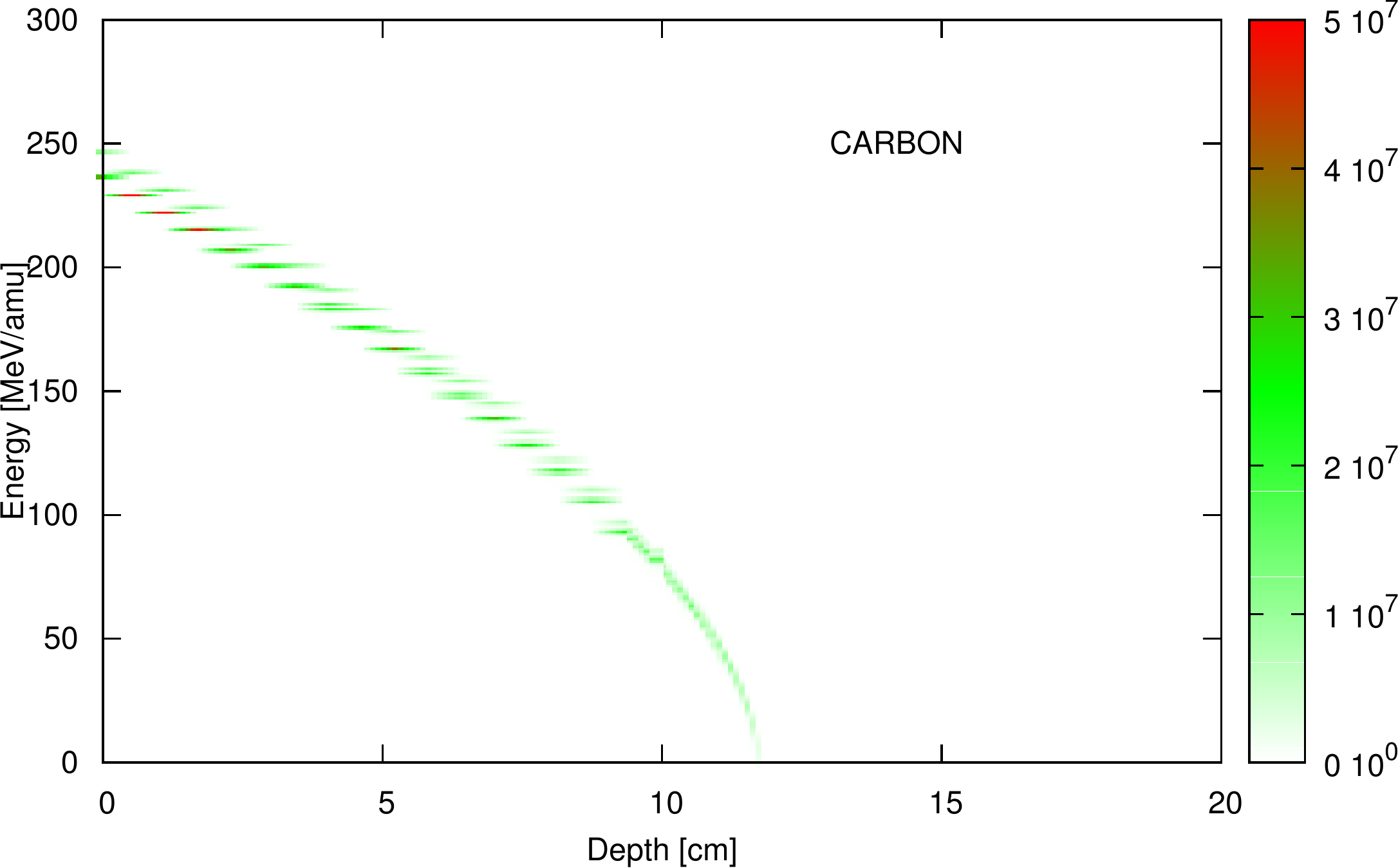}
 \caption{Fluences of primary and secondary ions primary carbon beam of
initial energy 270 MeV/amu in liquid water,  shown as a function of depth (along
the x-axis) and energy (along the y-axis). The entrance carbon beam dose is 2
Gy. On each panel energy-fluence spectra are shown for each ion species (protons
thru carbon). The fluence values for the different ions are represented by
different colours: red is the highest (range: $5 \cdot 10^7 \rm{cm}^{-2}$ to 
$6 \cdot 10^5 \rm{cm}^{-2}$ ), green - intermediate values, white background - no ions (fluence=0).}
 \label{fig:ch4_fragments}
\end{figure}

\section{Optimization of dose vs. depth distributions}

\subsection{The depth-dose profile optimization algorithm}
\label{subsec:ch4_optim}

The Spread-Out Bragg Peak (SOBP) is a depth dose distribution curve composed of
several pristine Bragg peaks to form a uniform dose distribution in the region
of interest (in proton radiotherapy) or a predefined shape of depth dose
distribution, designed to obtain uniform biological response over that region
(in carbon radiotherapy).

The SOBP is created as a sum of pristine Bragg peaks, produced by beams with
different initial energies and different intensities (fluences). The position
(depth) of the pristine Bragg peak maximum increases with initial energy and the
peak amplitude is proportional to the initial dose (or fluence). One of the
methods of passive beam shaping is application of a rotating energy modulator,
where the ion beam traverses absorber sectors, each of different thicknesses
placed on the modulator ring. The thickness of the absorber determines the
position of the pristine Bragg peak and the fraction of time the beam passes
through the given sector - the relative amplitude of the pristine beam component
in the SOBP.

Adjustment of the positions $\{p\}_i$ and heights $\{h\}_i$ of pristine Bragg peaks in
order to obtain a “flat” or constant dose over a given depth range can be
formulated as an optimization problem. Let us consider a continuous function
$f(x)$, representing dose (or survival) depth profile which is required to be flat
and equal to a value $C$ over a range of depths $x$ between $a$ and $b$. A measure of
flatness $M_{\rm{fltn}}$ of such a curve can be defined as:

\begin{equation}
M_{\rm{fltn}}(p_1,\ldots p_m, h_1, \ldots h_m) = \sum_{j=1}^n (f(x_j, p_1,\ldots p_m, h_1, \ldots h_m) - C)^2
\end{equation}

where $\{x\}_j$ forms a regular grid of $n$ points over the interval $[a,b]$
If $M_{\rm{fltn}}$ equals 0, function $f$ does not show any significant oscillations. If the
$\{x\}_i$ grid is dense enough, function $f(x)$ is flat and almost equal
to $C$ over the interval $[a,b]$.

To simplify calculations it can be assumed that positions of pristine Bragg
peaks ${p_i}$ also form a regular grid inside the interval [a,b]. We seek such
heights ${h_i}$ of the Bragg peaks for which the measure $M_{\rm{fltn}}$ of flatness is as
close to zero as possible.

The function $f$ describes a sum of pristine Bragg peaks with maxima at $p_i$ and
heights $h_i$:

\begin{equation}
f(x_j, p_1,\ldots p_m, h_1, \ldots h_m) = \sum_{i=1}^m h_i f( x_j, p_i)
\end{equation}

where $f(x,p)$ is dose at a depth $x$ in a pristine Bragg peak with a maximum at $p$ 
and height equal to 1.

Numerical problem of finding minimum of function $M_{\rm{fltn}}$ is well-defined as the minimized function is a
quadratic form with partial derivatives which can be calculated analytically:

\begin{equation}
\frac{\partial M_{\rm{fltn}}}{\partial h_i} = \frac{\partial}{\partial h_i} \sum_{j=1}^n ( \sum_{i=1}^m h_i f( x_j, p_i) - C)^2 =
\sum_{j=1}^n  2 \left( \sum_{i=1}^m h_i f( x_j, p_i) - C\right) f( x_j, p_j )
\end{equation}

Thus the gradient minimization method can be used to find the coefficients $\{p\}_i$.
The above method was implemented by the author in python language and added to
the libamtrack library. Dose in carbon ion pristine Bragg peaks was interpolated
from SPC data sets.

The procedure of generating SOBPs can be extended to find such coefficients of
pristine Bragg peaks that their linear combination gives any arbitrary
depth-dose profile $g(x)$ over the interval $[a,b]$. Here instead of a measure of
flatness we define another parameter $M_{\rm{prf}}$, which tells us how close is the linear
combination of pristine Bragg peaks to the desired profile $g(x)$:

\begin{equation}
M_{\rm{prf}} = \sum_{j=1}^n (f(x_j, p_1,\ldots p_m, h_1, \ldots h_m) - g(x_j) )^2
\end{equation}

where $\{x\}_j$ form a regular grid of $n$ points over the interval $[a,b]$
As in the previous case, such a numerical problem can be solved by gradient
minimization methods. This method was also implemented by the author in python
language and added to the libamtrack library. 
The author also implemented a version of this algorithm optimized for parallel
execution on machines with multiple processor cores. Optimization is based on
dividing the grid $\{x\}_j$ into as many parts as the number of cores in the
computer, and calculating parts of the sum in $M_{\rm{fltn}}$ or $M_{\rm{prf}}$ in parallel.
Calculations were performed by the author using the Cracow Cloud infrastructure
at IFJ PAN in Krakow \citep{IFJCloud_ref49} and at the Academic Computer Centre ACK Cyfronet AGH
in Krakow.

\subsection{A sample calculation of a flat dose vs. depth profile}
\label{subsec:ch4_sample}

In a sample calculation the algorithm described in the preceding paragraph was
used to calculate a combination of Bragg peaks of carbon ion beams of different
initial energies and fluences to produce a flat depth dose profile over the
range between 8 and 12 cm in liquid water.

\begin{figure}[h!]
 \centering
 \includegraphics[width=0.8\textwidth]{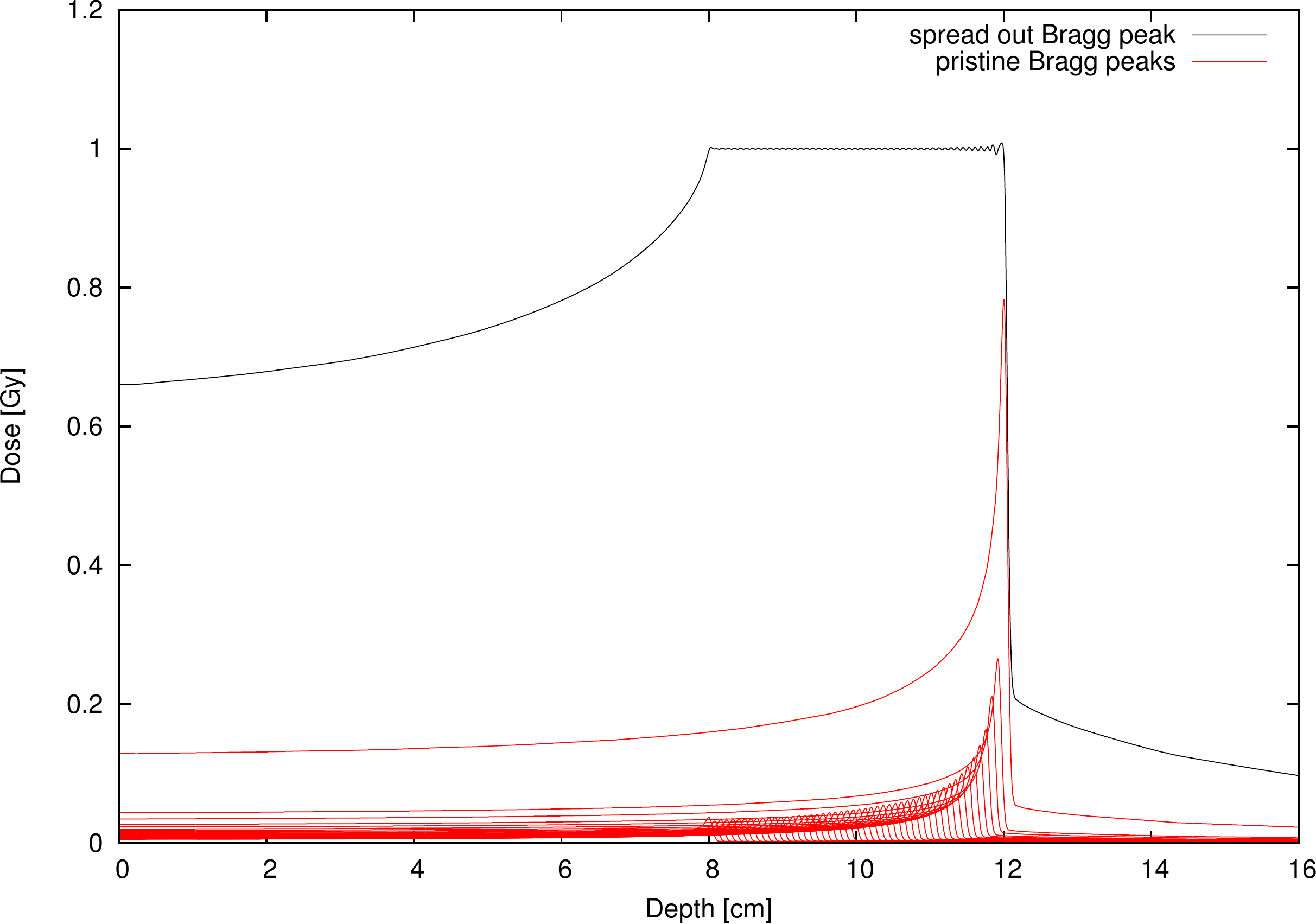}
 \includegraphics[width=0.8\textwidth]{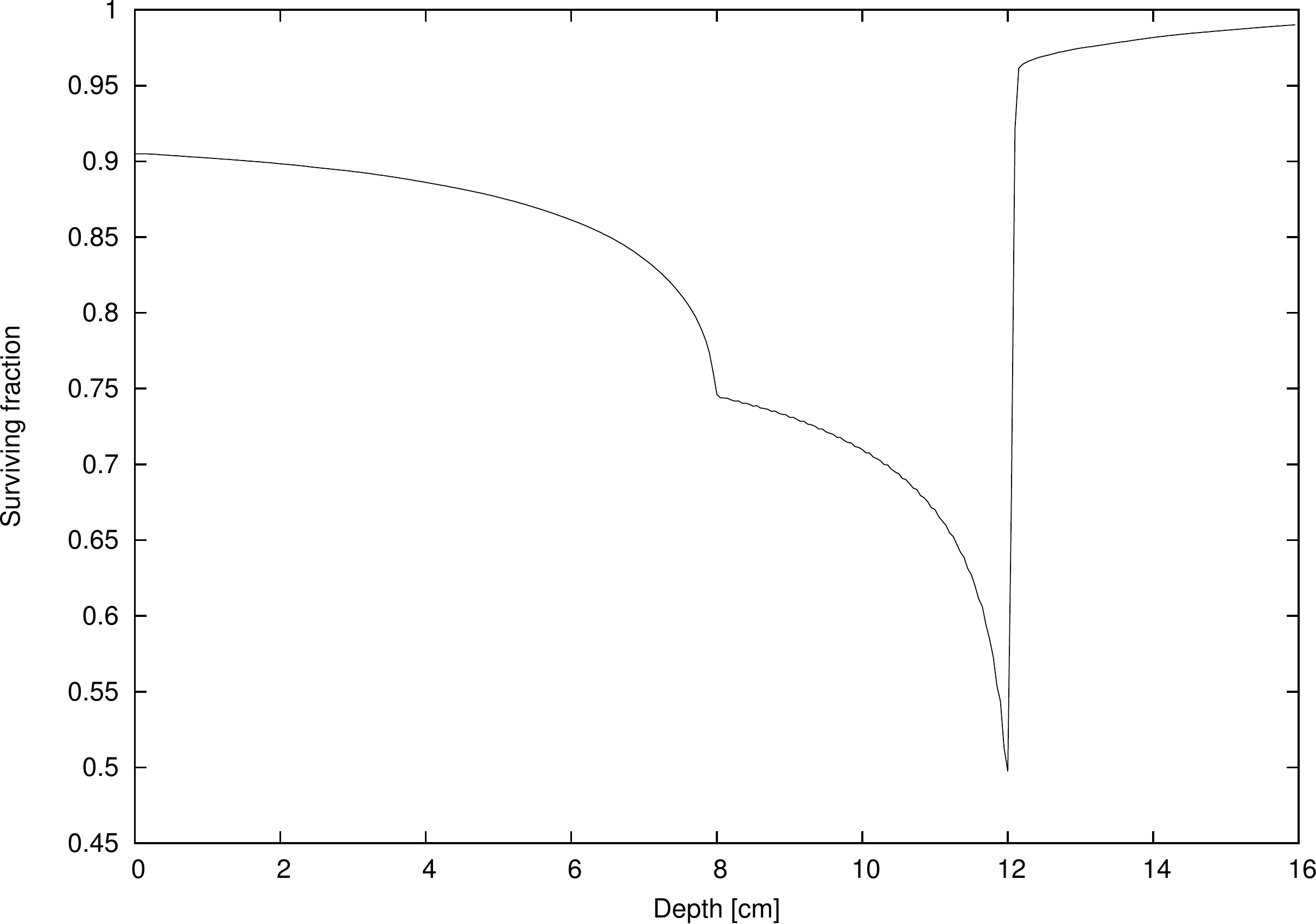}
 \caption{ Upper panel: A flat depth-dose distribution of 1 Gy over the depth
range 8-12 cm, obtained by summing the contributions of Bragg peaks of 49
pristine carbon beams of different initial energies and fluences (see Fig. \ref{fig:ch4_initial}).
Lower panel: Survival vs. depth of CHO cells irradiated by the carbon beam with
flat dose distribution shown in the upper panel.  Calculations of CHO cell
survival vs. depth were performed using the scaled Katz model, where CHO cells
were represented by model parameters  $m$ = 2.31, $D_0$ = 1.69 Gy, $\sigma_0$ =
$5.966 \cdot 10^{-11}$ m$^2$, and $\kappa$ = 1692.8 (see par. \ref{seq:ch3_fitting}). 
All calculations were performed for liquid water, using the libamtrack library.}
 \label{fig:ch4_depth_surv_dose}
\end{figure}

\begin{figure}[h!]
 \centering
 \includegraphics[width=0.8\textwidth]{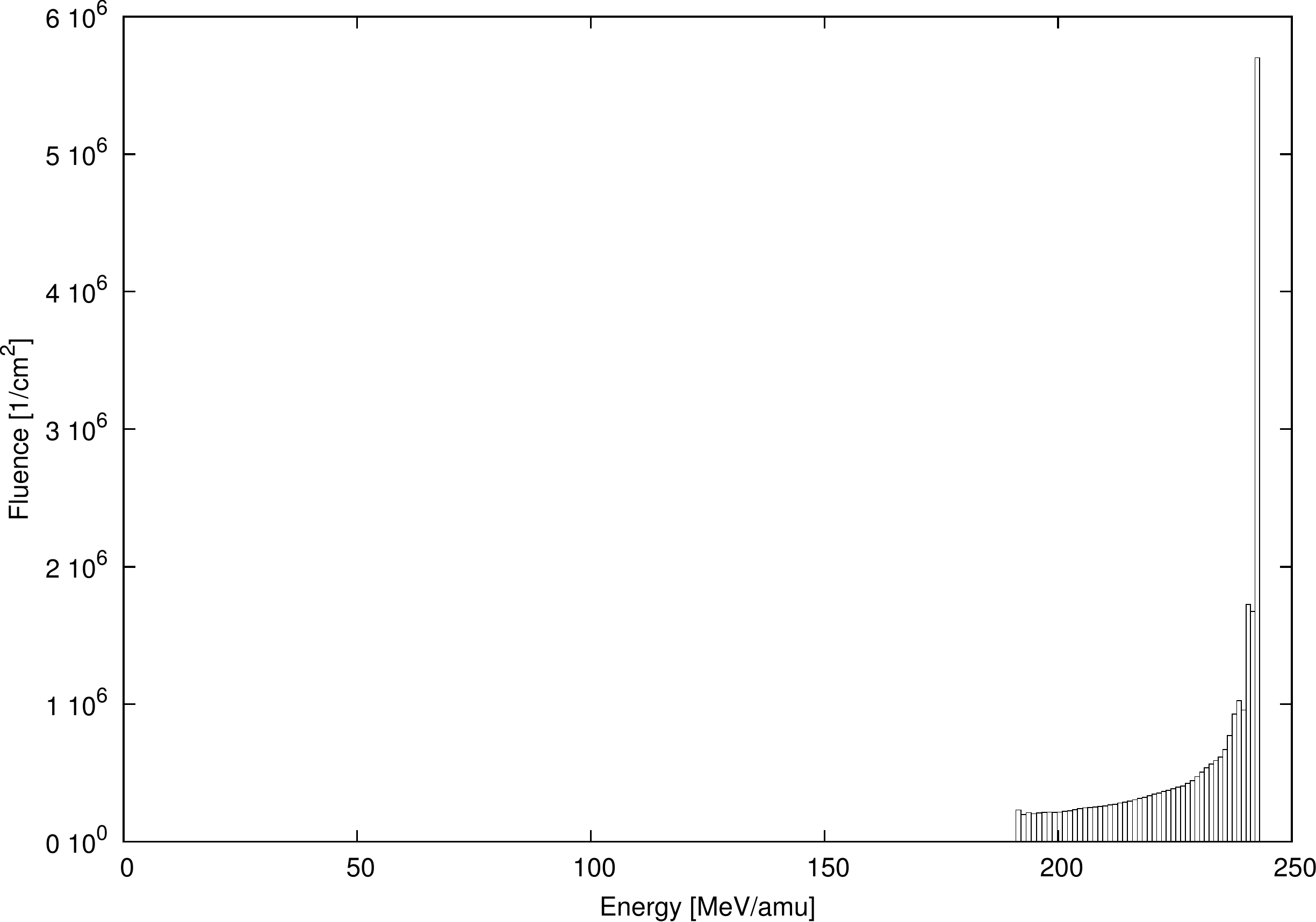}
 \caption{Initial energies and fluences of the 49 carbon ion beams required to
achieve the flat dose profile presented in the upper panel of Fig. \ref{fig:ch4_depth_surv_dose}. Each bar
corresponds to a single pristine Bragg peak of initial energy given on the
abscissa. Calculations performed using the libamtrack library.}
 \label{fig:ch4_initial}
\end{figure}

In the calculations pristine Bragg peaks were used, which show narrow shapes of
their Bragg peak (compared i.e. to pristine proton beams). The typical full
width of a carbon Bragg peak at half maximum is about 3 mm. In the sample
calculation 49 Bragg peaks were used to obtain a flat dose profile over a 40 mm
region spanning between 8 and 12 cm in depth. Maximum deviations from the
desired dose level of 1 Gy are below 1 (see Figure 4.4, upper panel) which is
acceptable within standards applied in clinical radiotherapy. Using more Bragg
peaks would lead to lower deviations. The experimentally observed “smearing out”
of  the initial energy of a pristine Bragg peak by inserting PMMA elements of
variable thickness into the beam also results in broader shapes of the Bragg
peaks and in decreasing any deviations from flatness of the dose profile.
A depth profile of cell survival was also calculated for this flat dose profile,
using the Katz model with parameters previously fitted in chapter 3 of this work
(cf. Fig. \ref{fig:ch3_fit}). The resulting profile is also shown in Figure \ref{fig:ch4_depth_surv_dose} (lower panel).
As could be expected, while the dose over the region between 8 and 12 cm is
constant, cell survival varies between 50\% and 75\%. This is due to changes with
depth of energy spectra in the SOBP of the carbon ions and of their fragments,
resulting in their  biological effectiveness increasing with depth.
To obtain a flat dose distribution over depths between 8 and 12 cm, carbon beams
of initial energies ranging between 191.5 MeV/amu and 242.5 MeV/amu have to be
applied with carefully adjusted fluence contributions, related to the dose
required over the flat part of the depth-dose distribution. The required initial
energy-fluence spectrum of these beams is shown in Figure \ref{fig:ch4_initial}. The right-most
peak of the highest fluence, corresponding to the component with the highest
energy represents the pristine carbon beam of the highest energy and range in
water of about 12 cm. The remaining components of lower fluence and lower
initial energies contribute to the dose distribution over lower depths.

\begin{figure}[h!]
 \centering
 \includegraphics[width=0.8\textwidth]{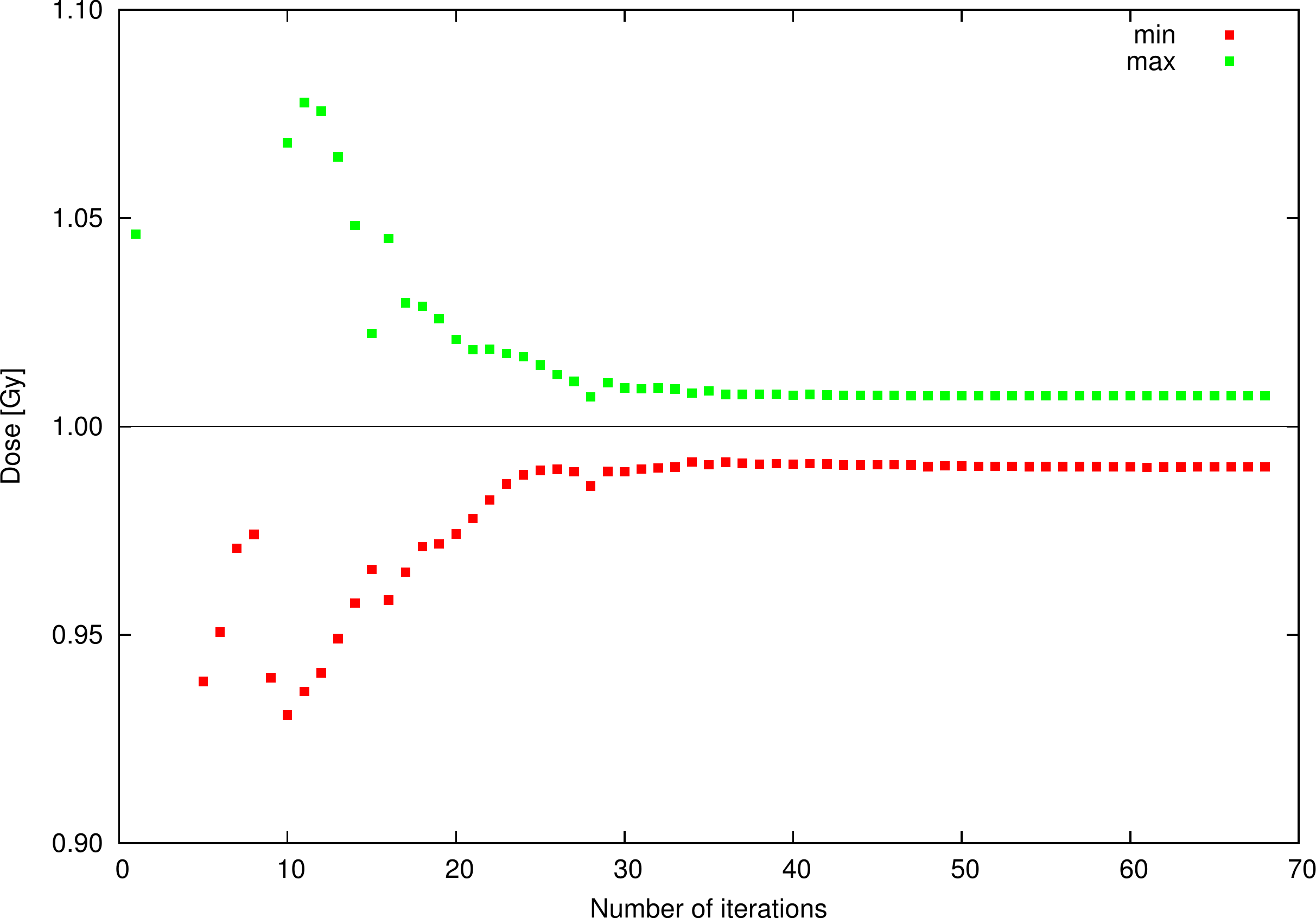}
 \caption{Convergence of the dose profile optimization algorithm. Points show the
minimum and maximum dose values over the depth of the flat region (8-12 cm)
after each iteration step. Calculations performed using the libamtrack library}
 \label{fig:ch4_convergence}
\end{figure}

To demonstrate the efficiency of the optimization algorithm used to achieve the
flat depth-dose distribution shown in Fig. \ref{fig:ch4_depth_surv_dose}, the degree of convergence vs.
number of iteration steps is shown in Figure \ref{fig:ch4_convergence}. Flatness of the dose
distribution was evaluated on a grid consisting of 200 points equally spaced
between 8 and 12 cm. The algorithm stopped after 68 steps at a minimum value of
$\chi^2$ =  0.000512. The maximum and minimum dose values found over the flat region
were 1.007351 and 0.990312, respectively, against the target value of 1 Gy, i.e.
deviating by less than 1\% . As may be seen in Fig. \ref{fig:ch4_convergence}, reasonable convergence
was obtained after about 35 iteration steps.

\section{Conclusions}

As the result of SHIELD-HIT10A Monte-Carlo simulations of the transport in water
of pristine carbon beams with initial energies ranging between 50 and 500
MeV/amu, a set of look-up files is SPC format containing energy-fluence spectra
of primary carbon ions and their fragments at various depths was available to
the author. In these simulations, the Bragg peak region was covered by a denser
grid to obtain better accuracy in the section where high dose gradients and
larger fluences of secondary ions arise. The author developed and implemented an
interpolation algorithm to estimate energy-fluence spectra at the desired beam
ranges for all contributing ion species. Using this algorithm, data could be
efficiently extracted from the look-up tables (as shown in Fig. \ref{fig:ch4_fragments}) 
and applied in further calculations.

The energy-fluence interpolation procedure was necessary for enabling the
pristine carbon beams to be combined into a spread-out Bragg peak configuration
in order to obtain a flat depth-dose distribution over a given depth range. An
optimisation algorithm was developed for this purpose by the author, whereby the
fluences and initial energies of a given number of pristine carbon beams could
be found which, when combined together, gave the required depth-dose profile
with satisfactory precision. The required dose profile can be made flat over a
pre-defined depth range, or assume any other shape. The developed optimisation
algorithm showed good convergence and computing efficiency. The accuracy of the
optimized solution, of better than 1\%, would fulfil the requirements of clinical
radiotherapy.

By suitably adding at pre-designed beam depths all ion contributions of the ion
beam combination configured to yield the desired depth-dose profile, it was
possible to implement the mixed-field calculation of cellular survival according
to the principles of the scaled version of the Katz model, developed in the
preceding parts of this thesis (Chapter 3). The best-fitted values of the four
parameters of Katz’s model, representing CHO (Chinese Hamster Ovary) cells were
applied in this mixed-field calculations for the carbon beam combination,
showing that applying a uniform (flat) dose profile to irradiate these cells
would result in a highly non-uniform depth distribution of cellular survival. 
The results obtained in this chapter will be further developed to find
combinations of pristine carbon beams of different energies and fluences to
yield the desired flat depth-survival distributions.

\mgrclosechapter

\chapter{Modelling the Depth Distribution of Cellular Survival}
In conventional radiotherapy with external beams of X-rays, $\gamma$-rays or
electrons, optimisation in therapy planning implies achieving a uniform dose
distribution in the treated volume, as uniform distribution of dose implies
uniform distribution of biological effect, i.e. cell survival or cell
inactivation (killing). It is evident, as illustrated, e.g. in Fig. \ref{fig:ch4_depth_surv_dose}, 
that optimisation in ion radiotherapy implies achievement of uniform distribution of
the biological endpoint (cell survival or killing) over the treated volume,
rather than uniform distribution of dose. Considering the strong and quite
complex dependence of cellular inactivation cross sections on ion
characteristics, such as LET or $z^{\star 2}/\beta^2$ (as demonstrated, e.g., in Fig. \ref{fig:ch3_inact_plateau}) and
the complex arrangement of pristine ion beams required to achieve uniform
depth-dose distribution (e.g., as shown in Fig. \ref{fig:ch4_depth_surv_dose}), a method needs to be
developed of finding a combination of pristine carbon beams, each of suitable
initial energy and fluence, which will result in obtaining a constant level of
cell survival over a given range of depths.

It is shown in this chapter how to develop such a method of optimising the
combination of pristine carbon beams in order to achieve a flat depth profile of
CHO (Chinese Hamster Ovary) cell survival over a given range of depths. The
scaled version of the Katz model (par. \ref{sec:ch3_scaled}) will be used and the CHO cells will
be characterised by the best-fitted values of model parameters representing this
cell line (par. \ref{seq:ch3_fitting}). The developed approach to modelling survival-depth
distributions will then be verified against published results of an experiment
in which CHO cells were irradiated at different depths by a combination of
pristine carbon beams, to achieve 20\% survival over a depth of 4 cm in water
\citep{mitaroff1998biological_ref66}. 
The effect of varying the input dose on SOBP flatness will next be
studied. By applying in these calculations the best-fitted values of model
parameters representing aerated (representing normal cells) and hypoxic
(representing cancer cells) V79 cells, the effect of cell oxygenation status on
the resulting depth-survival profiles will also be studied. 

\section{Calculation of cellular survival in a mixed ion field}

The linear combination of pristine Bragg peaks forming the SOBP or adjusted to
an arbitrary dose profile is well defined by a set of two parameters:
\begin{itemize}
 \item height $h_i$ of the pristine Bragg peak maximum, which is related to its initial
dose, $D_{\rm start}$ 
 \item position $p_i$ (depth) of the pristine Bragg peak maximum, which is related to its
initial energy $E_{\rm start}$
\end{itemize}

The fluence $F_i$ at given depth $d$ in a pristine Bragg peak of height equal to
unity ($h_i=1$) can be decomposed into the sum of fluences of the carbon ions and
of secondary fragments or, more generally, into a sum of fluences of ions of $Z$
ranging from 1 to 6 (for the six ion species involved, $Z = 1\ldots 6$) :

\begin{equation}
F_i(d) = \sum_{Z=1}^6 F_i( d, Z ) 
\end{equation}

The fluence of ions of charge $Z$ at depth $d$ is a sum of the ion energy-fluence
spectra:

\begin{equation}
F_i( d, Z ) = \sum_{j=1}^{n_Z} F_i( d, Z, E_j)
\end{equation}

where $n_Z$ is the number of components of the energy-fluence spectra of ion of charge $Z$.

Assuming that the linear energy transfer of an ion of charge $Z$ and energy $E$ is equal to
$L(Z,E)$ one may calculate the dose of ion $Z$ at depth $d$, as:

\begin{equation}
D_i( d, Z ) = \sum_{j=1}^{n_Z} \frac{1}{\rho} L(Z,E_j) F_i(d,Z,E_j)
\end{equation}

The dose $D_i(d)$ in a pristine Bragg peak at depth $d$ is then given by the
following equation:

\begin{equation}
D_i(d) = \sum_{Z=1}^6 D_i (d,Z) = \sum_{Z=1}^6 \sum_{j=1}^{n_Z} \frac{1}{\rho} L(Z,E_j) F_i(d,Z,E_j)
\end{equation}

In a linear combination of $N$ Bragg peaks, fluence and dose at depth $d$ are
calculated as follows:

\begin{equation}
F(d) = \sum_{i=1}^N h_i F_i(d) \qquad D(d) = \sum_{i=1}^N h_i D_i(d) 
\end{equation}

This may be expanded into:

\begin{eqnarray}
F(d) &=& \sum_{i=1}^N h_i \sum_{Z=1}^6 \sum_{j=1}^{n_Z} F_i(d,Z,E_j) \\
D(d) &=& \frac{1}{\rho} \sum_{i=1}^N h_i  \sum_{Z=1}^6 \sum_{j=1}^{n_Z} L(Z,E_j) F_i(d,Z,E_j) 
\end{eqnarray}

Or by changing the order of summation: 

\begin{eqnarray}
F(d) &=& \sum_{Z=1}^6 \sum_{j=1}^{n_Z} \sum_{i=1}^N h_i  F_i(d,Z,E_j) \\
D(d) &=& \frac{1}{\rho} \sum_{Z=1}^6 \sum_{j=1}^{n_Z} L(Z,E_j) \sum_{i=1}^N h_i F_i(d,Z,E_j)
\end{eqnarray}

At a depth $d$ we may thus split the beam into $n$ components (where $n=n_1+\ldots +n_6$), 
each related to ion of type $Z$ ($Z=1\ldots 6$), each of energy $E_j$ ($E_j = 1\ldots n_Z$), 
fluence $F_k(d)$, and dose $D_k(d)$ ($k=1\ldots n$), given by the following equation:

\begin{eqnarray}
F_k(d) &=& \sum_{i=1}^N h_i  F_i(d,Z,E_j) \\
D_k(d) &=& \frac{1}{\rho} L(Z,E_j) \sum_{i=1}^N h_i F_i(d,Z,E_j) 
\end{eqnarray}

Now, using the equations of the Katz scaled model with parameters: $m$, $D_0$,
$\sigma_0$, and $\kappa$, introduced in chapter 3, a method is provided to calculate
the fraction of surviving cells, $S(d)$, at depth $d$:

\begin{equation}
S(d) = \Pi_i(d) \Pi_{\gamma}(d)
\label{eq:ch5_mixed_survival}
\end{equation}

where:

\begin{equation}
 P_k = \begin{cases} 1 - \sigma_k(Z,E_j) / \sigma_0 &\text{if } \sigma_k( Z,E_j) \leq \sigma_0\\
0&\text{elsewhere}\end{cases}
\end{equation}

and

\begin{eqnarray}
\Pi_i(d) &=& \exp( - \sigma_0 \sum_{k=1}^n P_k F_k(d) )  \\
\Pi_{\gamma}(d) &=& 1 - \left(1 - \exp\left(- \frac{1}{D_0} \sum_{k=1}^n (1-P_k) D_k(d) \right)\right)^m
\end{eqnarray}

The above-described method was also used to calculate the survival vs. depth
profile shown in the lower panel of Fig. \ref{fig:ch4_depth_surv_dose} (Chapter 4).

\section{Calculation of survival vs. depth profile}
\label{sec:ch5_optim}

\subsection{The optimization algorithm}

The method of finding a linear combination of pristine Bragg peaks to form a
flat SOBP dose vs. depth profile, described in par. \ref{subsec:ch4_optim} of Chapter 4, can be
extended in order for the resulting survival profile to be constant (or flat) at
a given level of survival, $S$, over a given range of depths. The measure, Spf,
of survival profile flatness is then defined as follows:

\begin{equation}
\rm{Spf} = \sum_{j=1}^n ( S(x_j, p_1,\ldots p_m, h_1, \ldots h_m) - C)^2
\end{equation}

where $\{x\}_j$ forms a regular grid over the interval 
$[a,b]$ and $S(x_j,\ldots)$ is the survival level at depth $x_j$,
calculated using eq. \ref{eq:ch5_mixed_survival}

$S(x_j,p_1,\ldots p_m,h_1,\ldots h_m)$ is a non-linear function of $h_1,\ldots h_m$, 
so the gradient minimization algorithm is more time consuming than that of the dose profile
optimization problem as in this case the derivative of S needs to be evaluated
numerically.

Such an optimization algorithm was implemented by the author in python language,
along with the dose profile optimization algorithm (of par. \ref{subsec:ch4_optim}), as an
extension of the libamtrack library 

\subsection{A sample calculation of a flat survival vs. depth profile}

In a sample calculation the survival optimization algorithm was used to find a
combination of carbon Bragg peaks of different initial energies and fluences
which would give a flat depth survival profile at survival level 0.2 over the
range between 8 and 12 cm in liquid water. As the irradiated biological system,
Chinese Hamster Ovary (CHO) cells were selected, represented by four parameters
of Katz’s scaled model (see Chapter 3):  $m$ = 2.31,  $D_0$ = 1.69 Gy, $\sigma_0$
= $5.96 \cdot 10^{-11} \rm{m}^2$ , and $\kappa$ = 1692.8.

In this calculation pristine Bragg peaks were used, in a configuration similar
to that of par. \ref{subsec:ch4_sample}, consisting of 49 Bragg peaks placed on a regular grid
over a region spanning between depths of 8 and 12 cm.

\begin{figure}[ht!]
 \centering
 \includegraphics[width=0.8\textwidth]{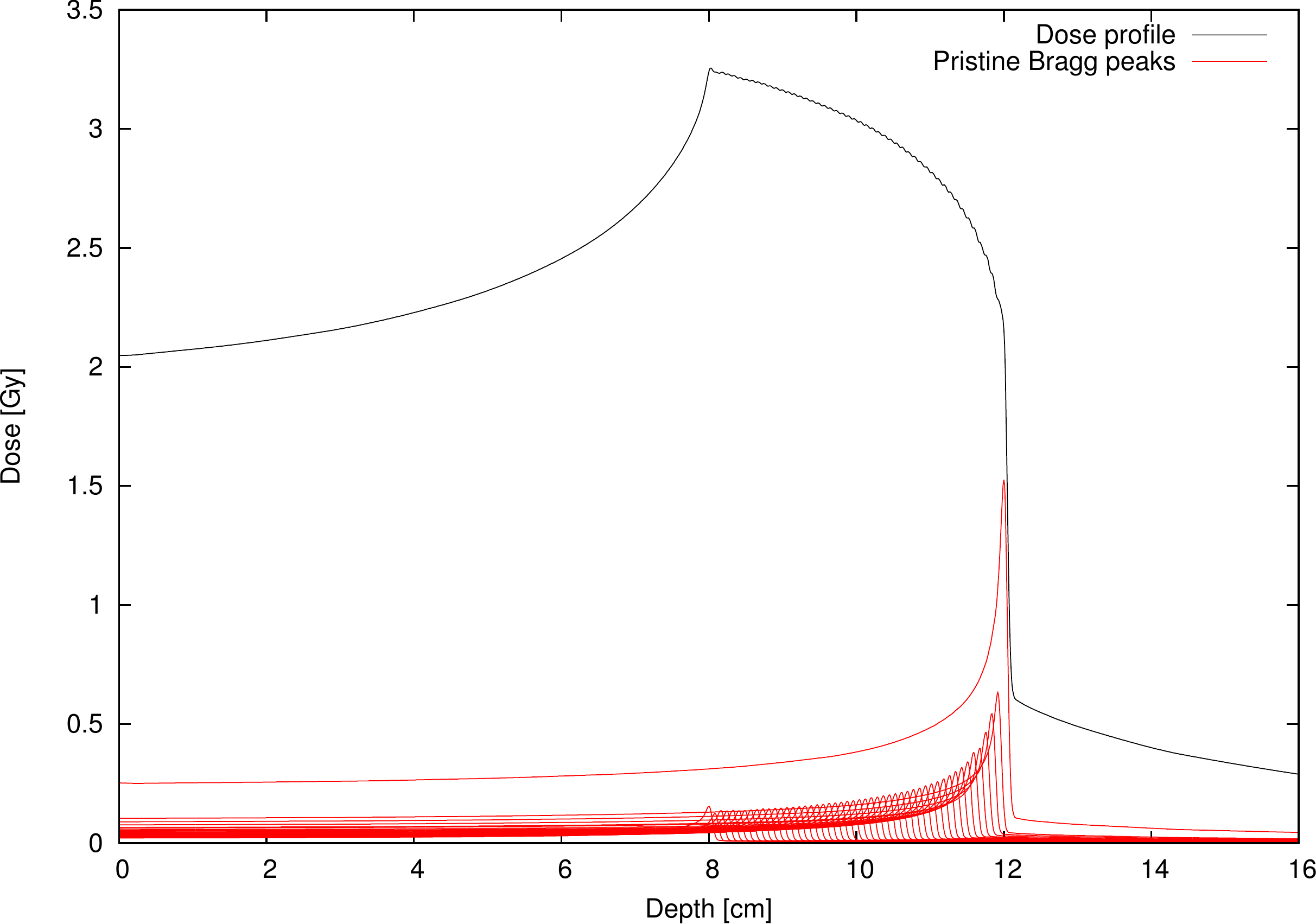}
 \includegraphics[width=0.8\textwidth]{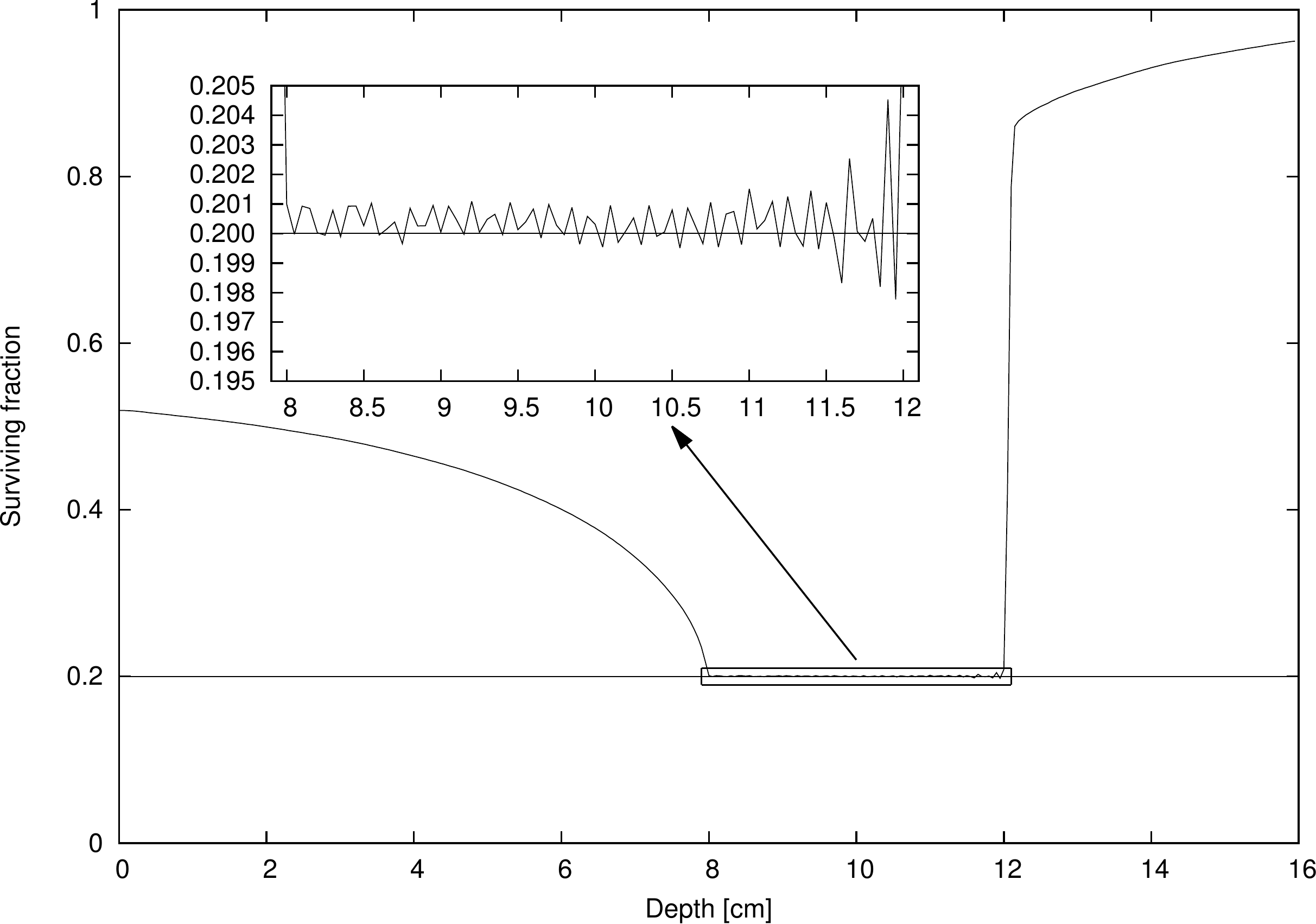}
 \caption{Upper panel: Depth-dose profile of a sum of 49 pristine Bragg peaks
with different initial energies and fluences  yielding a flat profile of
survival of CHO cells vs. depth at the survival level of 0.2 over depths between
8 and 12 cm, shown in the lower panel. CHO cells survival was calculated using
Katz’s scaled model. The model parameters representing CHO cell survival are:  $m$ = 2.31,  $D_0$ = 1.69 Gy, $\sigma_0$
= $5.96 \cdot 10^{-11} \rm{m}^2$ , and $\kappa$ = 1692.8. A magnified
inset of the flat survival region demonstrates small oscillations in the
survival level, due to the sharpness of pristine Bragg peaks. Calculations
performed in liquid water, using the libamtrack library.}
 \label{fig:ch5_optim_sample}
\end{figure}

The algorithm converged to a solution which gave a dose profile which decreased
with depth, as shown in the upper panel of Fig. \ref{fig:ch5_optim_sample}. At the depth of 8 cm a dose
of 3.25 Gy was needed for survival to decrease to the level of 0.2, while at the
end of the SOBP, at 12cm, only a dose of 2.16 Gy was required to obtain the same
survival level. The optimized survival vs. depth dependence is shown in the
lower panel of Fig. \ref{fig:ch5_optim_sample}. Maximum deviations from desired survival level of 0.2
were below 0.005 (see Fig. \ref{fig:ch5_optim_sample}) and were observed at the distal part of the
SOBP, while in central region deviations were of the order of 0.001. A similar
argument to that concerning dose profile optimization can be raised: the use of
a larger number of ion beams (Bragg peaks) and beam smearing will lead to
further smoothing of the depth-survival profile.

\begin{figure}[h!]
 \centering
 \includegraphics[width=0.8\textwidth]{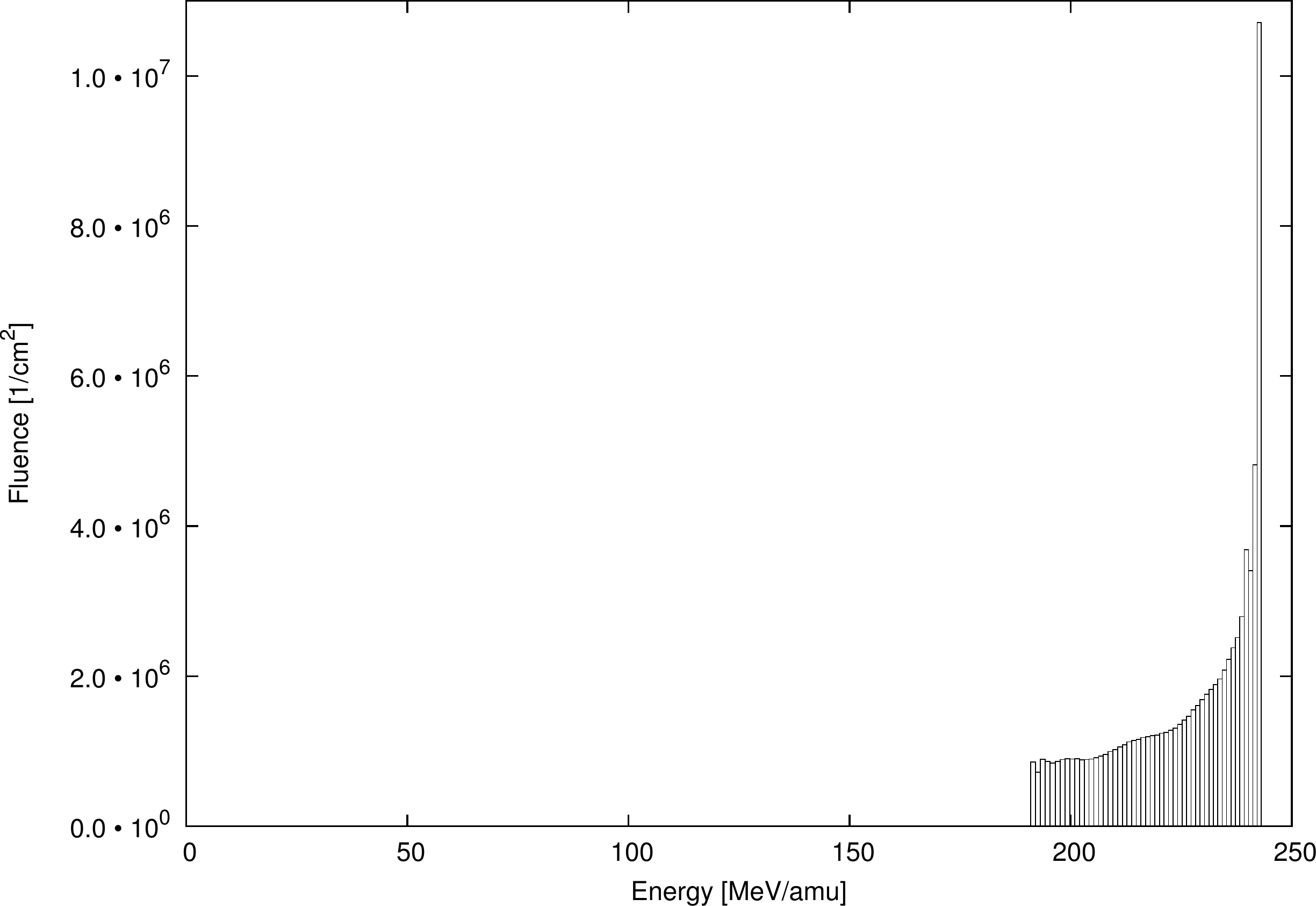}
 \caption{Initial energies and fluences of the 49 carbon ion beams required to
achieve the flat survival profile presented in the upper panel of Fig. \ref{fig:ch5_optim_sample}. 
Each bar corresponds to a single pristine Bragg peak of initial energy given on the
abscissa. Calculations performed in liquid water, using the libamtrack library.}
 \label{fig:ch5_initial}
\end{figure}

The initial energy-fluence spectrum of the ion beams is presented in Figure \ref{fig:ch5_initial}.
 The single peak at the energy of 242.5 MeV/amu corresponds to the pristine
carbon beam of the highest energy and range, contributing about 2 Gy to the dose
of at the end of SOBP. A high fluence of beams of energies between 191.5 MeV/amu
and about 210 MeV is required to deliver the dose of about 3 Gy in the proximal
part of the SOBP.

\begin{figure}[h!]
 \centering
 \includegraphics[width=0.8\textwidth]{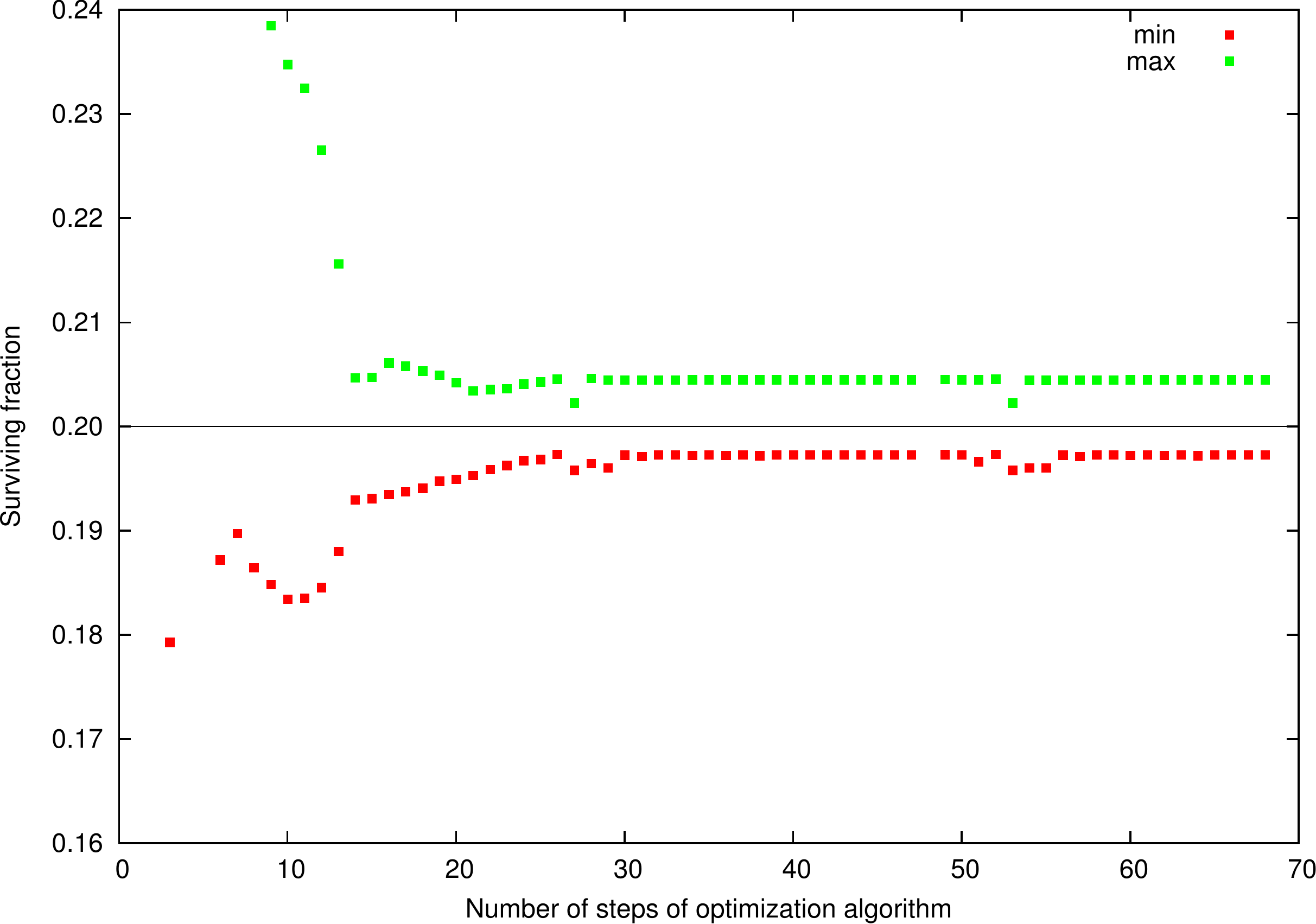}
 \caption{Convergence of the dose profile optimization algorithm. Points show
the minimum and maximum dose values over the depth of the flat region (8-12 cm)
after each iteration step. Calculations performed using the libamtrack library}
 \label{fig:ch5_convergence}
\end{figure}

Flatness of the survival profile was evaluated on a grid consisting of 1333
points equally spaced between 8 and 12 cm. The algorithm stopped after 68
iteration steps at a minimum value of $\chi^2$ =  0.000493. Maximum and minimum
survival levels found in the target region were 0.203155 and 0.195413
respectively, the relative deviation not exceeding 2.5\%. As may be seen in Fig. \ref{fig:ch5_convergence},
reasonable convergence is obtained after about 30 iteration steps,

\section{Comparison with a cell survival vs. depth experiment}
\label{sec:ch5_exp}

Mitaroff et al \citep{mitaroff1998biological_ref66} published results of a radiobiological experiment designed
to test the radiobiological models to be implemented in the TRiP98 TPS system at
GSI, Darmstadt. CHO K1 cell cultures in vitro were exposed at a range of depths
to carbon beams of initial energies ranging between  196 and 244 MeV/amu. An
early version of LEM (LEM I) was used to find such irradiation conditions
(entrance energy-fluence spectra) at which survival in the target region,
between 8 and 12 cm in depth, would be constant and equal to 0.2. Flasks with
cells were positioned in a water phantom at depths ranging between 1 and 19 cm,
separated by 0.5 or 1 cm. The measured values of survival of CHO cells in the
flasks were then compared with the planned survival level of 0.2 in the target
region.

\begin{figure}[h!]
 \centering
 \includegraphics[width=0.8\textwidth]{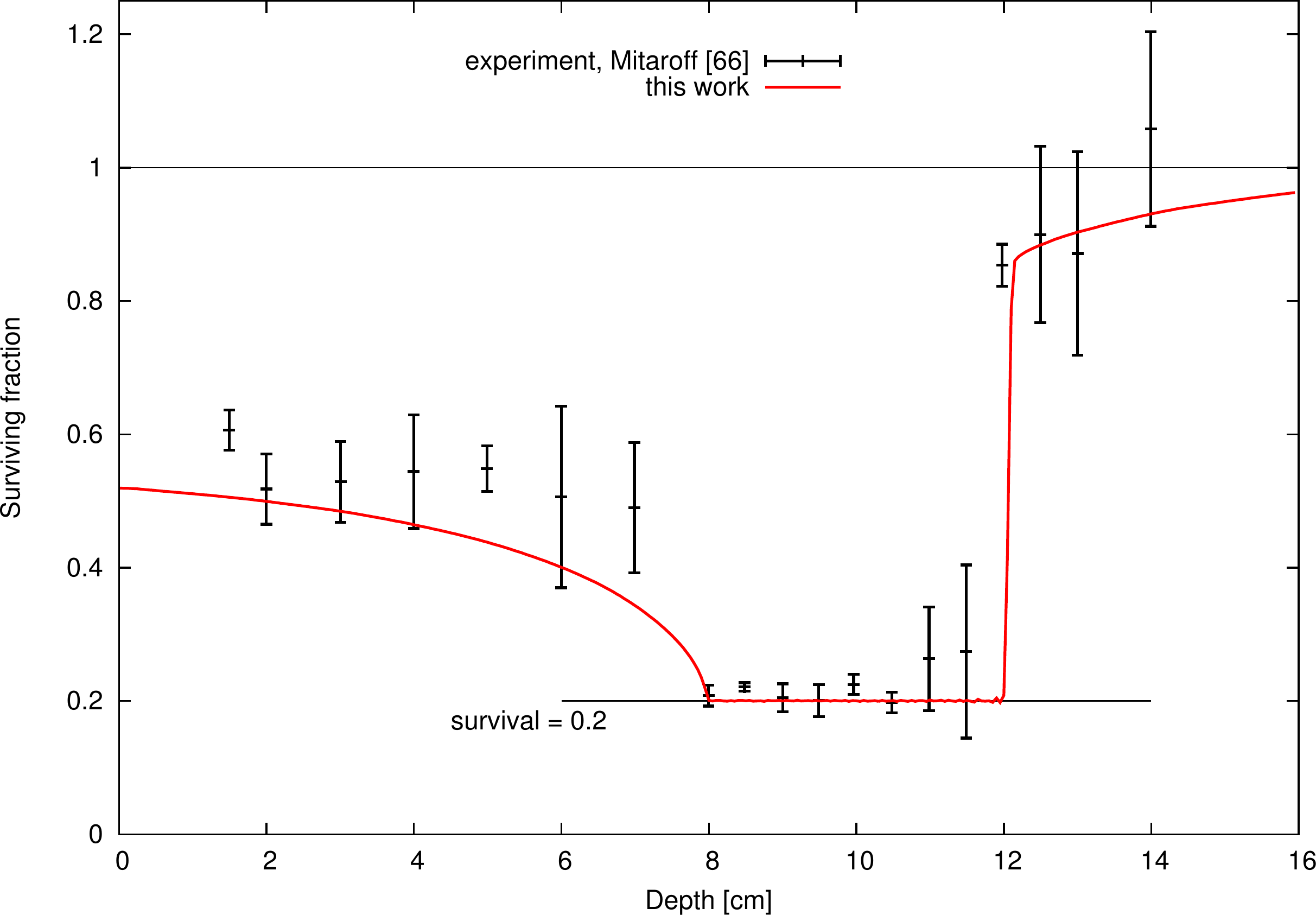}
 \caption{Survival of CHO cells vs depth: results of calculations using the
survival optimization algorithm and the scaled Katz model, where CHO cells are
represented by the model parameters:    $m$ = 2.31,  $D_0$ = 1.69 Gy, $\sigma_0$
= $5.96 \cdot 10^{-11}\ \rm{m}^2$ , and $\kappa$ = 1692.8 
(cf. Fig. \ref{fig:ch5_optim_sample}, lower panel),  calculations
performed using the libamtrack library, compared with experimental data
published by Mitaroff et al. \citep{mitaroff1998biological_ref66}.}
 \label{fig:ch5_exp}
\end{figure}

The scaled Katz model implemented the in libamtrack library was used to
calculate cell survival. The values of the best fitted parameters, found earlier
in par. \ref{seq:ch3_fitting}: chapter 3:  $m$ = 2.31,  $D_0$ = 1.69 Gy, $\sigma_0$
= $5.96 \cdot 10^{-11} \rm{m}^2$ , and $\kappa$ = 1692.8, were used in these calculations.
In Figure \ref{fig:ch5_exp} the optimized survival vs. depth profile of Fig. \ref{fig:ch5_optim_sample} (lower panel)
is compared with the experimental results of Mitaroff et al. \citep{mitaroff1998biological_ref66}. 
There appears to be satisfactory agreement between the results based on the Katz model and
experimental data of Mitaroff et al., at least over the target region.

\section{Comparison with LEM-based survival vs. depth calculations}

\begin{figure}[ht!]
 \centering
 \includegraphics[width=0.8\textwidth]{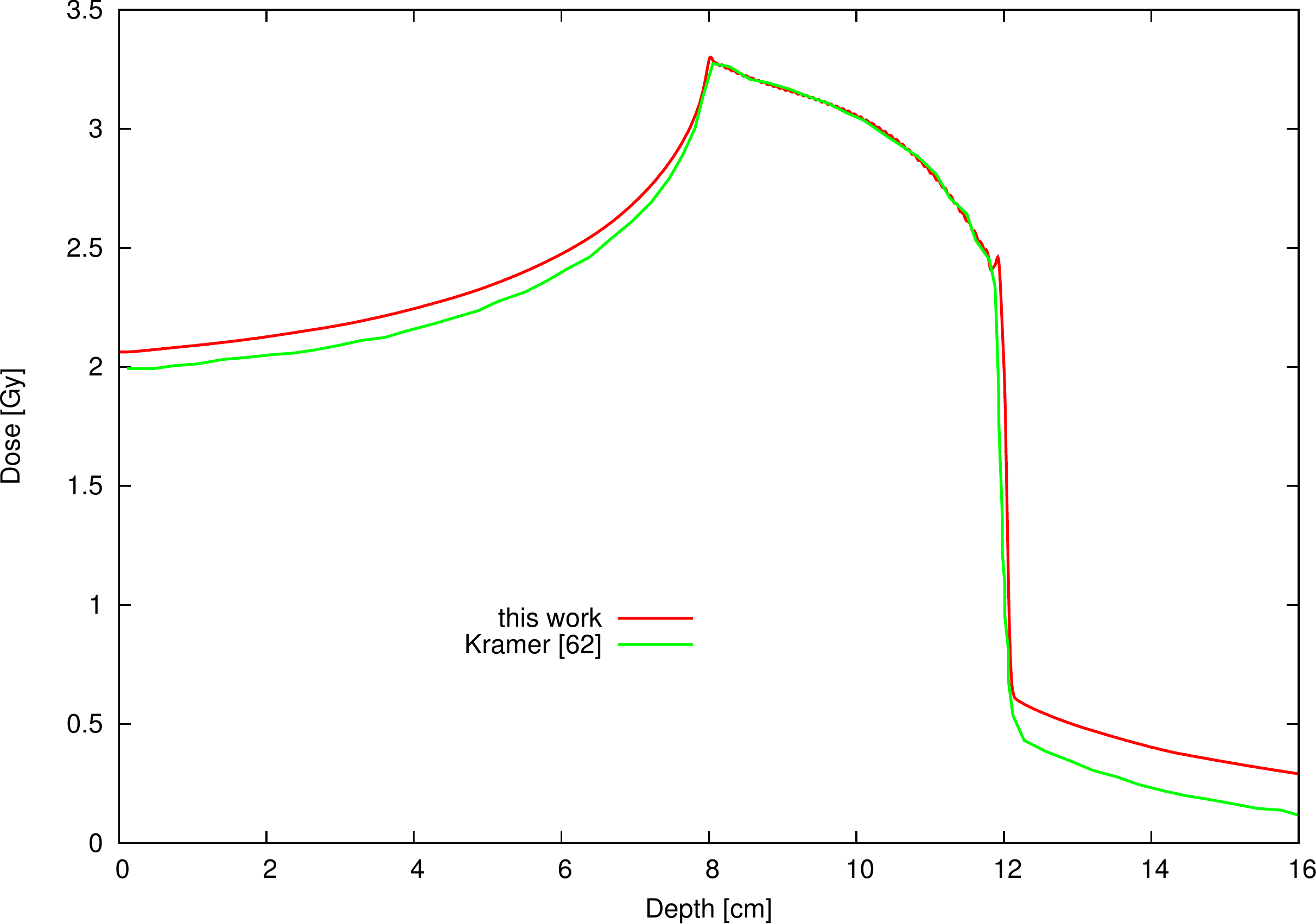}
 \caption{The depth-dose profile used in calculation of Kramer and Scholz
\citep{Scholz_ref62} (green line) compared with depth dose profile calculated with the aid of
the libamtrack library (red line). Profile marked with red line was adjusted to
agree with Kramer’s profile in depths range between 8 and 12 cm. Calculations
performed using the libamtrack library.}
 \label{fig:ch5_dose_scholz}
\end{figure}

In order to compare results of calculations of depth-survival profiles based on
Katz’s scaled model shown in Fig. 1 with results of calculations using LEM
published by Kramer and Scholz \citep{Scholz_ref62}, the depth-dose profile obtained in this
work (cf. Fig. \ref{fig:ch5_optim_sample}, upper panel) and that published by Kramer and Scholz are
compared in Fig. \ref{fig:ch5_dose_scholz}. To apply the depth-dose profile of Kramer and Scholz in
calculating the depth-survival dependence according to the present calculations,
a fourth-order polynomial, $D(z)$, was fitted to the dose profile of Kramer and
Scholz over the depths between 8 and 12 cm: 

\begin{equation}
D(z) = a_3 z^3 + a_2 z^2 + a_1 z + a_0
\end{equation}

where $a_0 = 20.7499$; $a_1 = -5.61387$; $a_2 = 0.607124$ and $a_3 = -0.0222678$.  

\begin{figure}[ht!]
 \centering
 \includegraphics[width=0.8\textwidth]{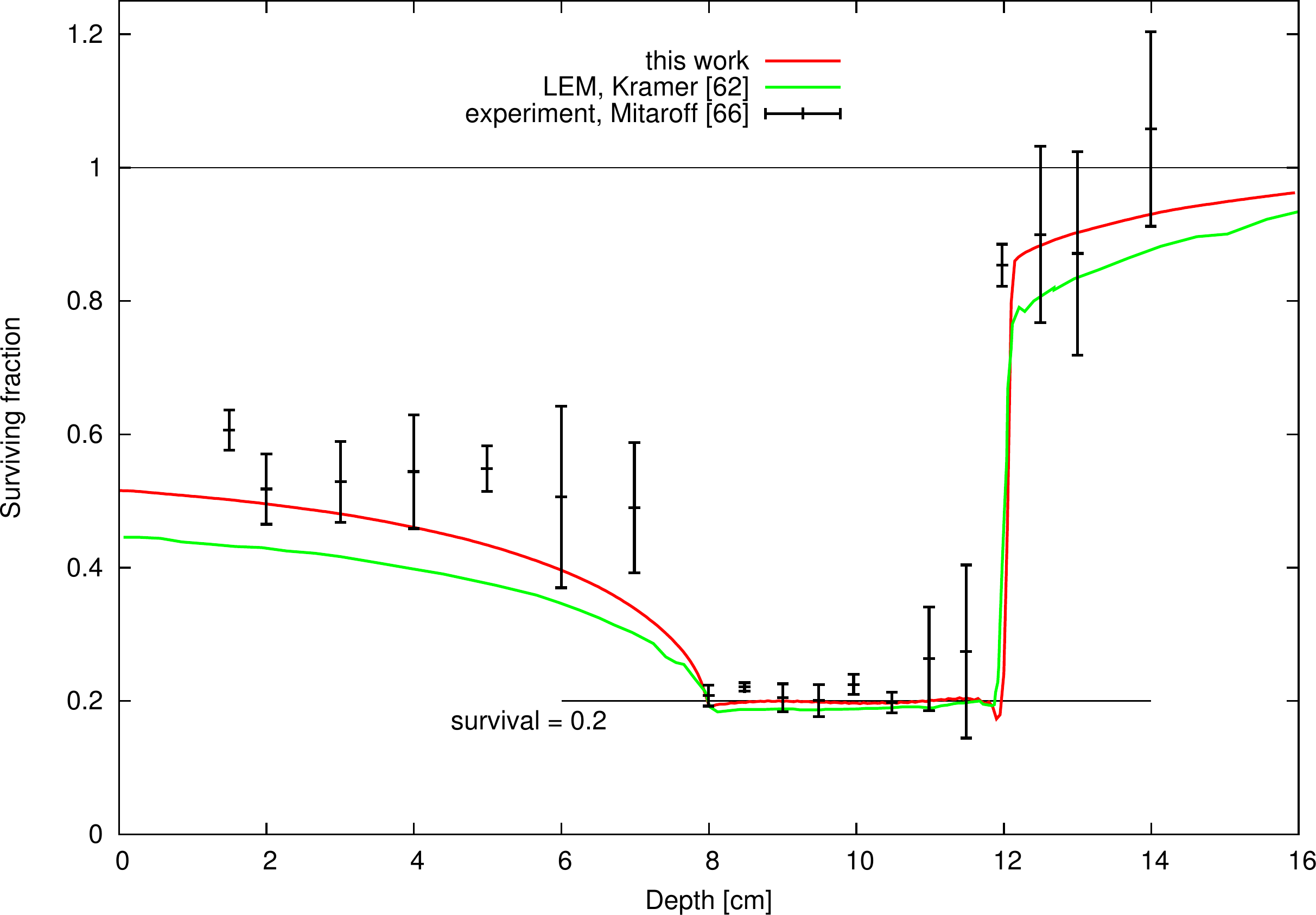}
 \caption{Survival of CHO cells vs depth: results of calculations using the
dose optimization algorithm (dose profile presented on figure \ref{fig:ch5_dose_scholz}) and the
scaled Katz model, where CHO cells are represented by the model parameters:    $m$ = 2.31,  $D_0$ = 1.69~Gy, $\sigma_0$
= $5.96 \cdot 10^{-11}\ \rm{m}^2$ , and $\kappa$ = 1692.8,  calculations
performed using the libamtrack library, compared with results of LEM
calculations (reproduced from Kramer et al. \citep{Scholz_ref62}).}
 \label{fig:survival}
\end{figure}

Next, by using the optimization procedure described in chapter 4, implemented in
libamtrack library, a linear combination of pristine Bragg peaks was found, such
that the dose profile described by $D(z)$ was maintained. The resulting depth-dose
profile is presented in figure 5.5 (red line). It agrees, as assumed, with the
dose profile used in LEM model over the depth range between 8 and 12 cm, but is
generally lower elsewhere. This discrepancy could be connected to different beam
model used in LEM model.
Comparison between the depth-survival dependences: calculated in this work (cf.
Fig. \ref{fig:ch5_optim_sample} and Fig. \ref{fig:ch5_exp}) and that published by Kramer and Scholz
is shown in Fig. \ref{fig:survival}.

The depth-survival dependences calculated using Katz’s model and LEM show good
agreement with experimentally measured CHO cell survival over the depth range
between 8 and 12 cm. Over the  entrance channel and behind, both calculations
tend to predict lower survival rates than those measured experimentally: by some
5-15\% in the case of calculations made in this work, and by some 5-30\% for
LEM-based calculations. Interestingly, the Katz model-based calculations show a
higher entrance dose to achieve the required 0.2 survival level in the target
area than do the LEM-based calculations. Yet, outside the target region, the
Katz model-predicted cell survival appears to be higher than that resulting from
LEM-based calculations – and perhaps better representing the actually measured
CHO cell survival. A more detailed comparison, at 1.5 cm depth, is given in
Table \ref{tab:ch5_dose_surv}. This difference may be related both to the different assumptions
concerning beam transport and to the differences in the radiobiological models
used.

\begin{table}[!h]
\begin{tabular}{m{0.2\textwidth}m{0.35\textwidth}m{0.35\textwidth}}

\hline
\textbf{} & {Dose [Gy]} & {Cell survival} \\
\hline

LEM & 2.03 & 0.432 \\

Katz & 2.11 & 0.502 \\

\hline
\end{tabular}
\caption{ Cell survival corresponding to dose calculated at 1.5cm depth, using the
Katz model and LEM.}
\label{tab:ch5_dose_surv}
\end{table}

\section{Dependence of survival vs. depth on beam entrance dose}
\label{sec:ch5_scaled}

An interesting question in carbon ion beam radiotherapy is to what extent is
scaling with dose valid? Assuming over the target region a flat depth-survival
profile at a given survival level, how will the survival level over that region
vary on varying the beam entrance dose? And how flat will this changed depth-
survival profile remain on varying the beam entrance dose? 

\begin{figure}[h!]
 \centering
 \includegraphics[width=0.8\textwidth]{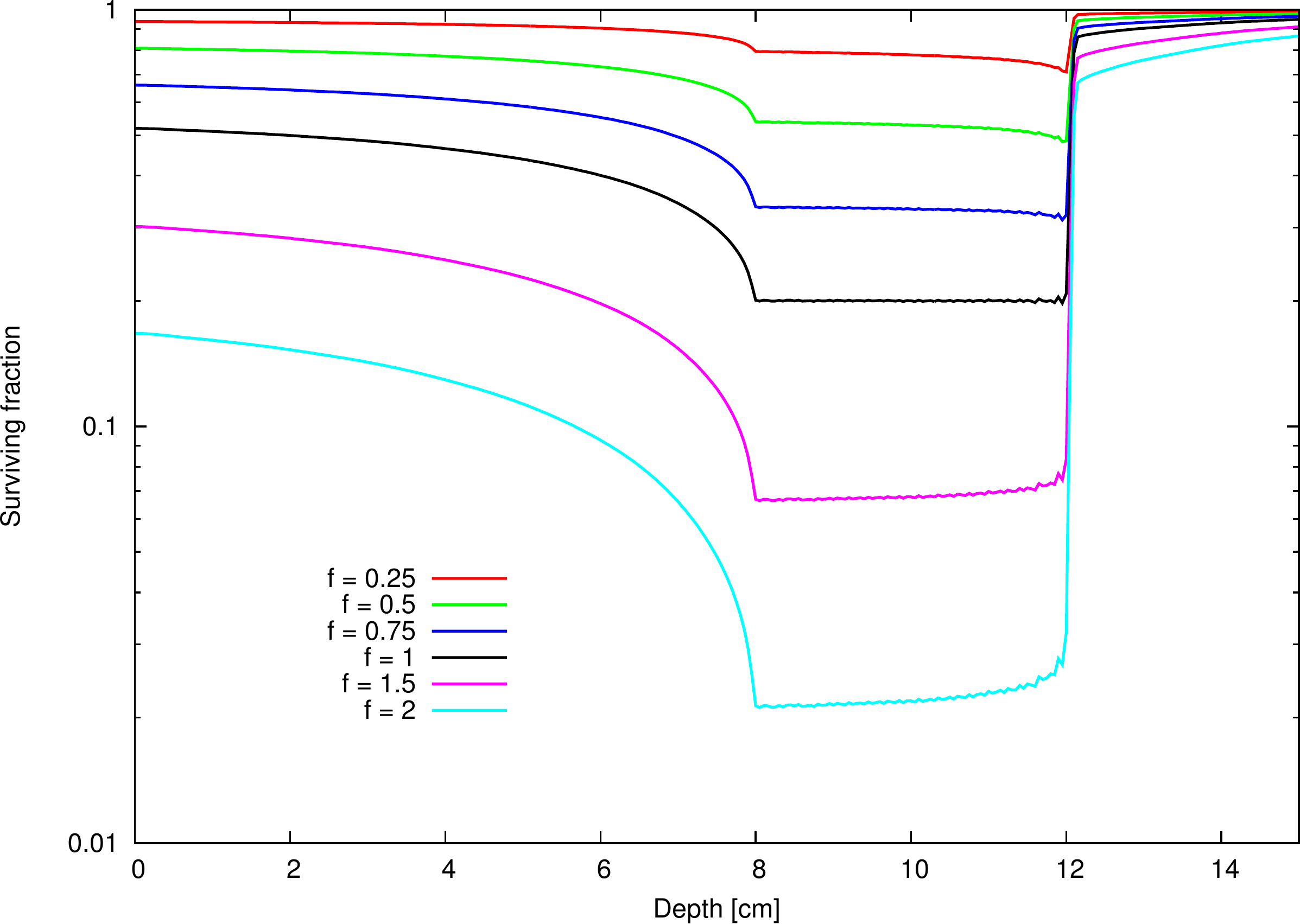}
 \caption{Variation of depth-survival profiles with beam entrance dose. A CHO
survival vs.  depth profile was designed to give a flat survival level of 0.5
between 8 and 12 cm depths (scaling factor, $f$=1, black full line). Next, the
input energy-fluence spectrum of the beam was re-scaled by factors: 0.25, 0.5,
1.5 and 2, and the respective depth-survival profiles were re-calculated
(coloured full lines). For the Katz model parameters representing CHO cells,
used in these calculations, see Fig. \ref{fig:ch5_optim_sample}. 
Calculations performed using the libamtrack library. .}
 \label{fig:ch5_scaling}
\end{figure}

To investigate this matter, a survival vs. depth profile was designed to yield a
flat survival level of 50\% (0.5) over the depth region between 8 and 12 cm, as
plotted in Fig. \ref{fig:ch5_scaling}. ($f$=1). Next, the initial energy-fluence spectrum (not
shown) was multiplied by factors: 0.25, 0.5, 1.5 and 2.0 and entered into the
survival profile calculations. The resulting depth-survival dependences are
plotted in Fig. \ref{fig:ch5_scaling} with values of their respective scaling factors. 
On increasing the beam entrance dose by a factor of two ($f$=2) the survival level
over the target region decreased from 0.5 (50\%) to about 0.022 (2.2\%) and the
survival level over the target region did not remain constant. A systematic
change of the “slope” of the of the depth-survival profile over the target
region may be observed with decreasing beam entrance dose.

\section{Dependence of survival vs. depth on cell oxygenation status}
\label{sec:ch5_oxyg}

Another interesting question is the dependence of the depth-survival profile on
the oxygenation status of the cell. One may assume that tumour cells may remain
in hypoxic conditions due to the insufficient supply of oxygen to the rapidly
growing tumour. Typically, hypoxic cells are more radioresistant. It is
therefore interesting to investigate the effect of the cell oxygenation status
on the design of the depth-survival profile.

A systematic study of the radiobiological parameters of Chinese hamster V79
cells in aerobic and hypoxic conditions was performed by a Japanese group at
NIRS and published by Furusawa et al. \citep{furusawa_ref59}. Systematic measurements of survival
of V79 cells under aerobic or hypoxic conditions after their irradiation by the
following ion beams: helium (energies between 1.17 and 9.74 MeV/amu), carbon
(energies between 1.9 and 123 MeV/amu) and neon (energies between 7.7 and 124
MeV/amu). From these two data sets, Katz model parameters representing V79 cells
in aerobic and hypoxic conditions, fitted by Korcyl \citep{KorcylPhD_ref45} are shown in Table \ref{tab:ch5_oxic_hypo}

\begin{table}[!h]
\begin{tabular}{m{0.2\textwidth}m{0.16\textwidth}m{0.16\textwidth}m{0.16\textwidth}m{0.16\textwidth}}

\hline
\textbf{} & {$m$} & {$D_0$ [Gy]} & {$\sigma_0$} [m$^2$] & $\kappa$\\
\hline

Aerobic & 2.91 & 2.0504 & $5.06 \cdot 10^{-11}$ & 689 \\

Hypoxic & 3.22 & 5.26 & $5.529 \cdot 10^{-11}$ & 1002.2 \\

\hline
\end{tabular}
\caption{Best-fitted parameters of the Katz model, representing aerobic and
hypoxic V79 cells \citep{KorcylPhD_ref45}}
\label{tab:ch5_oxic_hypo}
\end{table}

\begin{figure}[h!]
 \centering
 \includegraphics[width=0.8\textwidth]{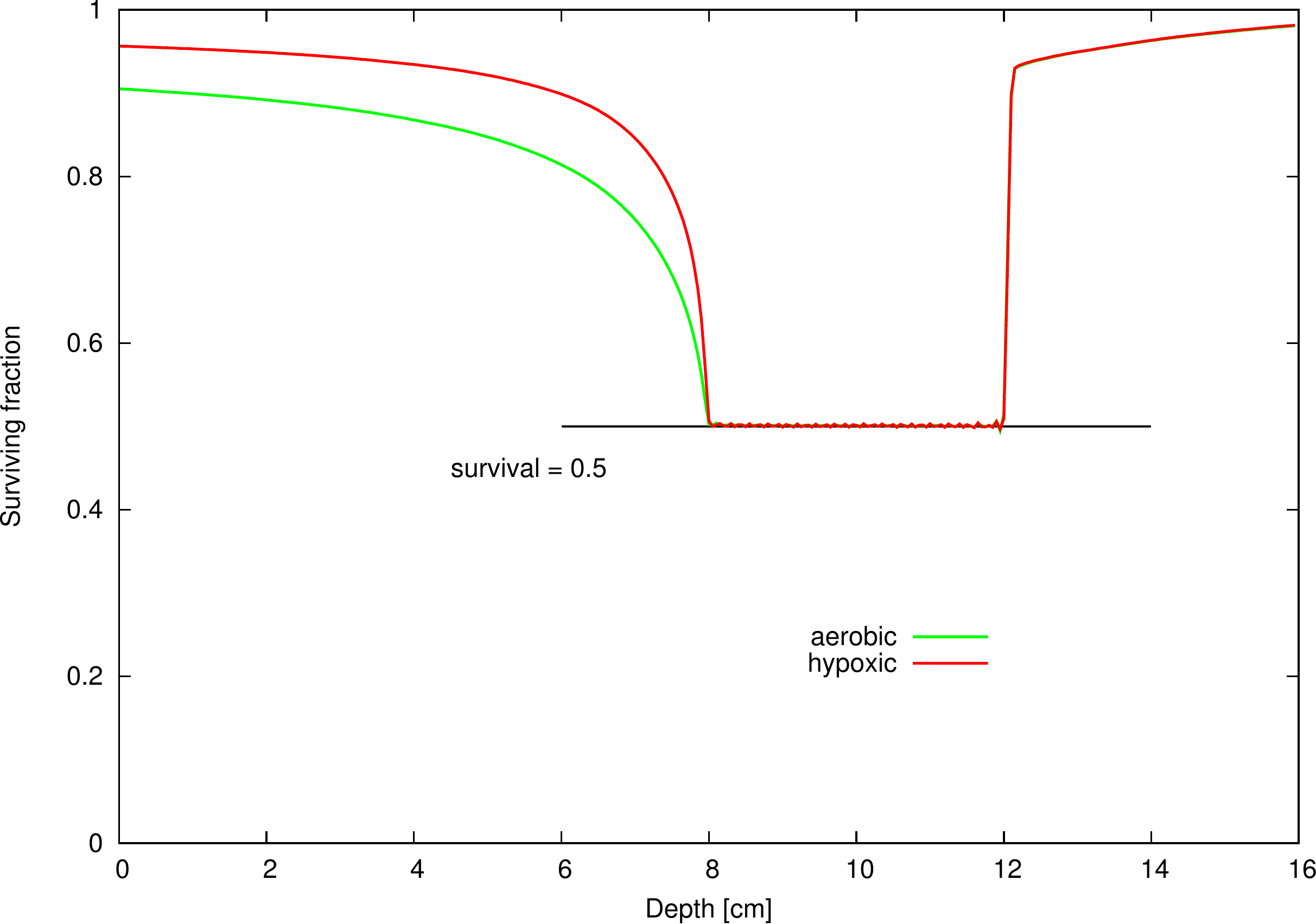}
 \caption{Cell survival depth-profiles calculated for aerobic and hypoxic V79
cells using Katz model (parameters listed in Table \ref{tab:ch5_oxic_hypo}).
For each case: aerobic and hypoxic two different beam configurations were used, 
each prepared in such
way that the designed survival was equal to 0.5 over the interval between 8 and
12 cm. Calculations performed using the libamtrack library.}
 \label{fig:ch5_oxyg_survival_both}
\end{figure}

Applying the approach described in par. \ref{sec:ch5_optim}, two carbon ion beam configurations
were prepared aimed at achieving a constant cell survival of the level of 0.5
over 8 and 12 cm depth, for V79 cells in aerobic or hypoxic conditions,
respectively. The respective calculated cell depth-survival profiles are shown
in figure \ref{fig:ch5_oxyg_survival_both}. The same Katz model parameters were used to calculate survival at
all depths, including the target region.

\begin{figure}[!h]
 \centering
 \includegraphics[width=0.8\textwidth]{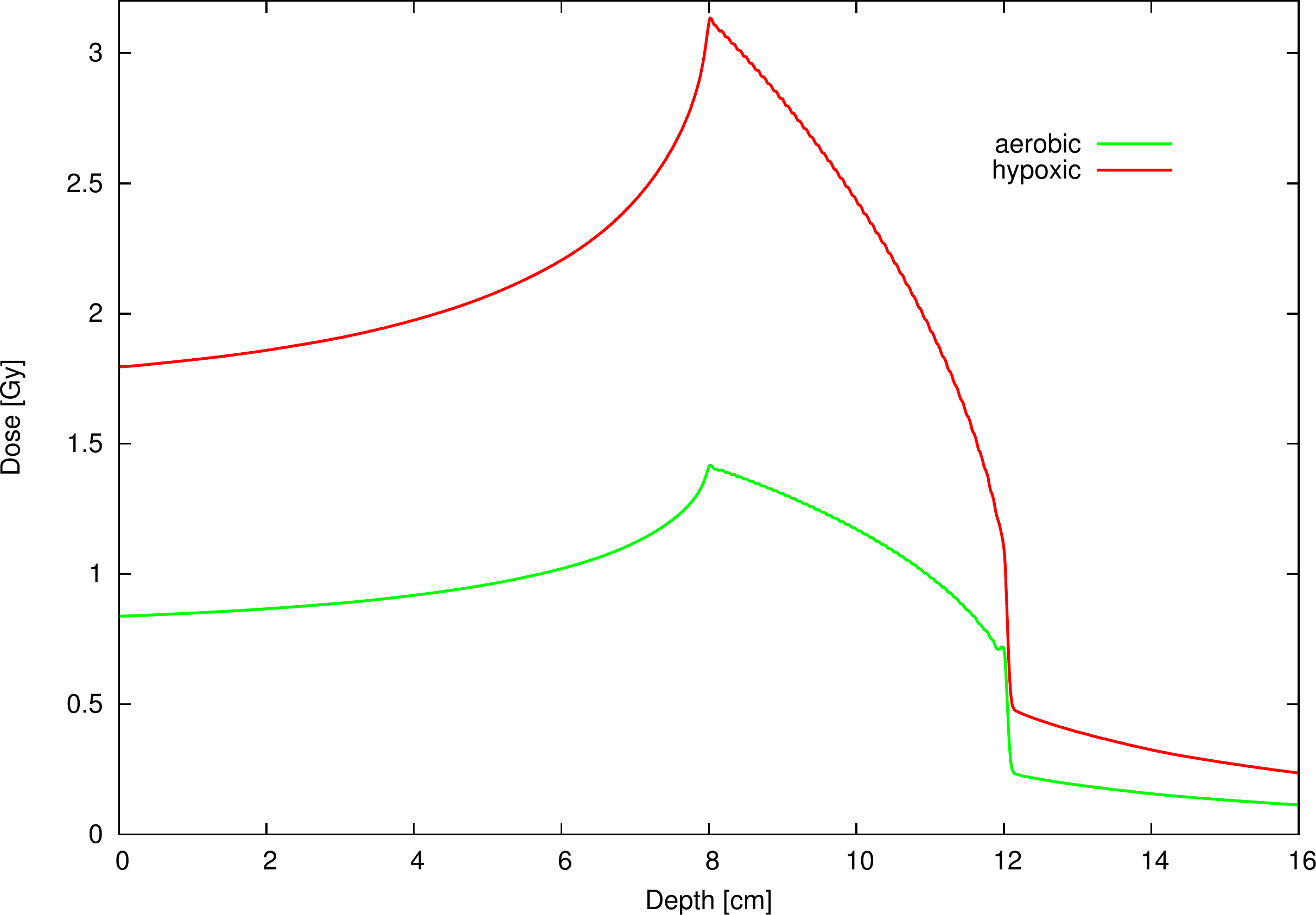}
 \caption{Dose depth-profiles related to cell survival profiles presented on
figure \ref{fig:ch5_oxyg_survival_both}. Calculations performed using the libamtrack library.}
 \label{fig:ch5_oxyg_dose}
\end{figure}

The depth-dose profiles required to achieve the depth-survival profiles of Fig.
\ref{fig:ch5_oxyg_survival_both}, are shown in Fig. \ref{fig:ch5_oxyg_dose}. As the V79 cells irradiated in hypoxic conditions
are more radioresistant than cells well-oxygenated (aerobic), one may observe in
Fig. \ref{fig:ch5_oxyg_survival_both} that to achieve the same survival of 50\% over the target region, about
twice as high entrance dose is required for the hypoxic cells than for aerated
cells. The difference between these two dose profiles can be attributed to the
LET-spectrum of the carbon beam varying with depth.

\begin{figure}[!h]
 \centering
 \includegraphics[width=0.8\textwidth]{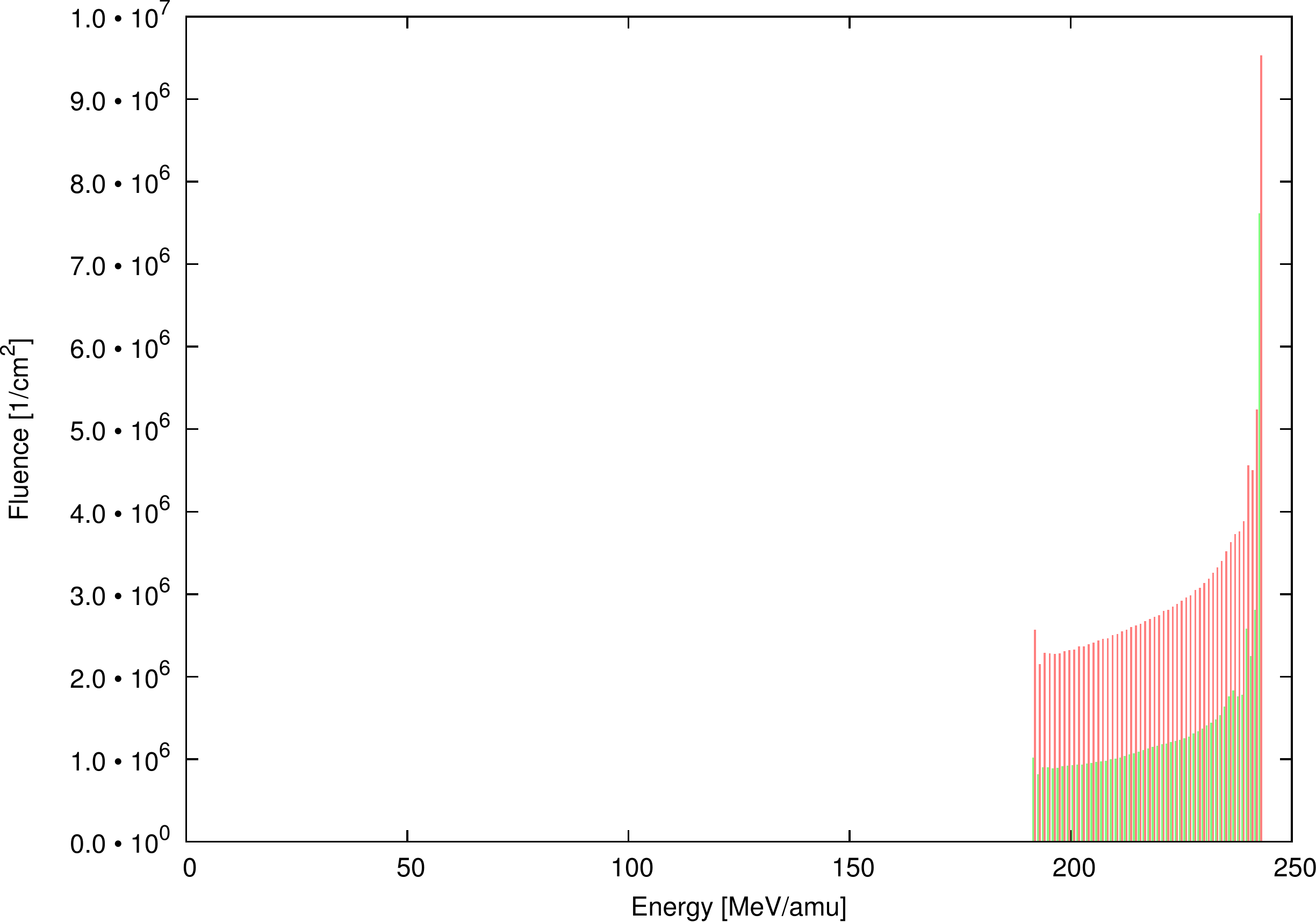}
 \caption{Initial energy-fluence spectra related to cell survival presented in
figure \ref{fig:ch5_oxyg_survival_both}. Calculations performed using the libamtrack library.}
 \label{fig:ch5_oxyg_init}
\end{figure}

The initial energy-fluence spectra of the two beams are shown in figure \ref{fig:ch5_oxyg_init}.
The beams are composed of 49 pristine Bragg peaks each, with initial energies
ranging between 191.5 MeV/amu and 242.5 MeV/amu.
 
\begin{figure}[p!]
 \centering
 \includegraphics[width=0.8\textwidth]{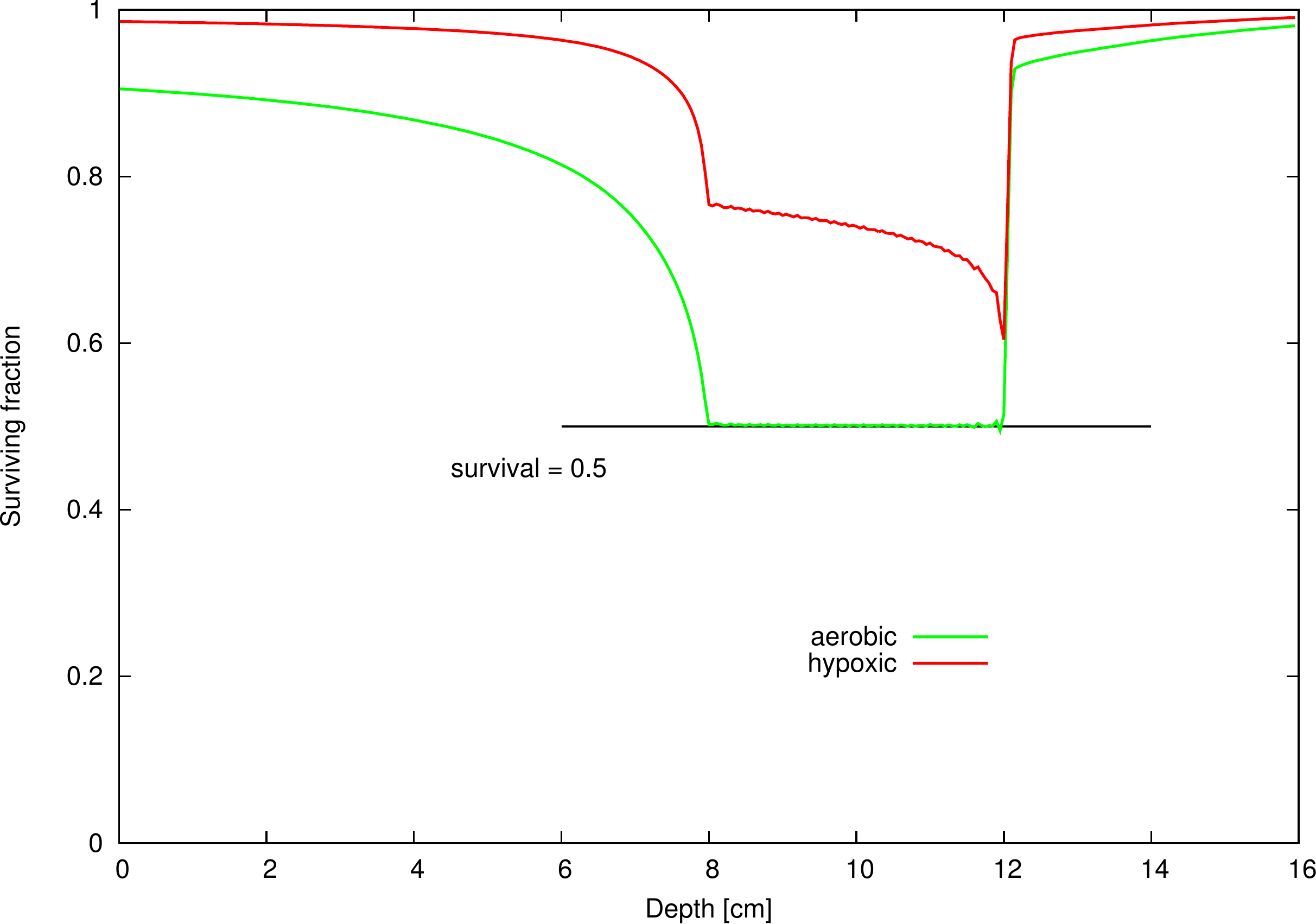}
\includegraphics[width=0.8\textwidth]{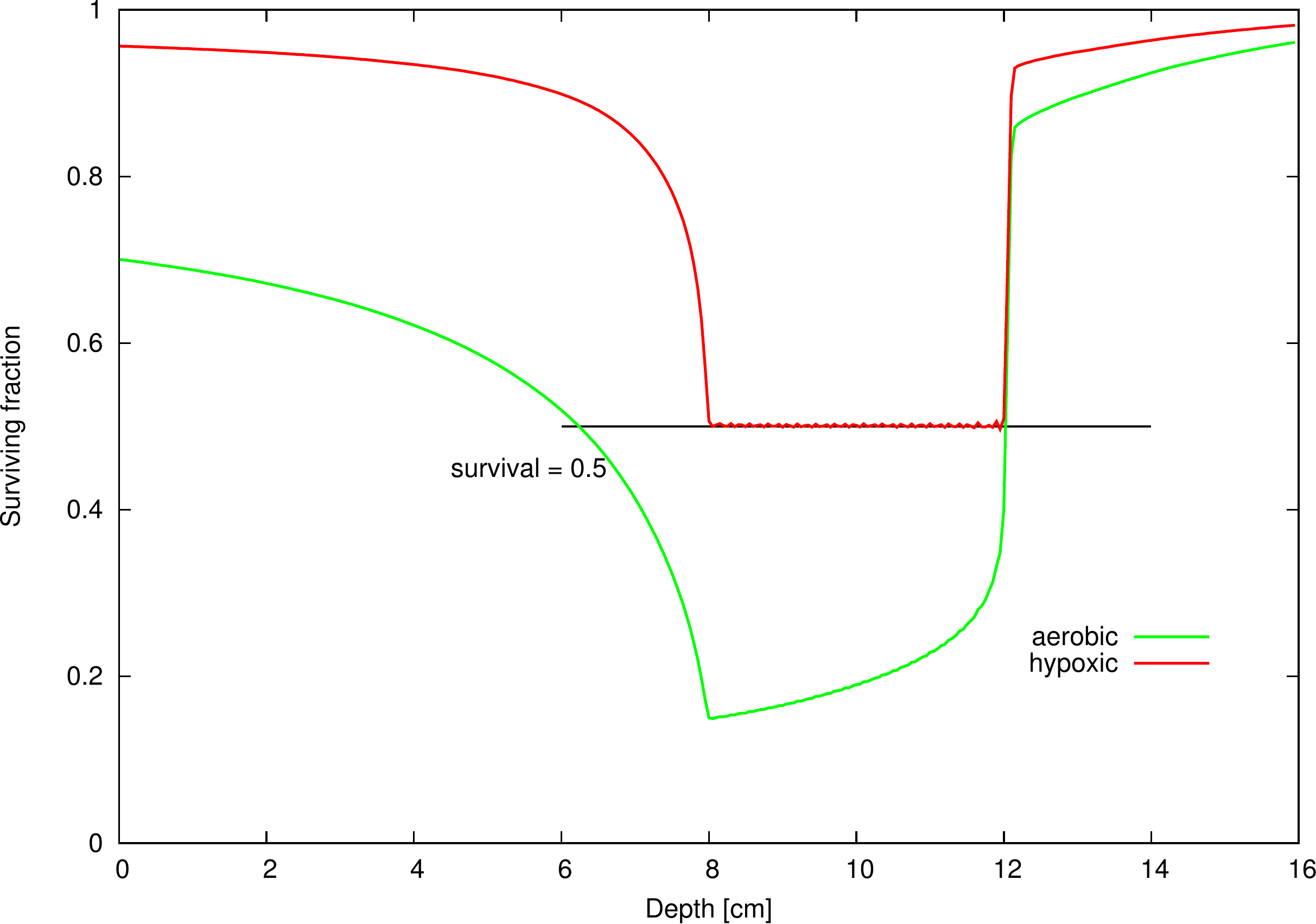}
 \caption{Cell survival depth-profiles calculated for aerobic and hypoxic V79
cells in two identical carbon ion beams. Initial beam configuration was prepared
for the survival of hypoxic cells to be 50\% (0.5) over the target region between
8 and 12 cm.}
 \label{fig:ch5_oxyg_sep}
\end{figure}

Beam configurations, planned for iso-survival of hypoxic and aerobic cells
respectively, were used to calculate the predicted depth-survival profiles in
the case where the same beam combination was used to obtain a flat 50\% survival
of aerobic cells or flat 50\% survival of hypoxic cells (Fig. \ref{fig:ch5_oxyg_sep}). 
In the first case, “cancer” (i.e. hypoxic) cells would not be sufficiently
treated; in the second, the “normal tissue” (aerobic) cells would show
over-exposure (or “complications”).  This example illustrates the predictive
capacity of the model calculation in optimising likely therapy situations,
provided that suitable representative model parameters to represent cells in
different oxygenation conditions, are available.

\section{Conclusion}

In this Chapter, all the elements discussed in earlier parts of this thesis have
been brought together and applied in a radiobiology-based approach to developing
a treatment planning system for carbon radiotherapy, albeit in one dimension
(depth) only. While the major part of this work has been performed by the
author, his collaboration with Steffen Greilich, Marta Korcyl and Pablo Botas is
gratefully acknowledged. All codes used in this work which he developed have now
been implemented in the libamtrack library.

The radiobiological model of Katz in its scaled version (Chapter 3), based on
the radial dose distribution (Chapter 2) has been applied to a model of a carbon
beam propagating through liquid water, as represented by 1-dimensional
energy-fluence spectra of the carbon beam and its secondary ions, obtained at
several depths, basing on Monte Carlo simulations (Chapter 4). Crucial to the
possibility of handling the approach presented in this chapter was the
libamtrack code library (Chapter 2) which contains all codes necessary for
performing the calculations. Application of interpolation techniques to
calculate the energy-fluence spectra over a regular grid (Chapter 4) and
development of numerical methods to optimize the distributions of depth-dose
(Chapter 4) and survival-depth (Chapter 5) dependences, was necessary. It is
very satisfying that the results of model calculations were able to closely
predict the results of a radiobiological experiment using CHO cells (par. \ref{sec:ch5_exp})
and that the optimised beam configuration in calculations representing this
experiment were found to closely agree with those evaluated independently by
other authors. This agreement suggested that predictions could be made of the
effects of varying the input dose (par. \ref{sec:ch5_scaled}) and of the cell oxygenation status
(par. \ref{sec:ch5_oxyg}) on the depth-survival profiles.

Thus, a quantitative model of a treatment planning kernel has been developed and
presented in this chapter, based on a highly efficient and predictive
radiobiological model, which enables quantitative predictions to be made of the
expected survival-depth dependences in a manner amenable for future development
into a carbon ion therapy planning system.

A more extensive discussion of the results obtained in this thesis is presented
in Chapter 6.

\mgrclosechapter

\chapter{Discussion and conclusions}
This final chapter contains a more detailed discussion of results obtained in
Chapters 2-5, a summary of this work, where key issues that have been resolved
by the author are listed, followed by brief conclusions and proposed future work
suggested by this thesis.

\section{Discussion}
Carbon ion radiotherapy is a new and fairly rare treatment modality. Only six
centres in the world are currently in operation, three in Japan (Chiba-NIRS,
Hyogo and Gunma) and single centres in Germany (HIT-Heidelberg), Italy
(CNAO-Pavia) and China (Lanzhou). As of March 2013, some 10 thousand patients
have been treated by carbon ions, most of them in Japan and some 1500 at the HIT and GSI
facility \footnote{\verb"http://ptcog.web.psi.ch/patient_statistics.html"}. Pioneering
work in the development of carbon radiotherapy began around 1997 at GSI
Darmstadt in Germany where about 600 patients were treated. The Local Effect
Model (LEM) was developed at that time at GSI \citep{Scholz1994_ref34} as the
radiobiological basis for the carbon ion treatment planning system, now in use
at HIT. Elements of LEM are also applied in the Japanese treatment planning
systems. The cellular Track Structure Theory (or the Katz model) was developed
earlier \citep{Butts1967}, but was believed to be less amenable to clinical
radiotherapy due to its reliance on the m-target rather than linear-quadratic
formalism. The rationale for this thesis was to investigate the possibility of
applying Katz’s radiobiological model in a carbon ion therapy planning system.
Attractive were the simplicity of the analytical formulation of the Katz model
and its well-known predictive power in describing RBE and cell survival in
vitro \citep{katz1994survey_ref88}.

Development of a TPS for clinical application is a major project, however the
basic features of such a system could be studied by developing its
one-dimensional kernel in which the scaled Katz model would be implemented as
its radiobiology component while the physical component would be supplied by a
data base of Monte Carlo-calculated beam transport calculations of carbon beams
and secondary ions in water, available to the author of this thesis. The
open-source libamtrack code library which has been co-developed by the author
would serve as the repository for the codes that were to be developed in the
course of this project, enabling free access to all these codes and results of
this research. Since the Katz model differs in many aspects from LEM 
\citep{Scholz_ref62}, \citep{scholz2004physical_ref68},  \citep{beuve2009formalization_ref89}, \citep{elsasser2010comments_ref90}, 
new insights to carbon therapy planning were
expected, especially since the LEM-based TPS in clinical use, e.g. at HIT in
Heidelberg is now a commercial product with no access to its code, nor is it
available for research purposes outside HIT.

The cellular Track Structure Theory (or the Katz model) uses many elements
shared by other amorphous track structure model approaches, such as the radial
distribution of average dose, $D(r)$, around the path of a heavy ion (par. 2.3)
which, combined with the response after uniform irradiation by a reference
radiation, given by the m-target formula (eq. 3.9), yields the inactivation
cross-section (eq. 3.12). The possibility of applying  scaling factors in this model appears to be closely related to the
selection of scalable average $D(r)$ formula (e.g., the
averaged equation of Zhang, see Table 2.2 and Appendices A and B), and to the
use of m-target formalism. Marta Korcyl in her Ph. D. thesis \citep{KorcylPhD_ref45}
proposed an efficient method of calculating the inactivation cross section based
on Zhang’s formula (see Chapter 2 and Appendix C). It was then possible to find
best-fitted values of model parameters from sets of published survival curves
for normal human skin fibroblasts \citep{korcyl2009track_ref86}.  Following a careful analysis of
the scaling properties of Katz’s model \citep{Korcyl2013_ref85}, the validity of the set
of simple analytic formulae (eq. 3.29-3.34), originally proposed by Katz, was
confirmed, the “track-width” approximations were re-calculated and the scaled
version of the  Katz model was implemented by the author in the libamtrack code
library. This scaled version of the Katz model, including Zhang’s radial dose
distribution formula, was first used by M. Korcyl to model relative
effectiveness of alanine and effect of heavy ion bombardment on E.Coli spores
\citep{Korcyl2013_ref85}.

This version of the scaled Katz model is also used in this thesis in all
calculations presented in Chapter 3 and Chapter 5. The author further developed
and implemented the parameter-fitting routine into the libamtrack library and
applied it to find best-fitting values of $m$, $D_0$, $\sigma_0$ and $\kappa$ representing
survival of Chinese Hamster Ovary (CHO) cells (eq. \ref{eq:ch3_bestfitted}) and their uncertainties
(eq. \ref{eq:ch3_bestfitted_uncert}), from a set of data published by Weyrather et al. \citep{Weyrather1999_ref24}
 (par. 3.9), to be used in calculations presented in Chapter 5.

 One may argue that no less than four parameters are required to describe the
variation of cellular survival curves after a fluence of ions of specified
charge Z and energy: two ($m$ and $D_0$) to describe the
“curvature” of the response (via the m-target expression) after doses of
reference radiation, one ($\sigma_0$) to give the purely exponential response (such as
that shown in Fig. 1.4) and one ($\kappa$) as a “mixing parameter” to generate
intermediate “curvatures” of the survival curves after ions of various charges
or energies. Taking this view, the linear-quadratic parameters ($\alpha$ and $\beta$ in eq.
1.11) are too few to provide the full description of cell survival after ion
doses (or fluences), without additional assumptions. Indeed, several such
assumptions are made in LEM, and a microdosimetry approach \citep{beuve2009formalization_ref89}
 is used rather than that of average dose assumed
in Katz’s model. A comparison was made by Paganetti and Goiten between the Katz
model and an early version of LEM (LEM I) with respect to V79 cell line
survival. Some discrepancies were found between calculations and experimental
data for irradiation by proton beams \citep{Paganetti2001}.

The Katz model has been applied in many areas, and recently in modelling the
risk from space radiation \citep{Cucinotta1997, cucinotta1999applications_ref87}
. Since galactic cosmic
rays are composed mostly of energetic protons (>200MeV), amorphous track models
are suitable for such studies.

The set of equations, eq. 3.29-3.34, was termed the “scaled Katz model” by the
author of this thesis and is used in all further calculations, due to their
simplicity and computational efficiency. The version of the model where
integration of the averaged $D(r)$ and of the cross section are performed
explicitly was termed the integrated version of this model (par. 3.6). Here,
“non-scalable” $D(r)$ formulae may be applied. In the integrated version, the
author proposes to replace $\kappa$ by the radius of the sensitive site, $a_0$ , as the
fourth model parameter. The integrated version of the model can also be used to
best fit model parameters, however best-fitted parameter values may in that case
be different from those fitted by the scaled version of the model. Application
of the integrated version of the Katz model may lead to interesting results, as
other $D(r)$ formulae, e.g. that of Geiss \citep{Geiss1997}, also used in, e.g., LEM
or other models \citep{Geiss1998} may then be applied for comparative
studies. However, the integrated version of Katz’s model is too
computing-intensive to be used in practical TPS development.

To evaluate the combined effect of a mixed-field (i.e. a field composed of track
segments of several ions of different charges, velocities and fluences) the Katz
model offers a well-defined analytical prescription (eq. 3.45-3.47), while in
LEM an “ansatz” for calculating the effective value of $\beta$ from a combination of $\alpha$
values representing cell survival curves after irradiation by the component
ions, multiplied by their respective dose contributions \citep{Rapid_ref63} is proposed. 
While a difference between the results of mixed-field
calculations of either model may then be expected, this was not verified.

The feature shared by the Katz model and by other amorphous models, such as LEM,
is the need to base model calculations on energy-fluence spectra of the ions in
the beam. In all calculations in this work, such energy-fluence spectra were
based on results of Monte Carlo simulations, as these spectra cannot be reliably
measured. Results of Monte Carlo simulations should closely match results of
available measurements, e.g. of depth-dose distributions. The precision of
Monte-Carlo generated beam profiles is good enough, such that dose deviations in
the plateau of numerically generated spread-out Bragg peaks is typically less
than 1\% which is a level satisfying clinical conditions for ion beam
radiotherapy.

Nuclear interaction models, implemented in SHIELD-HIT10A were recently updated
(compared to the previous version, the SHIELD-HIT08) and benchmarked against
experimental data, as reported by Armin et al \citep{ArminSPC_ref61}. Most of
these updates are relevant for carbon ion interactions with nuclei in the
target, and in that case agreement within two sigma with data is achieved \citep{Hansen-SHIELD_ref60}.

SHIELD-HIT10A, when compared with other Monte-Carlo codes, such as Geant4 \citep{Geant4_ref26}
or Fluka \citep{Fluka_ref27} is able to adequately reproduce fragment yields of ions lighter
than lithium. For boron and beryllium Fluka and Geant4 show better agreement
with data . It has to be stated that none of these codes alone
could be recommended to best cover all ion types and ion energy ranges
considered in carbon ion radiotherapy. 
SHIELD-HIT 10A Monte Carlo transport calculations of 50-500 MeV/amu carbon beams
in water performed by Pablo Botas at DFKZ (Heidelberg) served as the initial
input to beam transport modelling in this work. A suitable energy-fluence
spectra data base, in the format of SPC files, was implemented by the author in
the libamtrack library (par. 4.1.2). An algorithm was developed by the author in
order to estimate fluence-energy spectra at desired intermediate depths of
carbon ion beams of different initial energies (par. 4.1.3). The algorithm used
for this purpose is similar to the bilinear interpolation of energy-fluence
spectra used by Kramer and Scholz in LEM I \citep{Scholz_ref62}.
In 2006 a description of an improved version of LEM was published \citep{Rapid_ref63} which included a more efficient
derivation of tabularized data and an improved data access algorithm, based on a
lookup-table containing pre-calculated values of alpha and beta parameters of
the linear-quadratic model at different depths.

As an illustration of the complex interactions of carbon beams in water, it is
interesting to note the difference between the contribution of the primary and
secondary ions in a pristine 270 MeV/amu carbon beam to the depth-dose (Fig.
4.1) and to depth-fluence (Fig. 4.2) distributions, as presented in cumulative
and differential forms (upper and lower panels in these figures). While the
fluence of secondary protons from nuclear reaction clearly dominates that of the
primary carbon ions, the major contribution to the dose is still from the
primary carbon component of the beam. Also, interesting is the decrease of the
fluence of the primary carbon beam with depth (Fig. 4.2, lower panel), caused by
nuclear reactions and scattering.

The author’s development of the interpolation algorithm (par. 4.1.3) was
essential to achieve linear superposition of energy-fluence spectra within
regular steps in depth, as shown in Fig. 4.3, in preparation for spread-out
Bragg peak calculations where such a linear superposition of pristine carbon
beams of given initial energies and fluences was to yield the required flat
depth-dose distribution over a given depth region. The optimizing routine
implemented for this purpose is also able to find the optimum solution for
desired depth-survival dependences of shapes other than flat.  By applying an
inverse optimization algorithm developed by the author ( par. 4.2.1), in the
example shown, a flat dose of 1 Gy over the depth region 8-12 cm (Fig. 4.4,
upper panel) was achieved by a linear superposition of pristine carbon beams of
initial energies and fluences shown in Fig. 4.5, where the beam of the highest
energy clearly dominates. Good convergence of the developed optimization
algorithm (within about 30 steps, as illustrated in Fig. 4.6) attests to its
computation efficiency. Application of a mixed-field calculation of the Katz
(scaled) model using cellular parameters representing CHO cells to this
dose-depth profile results in a highly non-uniform distribution of cellular
survival (Fig. 4.4, lower panel).

This example clearly illustrates the basic problem of carbon ion beam
radiotherapy: achieving a uniform dose distribution over a given target region
will not result in a uniform distribution of cellular survival over that region.
This is due to the complex variation with depth of biological effectiveness
(RBE) of the carbon beam and its secondaries.

The presented dose profile optimization algorithm does not take into account
many aspects relevant for treatment planning system: plan robustness, multiple
beams, complex treatment volume shape, and presence of organs at risks. Its main
goal was to show the possibility of modifying the entrance energy-fluence
spectra of carbon ion beams in order to obtain the desired depth-dose profile.
Linearity of $\chi^2$ minimization in that case is exploited in the gradient
algorithm, which is an approach similar to that used in the TRiP98 planning
system, where a conjugate gradient algorithm is used \citep{kramerPlanning_ref65}.

The simple analytic calculation of mixed-field irradiation in the Katz model
(par. 3.10) was applied to the set of interpolated energy-fluence spectra,
resulting in the algorithm given in par. 5.1, to which another optimisation
algorithm (par. 5.2.1) was developed by the author. Results of sample
calculation: a flat depth-survival over a selected depth region, the
corresponding non-uniform depth-dose distribution (Fig. 5.1) and the initial
energy-fluence spectrum (Fig. 5.2), have been achieved efficiently, within about
30 iteration steps of the optimizing routine (Fig. 5.3). This example of a
calculation of a flat survival-depth profile (20\% survival over depths 8-12 cm)
with Katz model parameters representing CHO cells was chosen deliberately to
verify the model prediction against published results of a radiobiological
experiment \citep{mitaroff1998biological_ref66} .  The comparison is shown in Fig. 5.4, where
agreement between the levels of survival measured outside the flat region and
those predicted by the calculation are very satisfactory, considering the
experimental uncertainties and the difficulty of the experiment itself. The
calculated depth-dose profile to achieve the flat depth-survival dependence of
Fig. 5.3 is shown together with a similar profile published by Kramer and Scholz
\citep{Scholz_ref62}, used to verify LEM calculations against the
experiment of Mitaroff et al.  For additional verification, the author fitted a
polynomial to the appropriate part of the profile published by Kramer and Scholz
(Fig. 5.5) and re-calculated the depth-survival dependence. Results of this
calculation and the published results of LEM calculations [Kramer and Scholz,
200] are compared with the experimental data in Fig. 5.6. Interestingly, while
in the author’s calculation using the Katz model, a higher entrance dose is
required than that in the LEM calculation, the Katz model-predicts survival
levels which are systematically higher outside the flat region than those
calculated by LEM (see Fig. 5.6 and Table 5.1). Since in either model the beam
transport calculation is intimately and non-linearly tied up with the
radiobiology calculation, it is not possible to decide whether the source of
this difference lies in differences between the beam or the radiobiology
components of these two model calculations. One should add that this
intercomparison is very limited in scope, as only one cell line was studied and
a very simple target location was investigated. Broader studies could shed more
light on the quality of predictions of LEM and the Katz models.

Encouraged by this result, the dependence of the survival-depth curve on the
input dose (or fluence) was studied, also applying CHO cell parameters (Fig.
5.7), to demonstrate that not only does the survival level over the flat region
depend non-linearly on beam entrance dose, but so do the slope and flatness of
the survival vs. depth curve. In another example, where cellular parameters
representing aerated or hypoxic V79 cells fitted by Korcyl [Korcyl, 2012] were
applied (par.5.6), it was found that quite different depth profiles (Fig. 5.9)
and initial energy-fluence spectra (Fig. 5.10) are required to achieve the same
50\% survival vs. depth profiles over their flat regions (Fig. 5.8). It was
observed, as expected that less radiosensitive hypoxic V79 cells require higher
input fluence than aerobic cells, but also that the input fluence does not scale
equally with depth. As can be seen from Fig. 5.10, hypoxic cells require about
2.5 times higher fluence of lowest energy carbon ions and about 1.25 times
higher fluence of carbon ions of the highest energy than aerobic V79 cells.

The impact of rescaled input fluence on predicted cell survival level was
investigated in chapter 5.3. As was expected, the predicted survival did not
scale uniformly with input fluence. The survival profile did not remain constant
as the initial fluence was rescaled by factors ranging from 0.25 to 2. Largest
deviations are observed in the distal region of the SOBP. This observation
suggests that in carbon ion beam treatment one cannot introduce a universal
physical depth-dose profile to be applied in preparing plans with different dose
in the target region.

The possibility of adjusting carbon ion beam treatment plans according to
oxygenation distribution in the tumour volume could lead to increased tumour
control. As was shown in Chapter 5 (Fig. 5.11) this cannot be realized by simply
increasing the dose by a constant factor, but has to be handled on the basis of
the survival optimization algorithm. Another attempt to solve that problem is
the so called LET-painting: a method to reshape the LET distribution in the
carbon beam, while maintain a given dose profile, described in \citep{bassler2010dose_ref70}.

The above comment not only illustrates the general feature of the developed
calculation - that the shape of the resulting survival vs. depth profile is
strongly affected by the values of cellular parameters applied, but it also
demonstrates the likely difficulty in finding optimum conditions in carbon
therapy to correctly treat tumour cells (here represented by anoxic V79 cell
parameters) and healthy tissue cells (aerated V79 cells) – as shown in Fig.
5.11.

The inverse planning procedure, recognized usually as inverse planning is a
crucial component of any Treatment Planning System in hadron therapy. All
present treatment planning systems incorporate inverse planning procedures \citep{Scholz_ref62}. 
As most of the radiobiological models exploit the linear-quadratic
dose response model, biological dose optimization algorithms also follow this
approach by incorporating alpha-beta formalism within the optimization
procedure. The optimization procedure presented in this thesis is based on a
different, multi-target dose-response model. The differences between m-target
and alpha-beta models were widely discussed in the literature \citep{katz2009parameter_ref67}, \citep{scholz2004physical_ref68}.  The
li\-ne\-ar-quad\-ra\-tic approach has wider acceptance in clinical practice, but there
is yet no solid proof of the superiority of one approach over another.

In the core of developed cell survival optimization algorithm lays the
non-linear minimization problem. In this work it can however be easily tackled
as Katz’s scaled model is fully analytical. As may be seen in Fig. 5.3, the
survival optimization algorithm converged in about 30 steps and relative
deviations of cell survival over the target region did not exceed 2.5 \%, an
acceptable level, with room for further improvement.

The calculation tool developed in Chapter 5 can then be accepted as the
one-di\-men\-sio\-nal kernel of a therapy planning system based on Katz’s cellular
track structure theory, to be further compared against systems based on other
biophysical models, such as LEM.

The cell survival optimization algorithm presented in this thesis is a step
towards implementing the Katz model in a Treatment Planning System. Availability
of the open-source code within the libamtrack library may facilitate the
implementation of the developed kernel in a realistic TPS.

The clinical situations in many aspects are far more complex than the presented
model calculation of a one-dimensional passively shaped beams in a homogenous
environment. TPS algorithms are required to handle many aspects which remain
outside the scope of this thesis: multiple field optimization, complex patient
geometry or plan robustness.

The presented approach is limited to one dimension (along the beam depth). It
would be possible to extend it to a full 3-D beam model by repeating the Monte
Carlo simulations of carbon ion beams with increased statistics and to save such
data in a look-up table indexed not by depth z only, but by position (x,y) at
depth z. In such an approach various configurations could be studied: pristine
Bragg peaks emitted from a point-source, mono-energetic pencil beams, clinical
pencil beams and also broad beams of carbon ions. However, Kramer and Scholz
\citep{Rapid_ref63} reported a much simpler approach, incorporated in the TRiP98 treatment
planning system, namely that carbon ion pencil beam energy-fluence spectra are
tabulated along the Z-axis and an analytical Gaussian function is used to
describe the beam profile in X and Y-axes. Such a beam model was applied in the
GSI raster scan system and is presumably used in the clinical TPS at HIT.
Analytical beam models used to describe pencil beams 
 appear to describe well the total dose deposited by the beam,
but their consistency in representing beam fragmentation has not yet been
proven.

Full three-dimensional beam models are used at the Heidelberg Ion Therapy Center
(HIT) where a FLUKA Monte-Carlo code coupled with LEM \citep{FlukaHIT_ref64} has been integrated
within their treatment planning system. Using the FLUKA code, energy-fluence
spectra calculated at HIT are saved into plain files in SPC format and used in
the treatment planning system. The choice of SPC file as energy-fluence spectra
file format in libamtrack library was made to facilitate possible integration of
this library with existing treatment planning system (TRiP98, Syngo PT planning)
and Monte Carlo codes (SHIELD-HIT, FLUKA). Such integration would allow direct
comparison of treatment plans prepared using different radiobiological models.

\section{Summary and conclusions}
The general aim of this work was to develop and test the basic algorithms of a
kernel of a future therapy planning system for carbon ion radiotherapy, using in
its radiobiology component the cellular track structure model of Katz and
applying as its physical component a realistic Monte Carlo-generated data base
describing transport in water of carbon beams of various initial energies,
available to the author. Using this data set it was possible to simulate the
formation of the spread-out Bragg peak structure and to evaluate the
energy-fluence spectra of all generations of secondary ions up to the energy of
the primary carbon ions, at all beam depths. It was desirable to gather all
necessary codes and data in an open-source research code library

For this purpose, the author, in collaboration with Steffen Greilich and other
colleagues at the DKFZ and Aarhus research centres, developed a computer library
of codes - the libamtrack library. This open-source library is generally
available to all users. Subroutines of the library can be downloaded, edited or
modified and incorporated in other software. Together with the library,
implemented in ANSI C language, a set of wrapping methods is provided, making it
possible to use it in various computing languages or in numerical simulation
tools. Some of the library functions have now been incorporated into a web
interface, the libamtrack WebGUI \footnote{ \verb|http://webgui.libamtrack.dkfz.org/test|},
where users connected to the Internet can perform some basic calculations using
their web browsers. The libamtrack library is presented in Appendix C.

The author verified several aspects of the Katz model, notably the relationship
between the “integrated” version of the model which requires sequences of
numerical integrations to calculate its output, and its much faster analytical
or “scaled” version which exploits the scaling properties of this model with
respect to some of its parameters. In some parts of work in this area, the
author collaborated with Marta Korcyl \citep{KorcylPhD_ref45}. The author also performed a detailed
analysis of the parameter-fitting procedures in the “integrated” and analytical
representation of the Katz model, developing his algorithm of a fitting
procedure and implementing it as a tool in the libamtrack library.

Next, the author verified the consistency of the physical carbon ion beam model
to ensure that it reflected the major physical processes of interactions of
energetic carbon ions with water. SHIELD-HIT Monte Carlo transport calculations
of carbon ions in water were originally performed by Pablo Botas in
collaboration with DKFZ, Heidelberg. The author adapted the result data sets to
be handled by the libamtrack library routines and co-developed the algorithms of
data extraction from these data sets, to be used as input for Katz’s cellular
track structure model calculations and for handling and presenting results of
these calculations.

As a benchmark of the carbon TPS elements under development, an algorithm to
optimise beam properties in order to obtain constant levels of survival over the
required depth was developed and implemented by the author. Here, the
radiobiological model and beam model had both to be optimized to work correctly
in the minimization algorithm. As a result of these studies, the author
developed and implemented a general tool for adjusting the parameters of a
one-dimensional carbon beam in such a manner that a pre-selected constant
survival level could be achieved over a given range of beam depths.

The consistency of the physical and radiobiological components of the developed
TPS elements was next verified by the author against published results of a
radiobiological experiment involving measurement of the survival levels of
Chinese Hamster Ovary (CHO) cells placed at different beam depths and irradiated
by a pre-designed set of carbon beams of energies ranging between 196 and 244
MeV/amu \citep{mitaroff1998biological_ref66}. The planned level of survival was 20\% over 8-12
cm depths, and the experiment was designed to verify an earlier version of the
LEM. By fitting the Katz model cellular parameters to this cell line and
applying the benchmark optimisation calculations, the author was able to
consistently represent the results of this experiment to within 15\% relative
difference, well in agreement with the results of LEM calculations.

Finally, using the tools developed by the author, the effect of varying the
input beam fluence and varying the cellular parameters in the Katz model to
represent aerated (healthy tissues) or hypoxic (tumour cells), in a study of the
respective survival-depth dependences, the author showed in a predictive manner
the difficulties which may arise in achieving correct optimisation of such
dependences.\newline

The following overall conclusions can be drawn:

\begin{itemize}
 \item The general objective of this work - to develop and test the basic algorithms of
a kernel of a future therapy planning system for carbon ion radiotherapy, using
in its radiobiology component the cellular track structure model of Katz and
applying as its physical component a Monte Carlo-generated data base describing
transport in water of carbon beams of various initial energies, available to the
author - was successfully accomplished.
\item In the course of this work, the author proposed improvements to the Katz model,
derived algorithms required to model the survival of cells in vitro by a
realistic carbon beam propagating through water and derived optimisation
routines required to achieve a pre-designed depth survival profile by the
inverse planning approach. Efficient optimization algorithms for achieving
desired depth-dose and survival-depth distributions were developed.
\item All codes developed by author in the course of this work have been implemented 
in the freely accessible libamtrack code library. 
\item The basic kernel algorithm was successfully verified against published
experimental data. Results of the author’s calculations were found to somewhat
differ from published results of LEM-based calculations. These differences may
reflect differences in modelling within the radiobiology or the physical
components of the Katz- and LEM-based approaches.
\item While the developed kernel of the carbon ion therapy planning system is
one-di\-men\-sio\-nal only, it can be useful as a tool for predicting the likely
outcome of various beam configurations and of irradiating various cell types, as
represented by their sets of radiosensitivity parameters of the Katz model.
\item The one-dimensional TPS kernel developed in this thesis could be further
extended to a 3-D calculation for use in realistic 3-D therapy planning systems
in carbon ion radiotherapy.
\end{itemize}

\section{Future Work}
Many avenues could be followed in continuing this thesis:
A comprehensive database of the radiobiological data was published \citep{sorensen2011vitro_ref78},
presented in the form of the alpha and beta coefficients, fitted using
linear-quadratic model. Such data could serve as an initial stage for studies of
the Katz model predictions for other cells or endpoints than CHO or V79 cells of
different oxygenation status, studied in this work.

A very interesting and promising area is in inter-comparison studies of various
radiobiological models: Katz, LEM and other models. Only  a limited number of
paper has been published dealing with this issue and no reasonable conclusions
have yet been drawn.

The approach presented in this work was limited only to carbon ions. It would be
interesting to follow the idea of Cucinotta to calculate cell
survival in proton beams using the Katz model approach (scaled or integrated)
and the latest versions of LEM. Such comparisons could provide better
understanding of the applicability of constant RBE equal to 1.1 in proton
radiotherapy. Exploiting the predictive power of the Katz model, predictions of
ion radiotherapy using ions lighter than carbon (He, Li, Be, B or N) could also
be studied.

The scaled version of the Katz model can be based only on radial dose
distribution formulae which have particular  “scaling properties”. In the
integrated version of the Katz model any radial dose distribution formula may be
applied, such as e.g., Geiss’s $D(r)$ formula, also used in LEM. At this stage
there are at least three features of the Katz model in which it differs from
LEM: radial dose distribution, description of the reference radiation survival
profile ($m$-target model) and the concept of inactivation cross-section. By using
Geiss’s $D(r)$ formula in the integrated version of Katz’s model, the number of
distinctive features could be reduced to two, making inter-comparisons easier.
One could also go one step further and make more sophisticated changes to the
Katz model, e.g., replacing the m-target approach by the linear-quadratic
formulation. In this case, a scaled and fast version of the model might be
difficult to construct, but expressing the prediction in linear quadratic
formalism could make it more appealing for physicians.
The Katz model is thought to incorrectly predict cell survival and detector
response for lighter ions, such as protons or alpha particles, a fault also
shared by LEM. This discrepancy might be related to the radial dose distribution
model incorporated in these models. Further improvement in this area is still
possible, but requires more detailed studies. It is also important because among
fragmentation products produced by carbon ion beam protons have the highest
fluence.

Finally, some technical work could be performed to further improve the
efficiency of the present algorithms contained in the libamtrack library, to
bring them up to industry standards.

\mgrclosechapter

\chapter*{Appendix A (Radial Dose distribution and Electron range models)}
\section*{Delta electron range formulae}

Delta electron range $r_{\max}$ models implemented in the libamtrack library are summarized in table \ref{tab:ch2_ermodels}. 
Here a detailed listing of formulae is presented. The maximum delta electron range $r_{\max}$ can be expressed 
as a function of the ion energy $E$, or as a function of maximum delta electron energy $\omega$.

\subsection*{Formulae of Geiss and Scholz}

Two models, described as Geiss and Scholz encompass ion energy $E$, but yield different coefficients for Geiss:

$$
r_{\max} = 4 \cdot 10^{-5} \left(\frac{\text{E}}{\text{MeV}}\right)^{1.5} \text{cm}
$$

than for Scholz:

$$
r_{\max} = 5 \cdot 10^{-6} \left(\frac{\text{E}}{\text{MeV}}\right)^{1.7} \text{cm}
$$

\subsection*{Formula of Butts and Katz}

The model of Buttz and Katz shows a linear dependence on the delta electron energy $\omega$:

$$
r_{\max} = 10^{-6} \frac{\omega}{\text{keV}} \text{cm}
$$

\subsection*{Formula of Waligórski}

Model of Waligórski shows a power dependence on the delta electron energy $\omega$,
with exponent $\alpha$ which is taken to be 1.079 for $\omega$ < 1keV and 1.667 elsewhere:

$$
r_{\max} = 6 \cdot 10^{-6} \left(\frac{\omega}{\text{keV}}\right)^{\alpha} \text{cm}
$$

\subsection*{Formula of Tabata}

Tabata's formula incorporates a more complicated dependence of $r_{\max}$ on $\omega$:

$$
r_{\max}  = a_1 \left( 1/a_2 \ln ( 1 +  a_2 \omega / mc^2) - \frac{a_3 \omega /mc^2}{1 + a_4( \omega/mc^2)^{a_5}} \right)
$$

where:

\begin{eqnarray}
&& a_1 = b_1 A / Z^{b_2} \nonumber\\
&& a_2 = b_3 Z\nonumber\\
&& a_3 = b_4 - b_5 Z \nonumber\\
&& a_4 = b_6 - b_7 Z\nonumber\\
&& a_5 = b_8 / Z^{b_9}\nonumber
\end{eqnarray}

and $b_i$ are constants dependent on the material in which range is calculated. 
If the material is a mixture of chemical elements, then $Z$ and $A$ need to be exchanged by average values. 
Values of the $b_i$ coefficients are as follow:

\begin{eqnarray}
&& b_1 = 0.2335 \left[\frac{\text{g}}{\text{cm}^2}\right] \nonumber\\
&& b_2 = 1.209 \nonumber\\
&& b_3 = 1.78 \cdot 10^{-4} \nonumber\\
&& b_4 = 0.9891 \nonumber\\
&& b_5 = 3.01 \cdot 10^{-4} \nonumber\\
&& b_6 = 1.468 \nonumber\\
&& b_7 = 1.18 \cdot 10^{-2} \nonumber\\
&& b_8 = 1.232 \nonumber\\
&& b_9 = 0.109 \nonumber
\end{eqnarray}

\section*{Radial dose distribution formulae}

Radial dose distribution formulae implemented in the libamtrack library are summarized in table 1.2. 
Here a detailed listing of these formulae is presented.

\subsection*{Formula of Zhang}

$$
D(r) = C_1 \frac{z^{\star 2}}{\beta^2}\frac{1}{\alpha}\frac{1}{r} \frac{1}{r+\theta(I)}\left(1-\frac{r+\theta(I)}{r_{\max}+\theta(I)}\right)^{\alpha^{-1}}
$$

Where $C_1 = N e^4 / mc^2 (4 \pi \varepsilon_0)^2$ ($N$ - electron density of the material, $e$ - electron charge, $m$ - electron mass, $c$ - speed of light, $\varepsilon_0$ - electrical permittivity of vacuum).
$\theta(I)$ is the range of delta electrons of energy equal to ionization potential $I$.

\subsection*{Formula of Katz}

Applying in Zhang's formula $\alpha = 1$ and $I = 0$, one obtains Katz's formula:

$$
D(r) = C_1 \frac{z^{\star 2}}{\beta^2}\frac{1}{r} \left(\frac{1}{r}-\frac{1}{r_{\max}}\right)
$$

\subsection*{Formula of Geiss}

$$
D(r)=\begin{cases}C_2&\text{if $0<r<a_0$,}\\
\frac{C_2}{r^2}&\text{if $a_0\le r\le r_{\max}$}\\0&\text{elsewhere}\end{cases}
$$

In the formula of Geiss, the $C_2$ constant is taken to such value that average total dose deposited around single track yields stopping power value:

$$
C_2 = \frac{L}{\pi \rho \left( a_0^2 + 2 \ln\left(r_{\max}{a_0}\right) \right)}
$$

\subsection*{Formula of Cucinotta}

$$
D(r)=\begin{cases} C_1 \frac{z^{\star 2}}{\beta^2} f_S(r) f_L(r) \frac{1}{r^2} + C_3 \frac{\exp(-r/2d)}{r^2}&\text{if $0 \le r \le r_{\max}$}\\0&\text{elsewhere}\end{cases}
$$

where:

$$
f_S(r) = \left( \frac{r_0}{r} + (0.6 + 1.7 \beta + 1.1 \beta^2) \right)^{-1}
$$

where $r_0 = 1 \text{nm}$, and

$$
f_L(r) = \exp\left(- \frac{r^2}{ (0.37 \cdot r_{\max})^2} \right)
$$

and

$$
d = \frac{\beta}{2} \frac{h c}{2 \pi \omega_r}
$$

where:

$$
\omega_r = 13 \text{eV}
$$

In a manner similar to that in Geiss's formula, the $C_3$ constant is taken to such value that $\rho \int_0^{r_{\max}} 2 \pi r D(r) dr = L$ holds:

$$
C_3 = \frac{ \frac{L}{2 \pi \rho} - C_1 \frac{z^{\star 2}}{\beta^2}  \int_0^{r_{\max}} f_S(r) f_L(r) \frac{dr}{r} }{ \int_0^{r_{\max}} \exp(-r/2d) \frac{dr}{r} }
$$

\mgrclosechapter

\chapter*{Appendix B (Extended target calculations)}

Knowing the formula for the dose $D(r)$ delivered by delta electrons at a point at a distance $r$ from the 
ion track one may also write the formula for averaged dose $D_{\text{ext}}(t,a_0)$ delivered in a thin cylindrical volume 
of radius $a_0$ at a distance $t$ from the ion track. 
We will refer to the circle $S_t$ with radius $a_0$, at a distance $t$ from the ion track as the target. 
By neglecting volume thickness one may reduce this problem to 2-dimensional integration:

$$
D_{e} (t,a_0) = \frac{1}{|S_t|} \iint_{S_t} D(x,y) dx dy
$$

where $D(x,y)$ is the dose delivered to the point with coordinates $(x,y)$ 
( due to rotational symmetry one could easily calculate it as $D( \sqrt{x^2 + y^2} )$

By changing coordinates from Cartesian to polar, obtains:

$$
D_{e} (t,a_0) = \frac{1}{\pi a_0^2} \int_{t_{\min}}^{t_{\max}} D(r) \Phi( r, t, a_0)  dr
$$

Here $\Phi( r, t, a_0)$ denotes the length of an arc segment, centered around the ion track, 
of radius $r$, contained in a circle of radius $a_0$ at the distance $t$ from the ion track. 
$t_{\min}$ is the minimum distance of the ion track to the border of the target, which is equal to $t-a_0$ if the ion 
track is outside the target, or assumed to be 0 is ion track is inside or on the border of the target.
$t_{\max}$ is the maximum distance from the ion track to the border of the target, which is equal to $t+a_0$.

$\Phi$ could be calculated using following formula:

\begin{equation*}
\Phi( r, t, a_0 ) = \begin{cases} 2 \arctan \sqrt{ \frac{a_0^2 - (t-r)^2}{(r+t)^2-a_0^2} } & \text{if $r > |t-a_0|$,}
\\
\pi &\text{if $r \leq |t-a_0|$.}
\end{cases}
\end{equation*}

\mgrclosechapter

\chapter*{Appendix C (Software)}

The \emph{libamtrack} library, as an open-source project, is available for download from the webpage
\verb"libamtrack.dkfz.org". Detailed description of routines provided by the library is available 
in the reference manual provided on the project webpage. 
The libamtrack source codes, together with the set of scripts and database of the carbon ion 
beam spectra are also attached to this thesis and grouped in four folders:

\subsection*{libamtrack}
The source code of the libamtrack library is provided in the libamtrack directory. It contains:
\begin{itemize}
 \item compilation instructions
 \item \verb"src" subdirectory with source files (*.c) containing definition of all routines
 \item \verb"include" subdirectory with header files (*.h) containing declaration of all routines and documentation in doxygen format
 \item \verb"example" subdirectory with two sample codes written in C, showing usage of the libamtrack library
 \item \verb"wrapper" subdirectory with interface to the libamtrack library for Python language and R library
\end{itemize}

The \emph{libamtrack} library does not provide any executable file, as it was designed as a library -
a set of routines which can be invoked from any code provided by user. 
It can be compiled as a shared library under Linux and Windows operating system. 
Two sample codes are provided, which use the libamtrack library and can be compiled to an executable file: 
one shows how energy of the particle is calculated from its relative velocity beta and the second one 
produce an output which later can be used to prepare plots of electron range, radial dose distribution 
and stopping power for various configurations of the formulae used, particle and target material.
The easiest way to get familiar with the library is to use the R package. 
After installation of R one may easily install the libamtrack plugin, which enables user to use selected set of 
functions from libamtrack in the R environment. 
A sample session showing the usage of the R plugin by calculation of the radial dose distribution and 
maximum energy transfer to the delta electron:

\begin{verbatim}
> library("libamtrack")
This is libamtrack 0.5.3 'Green Wombat' (2012-04-27).
Type '?libamtrack' for help.
>      # Compute dose in several distances (from 1e-9 to 1e-4 m) of an 100 MeV/u 
>      # proton in water according to 'Cucinotta' distribution
>      AT.D.RDD.Gy(    r.m              = 10^(-9:-4),
+                      E.MeV.u          = 100,
+                      particle.no      = 60012,
+                      material.no      = 1,
+                      rdd.model        = 7,
+                      rdd.parameter    = c(1e-10, 1e-10),
+                      er.model         = 5,
+                      stopping.power.source.no = 1)$D.RDD.Gy
[1] 9.856050e+06 4.513804e+04 2.586443e+02 2.600422e+00 2.485355e-02
[6] 2.537419e-06
attr(,"Csingle")
[1] TRUE
>   # maximum energy transferred to delta electron by a 100 MeV/amu particle
> AT.max.E.transfer.MeV(E.MeV.u=100)$max.E.transfer.MeV
[1] 0.2309850
attr(,"Csingle")
[1] TRUE
\end{verbatim}

\subsection*{fitting-katz-cell-survival}
The script which aids in finding Katz model free parameters for which model prediction fits data best.
This script can be used by executing “find.py” file. The configuration is stored in the file “fit.cfg” and limited to the following items:
\begin{itemize}
\item input data folder with cell survival curves data
\item radial dose distribution formula to be used in Katz model
\item precision of the fitting algorithm
\end{itemize}
The output of the calculation will be stored in a separate directory, containing files with calculated parameters and data necessary to produce survival curve plots.

Codes performing necessary calculation are gathered in src directory.

Together with script example configuration and cell survival data is provided.

\subsection*{spc}
“spc” folder contains sample files with energy-fluence spectra of the carbon ion beam

\subsection*{carbon-sobp}
carbon-sobp contains scripts for finding coefficients of the linear combination of the Bragg peaks which gives certain dose or survival profile.
This tool can be used by executing “plot.py” script. All necessary input parameters need to be provided in a setup.cfg configuration file (an example configuration file is provided together with codes):
path to the folder with energy-fluence spectra files 
\begin{itemize}
\item range on which given profile is to be obtained
\item number of pristine Bragg peaks in the linear combination
\item choice of desired profile: either dose or survival
\item coefficients of polynomial defining desired profile
\item Katz model parameters (needed if survival is calculated)
\item parameters of grid on which profile accuracy is calculated
\item precision of minimization algorithm
\end{itemize}
The output of the calculation will be stored in a separate directory, containing files with calculated coefficients and data necessary to produce plots of dose (or survival profiles).

Codes performing necessary calculation are gathered in the src directory.

\mgrclosechapter


\bibliography{thesis}

\begin{thebibliography}{}

\bibitem[Agostinelli et~al., 2003]{Geant4_ref26}
Agostinelli, S., Allison, J., Amako, K.~e., Apostolakis, J., Araujo, H., Arce,
  P., Asai, M., Axen, D., Banerjee, S., Barrand, G., et~al. (2003).
\newblock Geant4-a simulation toolkit.
\newblock {\em Nuclear instruments and methods in physics research section A:
  Accelerators, Spectrometers, Detectors and Associated Equipment},
  506(3):250--303.

\bibitem[Andersen and Ziegler, 1977]{andersenZiegler_ref29}
Andersen, H. and Ziegler, J. (1977).
\newblock {\em Hydrogen stopping powers and ranges in all elements}.
\newblock Stopping and ranges of ions in matter. Pergamon Press.

\bibitem[{Attili} et~al., 2008]{attiliLEM_ref55}
{Attili}, A., {Russo}, G., {Marchetto}, F., {Bourhaleb}, F., {Ansarinejad}, A.,
  {Cirio}, R., {Cirrone}, P., {Donetti}, M., {Garella}, A., {Givehchi}, N.,
  {Giordanengo}, S., {Monaco}, V., {Pardo}, J., {Pecka}, A., {Peroni}, C.,
  {Rinaldi}, I., and {Sacchi}, R. (2008).
\newblock Evaluation of radiobiological effects of carbon ion beams: Mixed
  particle fields and fragmentation.
\newblock {\em Medical Physics}, 35:2935.

\bibitem[Barkas and Evans, 1963]{barkas1963nuclear}
Barkas, W. and Evans, D. (1963).
\newblock {\em Nuclear Research Emulsions: Techniques and theory}.
\newblock Pure and applied physics. Academic Press.

\bibitem[Bassler et~al., 2010]{bassler2010dose_ref70}
Bassler, N., J{\"a}kel, O., S{\o}ndergaard, C.~S., and Petersen, J.~B. (2010).
\newblock Dose-and {LET}-painting with particle therapy.
\newblock {\em Acta Oncologica}, 49(7):1170--1176.

\bibitem[{Battistoni} et~al., 2007]{Fluka_ref27}
{Battistoni}, G., {Cerutti}, F., {Fass{\`o}}, A., {Ferrari}, A., {Muraro}, S.,
  {Ranft}, J., {Roesler}, S., and {Sala}, P.~R. (2007).
\newblock The {FLUKA} code: description and benchmarking.
\newblock {\em AIP Conf.Proc.}, 896:31--49.

\bibitem[Berger et~al., 2005]{NISTESTAR}
Berger, M.~J., Coursey, J.~S., Zucker, M.~A., and Chang, J. (2005).
\newblock {ESTAR, PSTAR, and ASTAR: Computer Programs for Calculating
  Stopping-Power and Range Tables for Electrons, Protons, and Helium Ions
  (version 1.2.3)}.
\newblock In {\em NISTIR 4999}. National Institute of Standards and Technology.

\bibitem[Beringer and others (Particle Data~Group), 2012]{Passage_ref28}
Beringer, J. and others (Particle Data~Group) (2012).
\newblock Review of particle physics.
\newblock {\em Phys. Rev. D}, 86:010001.

\bibitem[{Bethe}, 1932]{bethe1932}
{Bethe}, H. (1932).
\newblock Bremsformel f{\"u}r elektronen relativistischer geschwindigkeit.
\newblock {\em Zeitschrift fur Physik}, 76:293--299.

\bibitem[Bethe, 1953]{Moliere_ref31}
Bethe, H.~A. (1953).
\newblock Moli\`ere's theory of multiple scattering.
\newblock {\em Phys. Rev.}, 89:1256--1266.

\bibitem[Beuve, 2009]{beuve2009formalization_ref89}
Beuve, M. (2009).
\newblock Formalization and theoretical analysis of the local effect model.
\newblock {\em Radiation research}, 172(3):394--402.

\bibitem[Butts and Katz, 1967]{Butts1967}
Butts, J.~J. and Katz, R. (1967).
\newblock Theory of {RBE} for heavy ion bombardment of dry enzymes and viruses.
\newblock {\em Radiation Research}, 30:855--871.

\bibitem[Chwastowski et~al., 2012]{IFJCloud_ref49}
Chwastowski, J., Grzymkowski, R., Kruk, M., Nabozny, M., Natkaniec, Z.,
  Olszewski, A., Sobocinska, Z., Sosnicki, T., Szostak, M., Syktus, P., et~al.
  (2012).
\newblock The {CC1} project--system for private cloud computing.
\newblock {\em Computer Science}, 13:2.

\bibitem[Cucinotta et~al., 1997]{Cucinotta1997}
Cucinotta, F., Wilson, J.~W., Shavers, M.~R., and Katz, R. (1997).
\newblock Calculation of heavy ion inactivation and mutation rates in radial
  dose model of track structure.
\newblock Technical report, NASA.

\bibitem[Cucinotta et~al., 1999]{cucinotta1999applications_ref87}
Cucinotta, F.~A., Nikjoo, H., and Goodhead, D.~T. (1999).
\newblock Applications of amorphous track models in radiation biology.
\newblock {\em Radiation and environmental biophysics}, 38(2):81--92.

\bibitem[Douglas and Fowler, 1976]{douglas1976}
Douglas, B. and Fowler, J. (1976).
\newblock The effect of multiple small doses of {X} rays on skin reactions in
  the mouse and a basic interpretation.
\newblock {\em Radiat Res}, 66(2):401--26.

\bibitem[Els\"asser et~al., 2008]{Elsasser2008_ref36}
Els\"asser, T., Kr\"amer, M., and Scholz, M. (2008).
\newblock Accuracy of the {L}ocal {E}ffect {M}odel for the prediction of
  biologic effects of carbon ion beams in vitro and in vivo.
\newblock {\em International Journal of Radiation Oncology*Biology*Physics},
  71(3):866 -- 872.

\bibitem[Els\"{a}sser and Scholz, 2007]{Elsasser2007b_ref35}
Els\"{a}sser, T. and Scholz, M. (2007).
\newblock Cluster effects within the local effect model.
\newblock {\em Radiation Research}, 167:319--329.

\bibitem[Els{\"a}sser and Scholz, 2010]{elsasser2010comments_ref90}
Els{\"a}sser, T. and Scholz, M. (2010).
\newblock Comments on "{F}ormalization and theoretical analysis of the local
  effect model" by {B}euve ({R}adiat. {R}es. 172, 394-402, 2009).
\newblock {\em Radiation research}, 173(6):855--856.

\bibitem[Els\"asser et~al., 2010]{Elsasser2010_ref37}
Els\"asser, T., Weyrather, W.~K., Friedrich, T., Durante, M., Iancu, G.,
  Kramer, M., Kragl, G., Brons, S., Winter, M., Weber, K.-J., and Scholz, M.
  (2010).
\newblock Quantification of the relative biological effectiveness for ion beam
  radiotherapy: Direct experimental comparison of proton and carbon ion beams
  and a novel approach for treatment planning.
\newblock {\em International Journal of Radiation Oncology*Biology*Physics},
  78(4):1177 -- 1183.

\bibitem[Fowler, 2010]{fowler201021years_ref73}
Fowler, J. (2010).
\newblock 21 years of biologically effective dose.
\newblock {\em British Journal of Radiology}, 83(991):554--568.

\bibitem[Fowler, 1964]{mtarget_ref54}
Fowler, J.~F. (1964).
\newblock Differences in survival curve shapes for formal multi-target and
  multi-hit models.
\newblock {\em Physics in Medicine and Biology}, 9(2):177.

\bibitem[Fowler, 1989]{fowler1989linear_ref72}
Fowler, J.~F. (1989).
\newblock The linear-quadratic formula and progress in fractionated
  radiotherapy.
\newblock {\em British Journal of Radiology}, 62(740):679--694.

\bibitem[Friedland et~al., 1998]{friedland1998partrac_ref82}
Friedland, W., Jacob, P., Paretzke, H.~G., and Stork, T. (1998).
\newblock Monte {C}arlo simulation of the production of short {DNA} fragments
  by low-linear energy transfer radiation using higher-order {DNA} models.
\newblock {\em Radiation research}, 150(2):170--182.

\bibitem[Furusawa et~al., 2000]{furusawa_ref59}
Furusawa, Y., Fukutsu, K., Aoki, M., Itsukaichi, H., Eguchi-Kasai, K., Ohara,
  H., Yatagai, F., Kanai, T., and Ando, K. (2000).
\newblock Inactivation of aerobic and hypoxic cells from three different cell
  lines by accelerated 3{H}e-, 12{C}- and 20{N}e-ion beams.
\newblock {\em Radiation research}, 154(5):485--496.

\bibitem[Geiss, 1997]{Geiss1997}
Geiss, O.~B. (1997).
\newblock {\em TLDs in Teilchenstrahlen (in German)}.
\newblock PhD thesis, Gesellschaft f\"ur Schwerionenforschung, Darmstadt,
  Germany.

\bibitem[Geiss et~al., 1998]{Geiss1998}
Geiss, O.~B., Kr\"{a}mer, M., and Kraft, G. (1998).
\newblock Efficiency of thermoluminescence detectors to heavy charged
  particles.
\newblock {\em NIM B}, 142:592--598.

\bibitem[Goodhead, 1987]{goodhead1987relationship_ref80}
Goodhead, D.~T. (1987).
\newblock Relationship of microdosimetric techniques to applications in
  biological systems.
\newblock {\em The dosimetry of ionizing radiation}, 2:1--89.

\bibitem[Greilich et~al., 2010]{libamtrack_ref40}
Greilich, S., Grzanka, L., Bassler, N., Andersen, C., and J\"akel, O. (2010).
\newblock Amorphous track models: A numerical comparison study.
\newblock {\em Radiation Measurements}, 45(10):1406 -- 1409.

\bibitem[Greilich et~al., 2013]{greilich2013efficient_ref91}
Greilich, S., Hahn, U., Kiderlen, M., Andersen, C., and Bassler, N. (2013).
\newblock Efficient calculation of local dose distribution for response
  modelling in proton and ion beams.
\newblock {\em arXiv preprint arXiv:1306.0185}.

\bibitem[Grzanka et~al., 2011]{grzanka2011application_ref43}
Grzanka, L., Greilich, S., Korcyl, M., J{\"a}kel, O., Walig{\'o}rski, M., and
  Olko, P. (2011).
\newblock The application of amorphous track models to study cell survival in
  heavy ions beams.
\newblock {\em Radiation Protection Dosimetry}, 143(2-4):232--236.

\bibitem[Gudowska et~al., 2004]{SHIELD-HIT_ref38}
Gudowska, I., Sobolevsky, N., Andreo, P., Belkic, D., and Brahme, A. (2004).
\newblock Ion beam transport in tissue-like media using the {M}onte {C}arlo
  code {SHIELD-HIT}.
\newblock {\em Physics in Medicine and Biology}, 49(10):1933.

\bibitem[Haettner, 2006]{Haettner1997}
Haettner, E. (2006).
\newblock {\em Experimental study on carbon ion fragmentation in water using
  GSI therapy beams}.
\newblock PhD thesis, Kungliga tekniska hogskolan Stockholm, Sweden.

\bibitem[Hall and Giaccia, 2006]{hall2006radiobiology}
Hall, E. and Giaccia, A. (2006).
\newblock {\em Radiobiology for the radiologist}.
\newblock Lippincott Williams \& Wilkins.

\bibitem[Hansen et~al., 2012]{Hansen-SHIELD_ref60}
Hansen, D.~C., L\"{u}hr, A., Sobolevsky, N., and Bassler, N. (2012).
\newblock Optimizing {SHIELD-HIT} for carbon ion treatment.
\newblock {\em Physics in Medicine and Biology}, 57(8):2393.

\bibitem[Hauptner et~al., 2006]{hauptner2006spatial_ref53}
Hauptner, A., Friedland, W., Dietzel, S., Drexler, G., Greubel, C., Hable, V.,
  Strickfaden, H., Cremer, T., Friedl, A., Kr{\"u}cken, R., et~al. (2006).
\newblock Spatial distribution of {DNA} double-strand breaks from ion tracks.
\newblock {\em Matematisk-fysiske Meddelelser}, 52:59--85.

\bibitem[Hawkins, 1998]{MKM_ref50}
Hawkins, R.~B. (1998).
\newblock A microdosimetric-kinetic theory of the dependence of the {RBE} for
  cell death on {LET}.
\newblock {\em Medical Physics}, 25(7):1157--1170.

\bibitem[Herrmann et~al., 2011]{Hermann-libamtrack_ref41}
Herrmann, R., Greilich, S., Grzanka, L., and Bassler, N. (2011).
\newblock Amorphous track predictions in libamtrack for alanine relative
  effectiveness in ion beams.
\newblock {\em Radiation Measurements}, 46(12):1551 -- 1553.

\bibitem[ICRU, 1983]{icru1983report}
ICRU (1983).
\newblock {\em Microdosimetry}.
\newblock Number~36 in ICRU report. International Commission on Radiation Units
  and Measurements.

\bibitem[ICRU, 1994]{icru49protons_ref30}
ICRU (1994).
\newblock {\em Stopping Powers and Ranges for Protons and Alpha Particles}.
\newblock Number~49 in ICRU report. International Commission on Radiation Units
  and Measurements.

\bibitem[ICRU, 2005]{icru73stopping}
ICRU (2005).
\newblock {\em Stopping of ions heavier than helium}.
\newblock Number~73 in ICRU report. International Commission on Radiation Units
  and Measurements.

\bibitem[ICRU, 2011]{icru85units}
ICRU (2011).
\newblock {\em Fundamental quantities and units for ionizing radiation}.
\newblock Number~85 in ICRU report. International Commission on Radiation Units
  and Measurements.

\bibitem[Janni, 1982]{Janni_ref25}
Janni, J.~F. (1982).
\newblock Proton {R}ange-{E}nergy {T}ables, 1 ke{V}-10 {G}e{V}, {E}nergy
  {L}oss, {R}ange, {P}ath {L}ength, {T}ime-of-{F}light, {S}traggling,
  {M}ultiple {S}cattering, and {N}uclear {I}nteraction {P}robability. {P}art
  {I}. {F}or 63 {C}ompounds.
\newblock {\em Atomic data and nuclear data tables}, 27:147.

\bibitem[Katz, 2003]{katz2009parameter_ref67}
Katz, R. (2003).
\newblock The parameter-free track structure model of {S}cholz and {K}raft for
  heavy-ion cross sections.
\newblock {\em Radiation research}, 160(6):724--728.

\bibitem[Katz and Cucinotta, 1999]{KatzTracks_ref48}
Katz, R. and Cucinotta, F. (1999).
\newblock Tracks to therapy.
\newblock {\em Radiation Measurements}, 31(1-6):379--388.
\newblock Proceedings of the 19th International Conference on Nuclear Tracks in
  Solids.

\bibitem[Katz et~al., 1971]{katz1971cellular_ref58}
Katz, R., Sharma, S., and Hamayoonfar, M. (1971).
\newblock Cellular inactivation by heavy ions, neutrons, and pions.
\newblock {\em Robert Katz Publications}, page~72.

\bibitem[Katz and Sharma, 1974]{KatzParamCollection_ref47}
Katz, R. and Sharma, S.~C. (1974).
\newblock Heavy particles in therapy: an application of track theory.
\newblock {\em Physics in Medicine and Biology}, 19(4):413.

\bibitem[Katz et~al., 1972]{Katz1972b}
Katz, R., Sharma, S.~C., and Homayoonfar, M. (1972).
\newblock The structure of particle tracks.
\newblock In Attix, F.~H., editor, {\em Topics in Radiation Dosimetry},
  chapter~6, pages 317--383. Academic Press, New York.

\bibitem[Katz and Varma, 1992]{katz1992radial_ref84}
Katz, R. and Varma, M.~N. (1992).
\newblock Radial distribution of dose.
\newblock In {\em Physical and Chemical Mechanisms in Molecular Radiation
  Biology}, pages 163--180. Springer.

\bibitem[Katz et~al., 1994]{katz1994survey_ref88}
Katz, R., Zachariah, R., Cucinotta, F.~A., and Zhang, C. (1994).
\newblock Survey of cellular radiosensitivity parameters.
\newblock {\em Radiation research}, 140(3):356--365.

\bibitem[Klein et~al., 2011]{Klein20111607_ref42}
Klein, F., Greilich, S., Andersen, C.~E., Lindvold, L.~R., and J{\"a}kel, O.
  (2011).
\newblock A thin layer fiber-coupled luminescence dosimeter based on
  {A}l$_2${O}$_3$:{C}.
\newblock {\em Radiation Measurements}, 46(12):1607--1609.

\bibitem[Kolb, 2010]{Kolb}
Kolb, C. (2010).
\newblock {\it Eine plattformunabh\"angige Benutzerschnittstelle f\"ur
  numerische Programmbibliotheken am Beispiel libamtrack.} {Bachelor's thesis,
  Hochschule Heilbronn}.

\bibitem[Korcyl, 2012]{KorcylPhD_ref45}
Korcyl, M. (2012).
\newblock {\em Track structure modelling for ion radiotherapy}.
\newblock PhD thesis, Jagiellonian University, Cracow, Poland.

\bibitem[Korcyl et~al., 2013]{Korcyl2013_ref85}
Korcyl, M., Grzanka, L., Olko, P., and Walig\'{o}rski, M. (2013).
\newblock On the scaling properties of the radial distribution of dose in
  katz's track structure model applied to 1-hit detectors.

\bibitem[Korcyl and Walig{\'o}rski, 2009]{korcyl2009track_ref86}
Korcyl, M. and Walig{\'o}rski, M.~P. (2009).
\newblock Track structure effects in a study of cell killing in normal human
  skin fibroblasts.
\newblock {\em International journal of radiation biology}, 85(12):1101--1113.

\bibitem[Kr\"amer et~al., 2000]{kramerPlanning_ref65}
Kr\"amer, M., J\"akel, O., Haberer, T., Kraft, G., Schardt, D., and Weber, U.
  (2000).
\newblock Treatment planning for heavy-ion radiotherapy: physical beam model
  and dose optimization.
\newblock {\em Physics in Medicine and Biology}, 45(11):3299.

\bibitem[Kr\"amer and Scholz, 2000]{Scholz_ref62}
Kr\"amer, M. and Scholz, M. (2000).
\newblock Treatment planning for heavy-ion radiotherapy: calculation and
  optimization of biologically effective dose.
\newblock {\em Physics in Medicine and Biology}, 45(11):3319.

\bibitem[Kr\"{a}mer and Scholz, 2006]{Rapid_ref63}
Kr\"{a}mer, M. and Scholz, M. (2006).
\newblock Rapid calculation of biological effects in ion radiotherapy.
\newblock {\em Physics in Medicine and Biology}, 51(8):1959.

\bibitem[Landau, 1944]{Landau1944_ref32}
Landau, L. (1944).
\newblock On the energy loss of fast particles by ionization.
\newblock {\em J.Phys.(USSR)}, 8:201--205.

\bibitem[L\"uhr et~al., 2012]{ArminSPC_ref61}
L\"uhr, A., Hansen, D.~C., Teiwes, R., Sobolevsky, N., J\"akel, O., and
  Bassler, N. (2012).
\newblock The impact of modeling nuclear fragmentation on delivered dose and
  radiobiology in ion therapy.
\newblock {\em Physics in Medicine and Biology}, 57(16):5169.

\bibitem[Mairani et~al., 2010]{FlukaHIT_ref64}
Mairani, A., Brons, S., Cerutti, F., Fasso, A., Ferrari, A., Kr\"amer, M.,
  Parodi, K., Scholz, M., and Sommerer, F. (2010).
\newblock The {FLUKA} {M}onte {C}arlo code coupled with the local effect model
  for biological calculations in carbon ion therapy.
\newblock {\em Physics in Medicine and Biology}, 55(15):4273.

\bibitem[Mitaroff et~al., 1998]{mitaroff1998biological_ref66}
Mitaroff, A., Kraft-Weyrather, W., Gei{ss}, O., and Kraft, G. (1998).
\newblock Biological verification of heavy ion treatment planning.
\newblock {\em Radiation and environmental biophysics}, 37(1):47--51.

\bibitem[Olko, 2002]{Olko2002_ref69}
Olko, P. (2002).
\newblock {\em Microdosimetric Modelling of Physical and Biological Detectors}.
\newblock Habilitation, Institute of Nuclear Physics, Krak\'ow, Poland.

\bibitem[Paganetti and Goitein, 2001]{Paganetti2001}
Paganetti, H. and Goitein, M. (2001).
\newblock Biophysical modeling of proton radiation effect based on amorphous
  track models.
\newblock {\em Int J Radiat Biol}, 77(9):911--928.

\bibitem[Paretzke, 1987]{paretzke1987radiation_ref79}
Paretzke, H. (1987).
\newblock Radiation track structure theory.
\newblock {\em Kinetics of nonhomogeneous processes}, pages 89--170.

\bibitem[Paretzke et~al., 1974]{paretzke1974moca_ref83}
Paretzke, H., Leuthold, G., Burger, G., and Jacobi, W. (1974).
\newblock Approaches to physical track structure calculations.
\newblock In {\em Fourth Symposium on Microdosimetry}, pages 123--140.

\bibitem[Paretzke, 1973]{paretzke1973comparison_ref81}
Paretzke, H.~G. (1973).
\newblock Comparison of track structure calculations with experimental results.
\newblock In {\em 4th symposium on microdosimetry}.

\bibitem[Pirruciello and Tobias, 1980]{bevelac1980_ref75}
Pirruciello, M. and Tobias, C. (1980).
\newblock Biological and medical research with accelerated heavy ions at the
  bevelac.
\newblock Report LBL-11220, Berkeley, CA: Lawrence Berkeley Laboratory.

\bibitem[Pshenichnov et~al., 2005]{Pshenichnov2005}
Pshenichnov, I., Mishustin, I., and Greiner, W. (2005).
\newblock Neutrons from fragmentation of light nuclei in tissue-like media: a
  study with the {GEANT4} toolkit.
\newblock {\em Physics in Medicine and Biology}, 50(23):5493.

\bibitem[Puck and Marcus, 1956]{puck1956action_ref76}
Puck, T.~T. and Marcus, P.~I. (1956).
\newblock Action of {X}-rays on mammalian cells.
\newblock {\em The Journal of experimental medicine}, 103(5):653--666.

\bibitem[Roth and Katz, 1980]{roth1980heavy_ref74}
Roth, R.~A. and Katz, R. (1980).
\newblock Heavy ion beam model for radiobiology.
\newblock {\em Radiation Research}, 83(3):499--510.

\bibitem[Russo, 2007]{RussoPhD_ref52}
Russo, G. (2007).
\newblock {\em Effetti biologici dell'irraggiamento con ioni carbonio}.
\newblock PhD thesis, Politecnico di Torino, Torino, Italy.

\bibitem[Scholz, 2001]{Scholz2001a}
Scholz, M. (2001).
\newblock {\em Grundlagen der biologischen Bestrahlungsplanung f\"ur die
  Schwerionen-Tumortherapie}.
\newblock Habilitation, Ruprecht-Karls-Universit\"at, Heidelberg, Germany.

\bibitem[Scholz and Kraft, 1994]{Scholz1994_ref34}
Scholz, M. and Kraft, G. (1994).
\newblock Calculation of heavy ion inactivation probabilities based on track
  structure, x ray sensitivity and target size.
\newblock {\em Radiation Protection Dosimetry}, 52(1-4):29--33.

\bibitem[Scholz and Kraft, 2004]{scholz2004physical_ref68}
Scholz, M. and Kraft, G. (2004).
\newblock The physical and radiobiological basis of the local effect model: {A}
  response to the commentary by {R}. {K}atz.
\newblock {\em Radiation research}, 161(5):612--620.

\bibitem[S{\o}rensen et~al., 2011]{sorensen2011vitro_ref78}
S{\o}rensen, B.~S., Overgaard, J., and Bassler, N. (2011).
\newblock In vitro {RBE}-{LET} dependence for multiple particle types.
\newblock {\em Acta Oncologica}, 50(6):757--762.

\bibitem[Tabata et~al., 1972]{Tabata1972}
Tabata, T., Ito, R., and Okabe, S. (1972).
\newblock Generalized semiempirical equations for the extrapolated range of
  electrons.
\newblock {\em Nuclear Instruments and Methods}, 103(1):85 -- 91.

\bibitem[Varma and Baum, 1980]{Varma1980Neon}
Varma, M.~N. and Baum, J.~W. (1980).
\newblock Energy {D}eposition in {N}anometer {R}egions by 377 {M}e{V}/{N}ucleon
  ${}^{20}$ {N}e {I}ons.
\newblock {\em Radiation Research}, 81:355--363.

\bibitem[Vavilov, 1957]{Vavilov1957_ref33}
Vavilov, P. (1957).
\newblock Ionization losses of high-energy heavy particles.
\newblock {\em Sov.Phys.JETP}, 5:749--751.

\bibitem[Walig\'{o}rski et~al., 1986]{Waligorski1986_ref44}
Walig\'{o}rski, M., Hamm, R., and Katz, R. (1986).
\newblock The radial distribution of dose around the path of a heavy ion in
  liquid water.
\newblock {\em International Journal of Radiation Applications and
  Instrumentation. Part D. Nuclear Tracks and Radiation Measurements},
  11(6):309 -- 319.

\bibitem[Wambersie et~al., 1994]{wambersie1994development_ref77}
Wambersie, A., Richard, F., and Breteau, N. (1994).
\newblock Development of fast neutron therapy worldwide: Radiobiological,
  clinical and technical aspects.
\newblock {\em Acta oncologica}, 33(3):261--274.

\bibitem[Weyrather et~al., 1999]{Weyrather1999_ref24}
Weyrather, K., Ritter, S., Scholz, M., and Kraft, G. (1999).
\newblock {RBE} for carbon track-segment irradiation in cell lines of differing
  repair capacity.
\newblock {\em International Journal of Radiation Biology}, 75(11):1357--1364.

\bibitem[Weyrather and Kraft, 2004]{Weyrather_ref51}
Weyrather, W. and Kraft, G. (2004).
\newblock {RBE} of carbon ions: {E}xperimental data and the strategy of {RBE}
  calculation for treatment planning.
\newblock {\em Radiotherapy and Oncology}, 73, Supplement 2(0):S161 -- S169.

\bibitem[Wilson et~al., 1946]{wilson1946radiological_ref71}
Wilson, R.~R. et~al. (1946).
\newblock Radiological use of fast protons.
\newblock {\em Radiology}, 47(5):487--491.

\bibitem[Zhang et~al., 1985]{Zhang1985}
Zhang, C., Dunn, D., and Katz, R. (1985).
\newblock Radial distribution of dose and cross-sections for the inactivation
  of dry enzymes and viruses.
\newblock {\em Radiation Protection Dosimetry}, 13:215--218.

\bibitem[Zhu et~al., 1997]{L-BFGS-B_ref46}
Zhu, C., Byrd, R.~H., Lu, P., and Nocedal, J. (1997).
\newblock Algorithm 778: {L-BFGS-B}: {F}ortran subroutines for large-scale
  bound-constrained optimization.
\newblock {\em ACM Trans. Math. Softw.}, 23(4):550--560.

\end{thebibliography}

\listoftables

\listoffigures

\end{document}